\documentclass{emulateapj}
\usepackage{apjfonts}

\slugcomment{Accepted by The Astrophysical Journal; June 19, 2008}

\shorttitle{Neutrino Cross Sections for Supernovae}
\shortauthors{Yoshida et al.}

\begin{document}


\title{
Neutrino-Nucleus Reaction Cross Sections for Light \\
Element Synthesis in Supernova Explosions
}


\author{Takashi Yoshida\altaffilmark{1}, Toshio Suzuki\altaffilmark{2,3},
Satoshi Chiba\altaffilmark{4}, Toshitaka Kajino\altaffilmark{5,6},
Hidekazu Yokomakura\altaffilmark{7}, Keiichi Kimura\altaffilmark{7}, 
Akira Takamura\altaffilmark{8}, Dieter H. Hartmann\altaffilmark{9}}

\altaffiltext{1}{National Astronomical Observatory of Japan, 
2-21-1 Osawa, Mitaka, Tokyo 181-8588, Japan; 
e-mail:takashi.yoshida@nao.ac.jp}
\altaffiltext{2}{Department of Physics, College of Humanities and Science,
Nihon University, Sakurajosui 3-25-40, Setagaya-ku, Tokyo 156-8550, Japan}
\altaffiltext{3}{Center for Nuclear Study, University of Tokyo,
Hirosawa, Wako-shi, Saitama 351-0198, Japan}
\altaffiltext{4}{Advanced Science Research Center, Japan Atomic
Energy Agency, 2-4 Shirakata-shirane, Tokai, Ibaraki 319-1195, Japan}
\altaffiltext{5}{National Astronomical Observatory, and The Graduate 
University for Advanced Studies, 2-21-1 Osawa, Mitaka, Tokyo 181-8588, Japan}
\altaffiltext{6}{Department of Astronomy, Graduate School of Science,
University of Tokyo, 7-3-1 Hongo, Bunkyo-ku, Tokyo 113-0033, Japan}
\altaffiltext{7}{Department of Physics, Graduate School of Science,
Nagoya University, Furo-cho, Chikusa-ku, Nagoya, Aichi 464-8602, Japan}
\altaffiltext{8}{Department of Mathematics, Toyota National College of 
Technology, Eisei-cho 2-1, Toyota, Aichi 471-8525, Japan}
\altaffiltext{9}{Department of Physics and Astronomy, Clemson University, 
Clemson, SC 29634, USA}



\begin{abstract}
The neutrino-nucleus reaction cross sections of $^4$He and $^{12}$C are
evaluated using new shell model Hamiltonians.
Branching ratios of various decay channels are calculated to evaluate 
the yields of Li, Be, and B produced through the $\nu$-process in supernova 
explosions.
The new cross sections enhance the yields of $^7$Li and $^{11}$B produced
during the supernova explosion of a 16.2 $M_\odot$ star model compared to
the case using the conventional cross sections by about $10\%$.
On the other hand, the yield of $^{10}$B decreases by a factor of two.
The yields of $^6$Li, $^9$Be, and the radioactive nucleus $^{10}$Be are
found at a level of $\sim 10^{-11} M_\odot$.
The temperature of $\nu_{\mu,\tau}$- and $\bar{\nu}_{\mu,\tau}$-neutrinos
inferred from the supernova contribution of $^{11}$B in Galactic chemical 
evolution models is constrained to the $4.3-6.5$ MeV range.
The increase in the $^7$Li and $^{11}$B yields due to neutrino oscillations
is demonstrated with the new cross sections.
\end{abstract}





\section{Introduction}

Supernova (SN) explosions constitute one of several production sites of the
relatively rare light elements Li, Be, and B. In SN environments these elements 
are produced through neutrino-nucleus reactions
\citep[the $\nu$-process;][]{de78,wh90}. 
Neutrinos of all flavors are emitted in large 
numbers from a proto-neutron star, created during core-collapse of 
massive stars and the subsequent supernova explosion.
Among the light elements, $^7$Li and $^{11}$B are abundantly produced 
through the $\nu$-process \citep{wh90,yt04,yk05,hk05}. 
Production of these light element isotopes in core-collapse supernovae 
(ccSNe) can contribute significantly to the increase in their abundances 
during Galactic chemical evolution \citep[GCE;][]{fo00,rl00,rs00}.

Cross sections for neutrino-nucleus interactions are some of the most
important data required to reliably estimate the $^7$Li and $^{11}$B
yields in supernovae.
The $\nu$-process cross sections have been evaluated for a wide range of
nuclear species in \citet{wh90}. The data are tabulated in Hoffman \& Woosley 
(1992, hereafter referred to as HW92)
\footnote{See http://www-phys.llnl.gov/Research/RRSN/nu\_csbr/neu\_rate.html}.
Since the evaluation by HW92, further development of shell model calculations
now enable us to more accurately evaluate these essential cross sections.

The $\nu$-process cross sections are often presented as a function of
neutrino temperature, based on averaging energy dependent cross 
sections over a Fermi-Dirac distribution of given temperature and chemical
potential (for simplicity often assumed to be zero). However, it is more
appropriate to consider the energy dependence as the primary information,
as studies of SN neutrino transport show that their spectra do not exactly 
follow Fermi-Dirac distributions with zero-chemical potential \citep[e.g.,][]{kr03}.
Furthermore, when considering neutrino oscillations in SNe, the spectra
are non-thermal after the neutrino flavors change, even if the Fermi-Dirac 
distribution approximates the spectra at the neutrino 
sphere reasonably well \citep[e.g.,][]{ds00,tw01}.


The main purpose of this study is the re-evaluation of neutrino-nucleus
reaction cross section for $^{12}$C and $^4$He using new shell-model
Hamiltonians. We evaluate the branching ratios of many decay channels 
for light element species.
Then, we evaluate the yields of the light elements, $^6$Li, $^7$Li, 
$^9$Be, $^{10}$Be, $^{10}$B, and $^{11}$B and discuss their production
processes. We re-estimate the allowed range of the neutrino temperatures 
derived from constraints on the SN contribution of $^{11}$B in GCE models 
\citep[following][]{yk05}. 
We also investigate the dependence of the neutrino oscillation parameters, 
i.e., mass hierarchy and the mixing angle $\theta_{13}$, 
on the $^7$Li and $^{11}$B yields using the new cross sections.

The paper is organized as follows. In \S 2 new cross sections for neutrino-
$^{12}$C reactions are derived using the SFO and PSDMK2 Hamiltonians.
New cross sections for neutrino-$^4$He reactions are evaluated using the
WBP and SPSDMK Hamiltonians, as also shown in this section.
The temperature dependence of the cross sections is discussed.
The supernova explosion model and the supernova neutrino models employed 
are introduced and explained in detail in \S 3.
The nuclear reaction network used in this study is presented briefly.
Light-element production mechanisms are discussed in \S 4.
The yields obtained using the new cross sections and the differences 
from those obtained with old cross sections are shown.
The dependence of the light element yields on neutrino chemical potential
is also discussed.
The dependence of the yields of $^7$Li and $^{11}$B on the neutrino
oscillation parameters, mass hierarchy and the mixing angle $\theta_{13}$
is shown in \S 5.
The dependence of neutrino oscillation parameters on the $^7$Li/$^{11}$B ratio,
the elemental abundance ratios of the light elements, is considered, and the 
possibility of constraining mass hierarchies and the mixing angle 
$\theta_{13}$ is evaluated. 
Other effects on flavor-exchange of neutrinos in supernovae are 
discussed in \S 6, and our conclusions are finally presented in \S 7.

\begin{figure*}[p]
\epsscale{1.05}
\plottwo{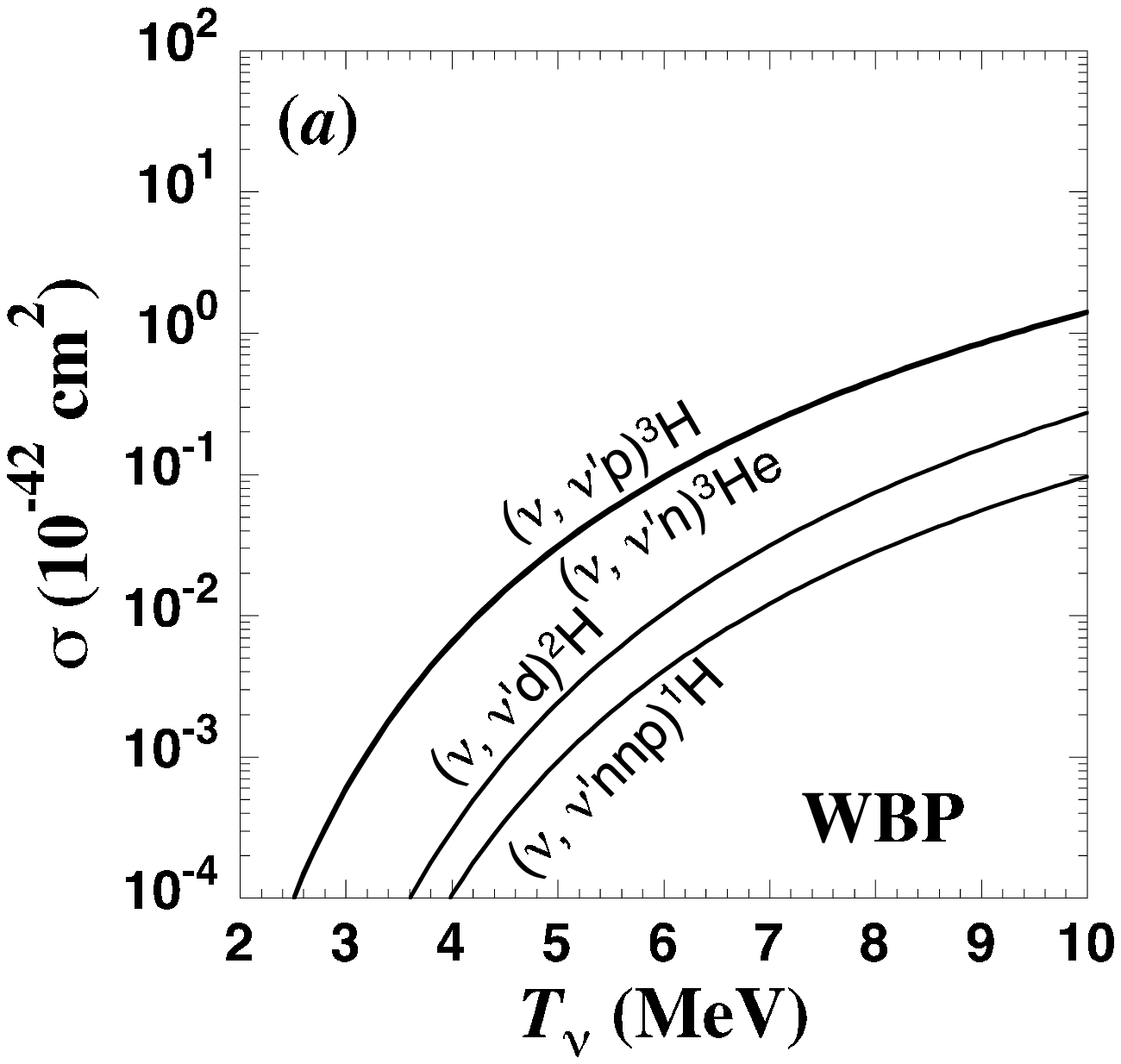}{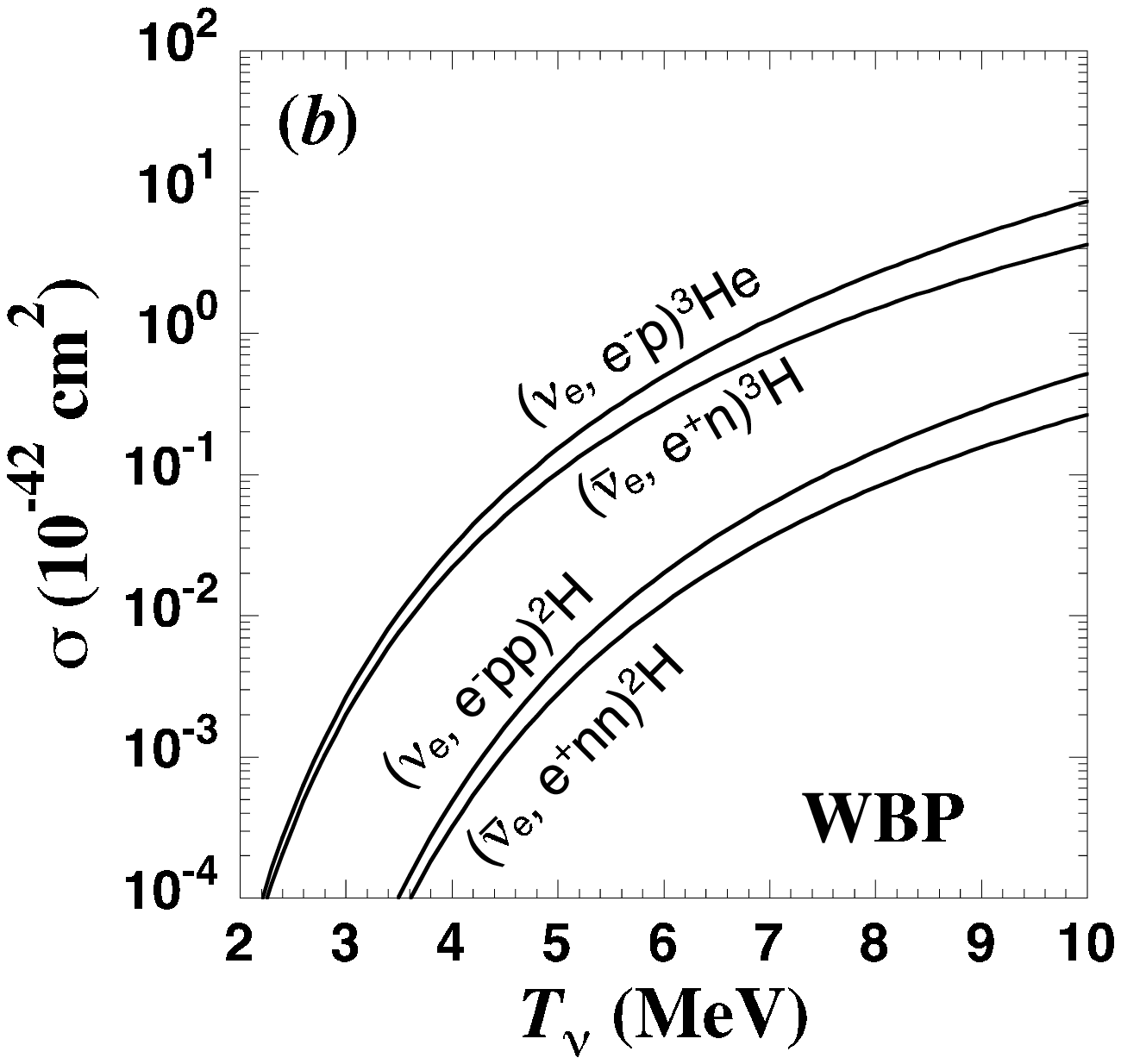}
\plottwo{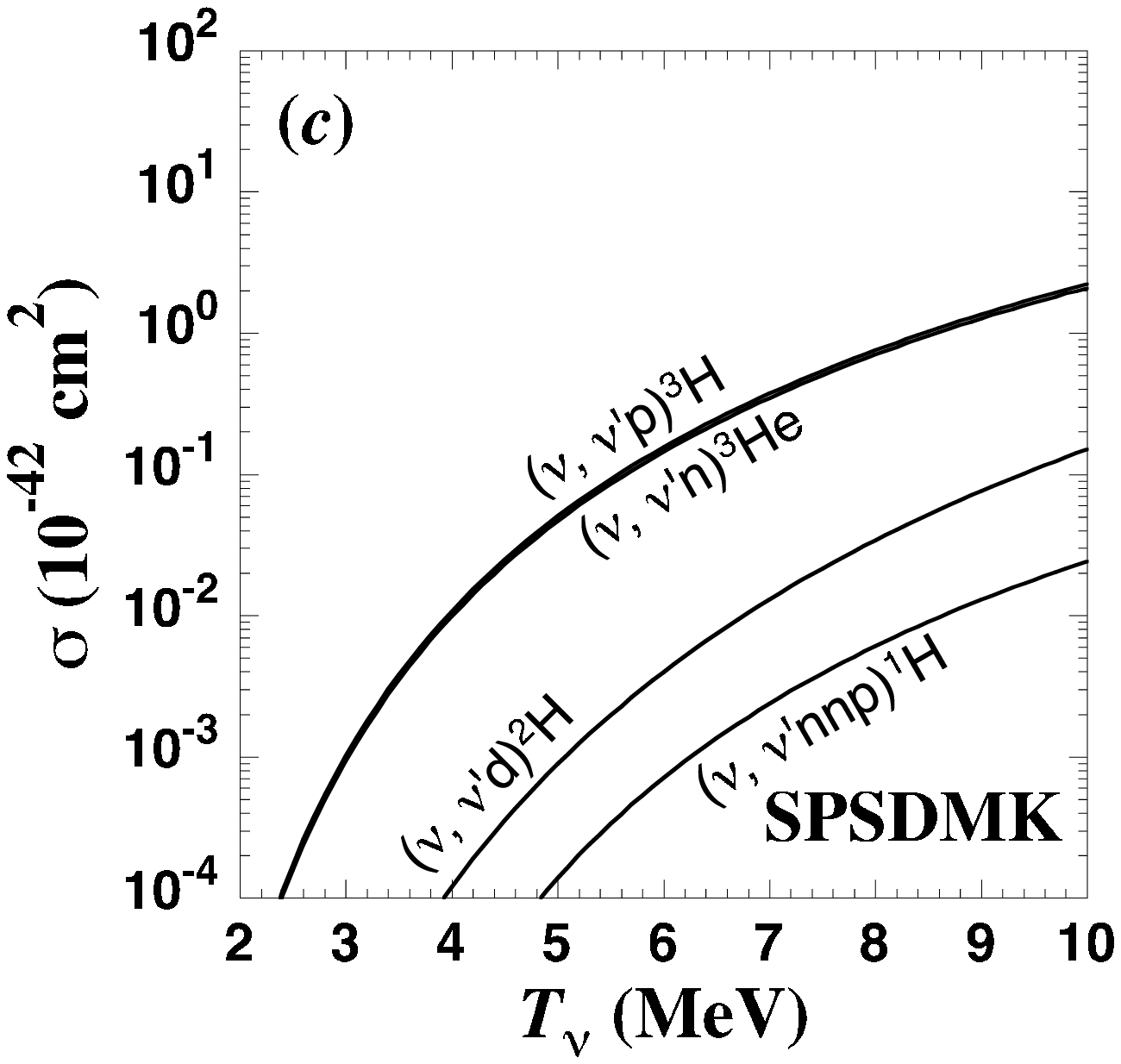}{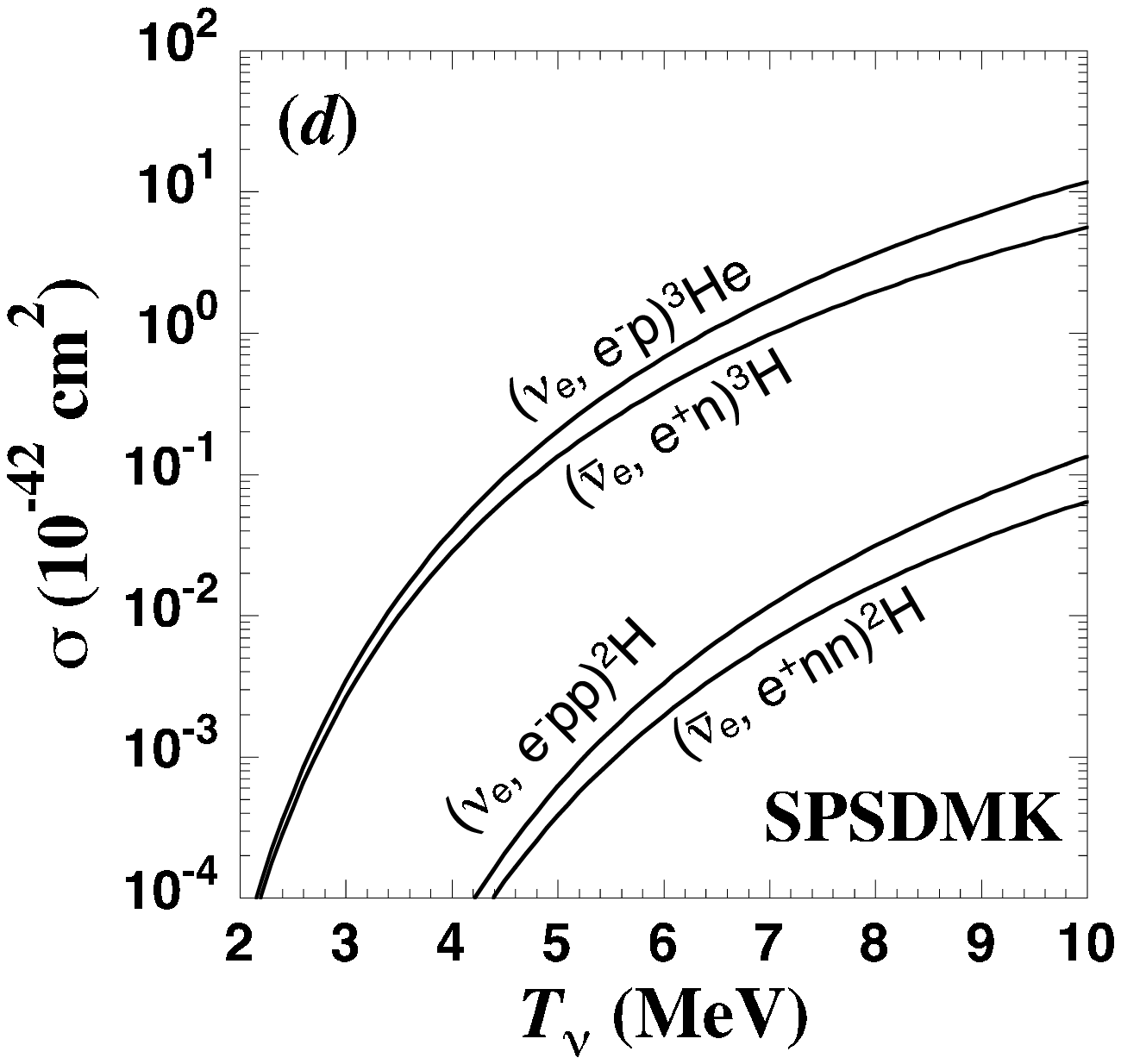}
\caption{
Cross sections for $^4$He as a function of neutrino temperature $T_\nu$.
The neutrino energy spectrum is assumed to follow a Fermi-Dirac distribution
with zero chemical potential. ($a$): neutral-current reactions with the WBP 
Hamiltonian, ($b$): charged-current reactions with the WBP Hamiltonians,
($c$): neutral-current reactions with the SPSDMK Hamiltonian,
($d$): charged-current reactions with the SPSDMK Hamiltonian.
}
\label{crosshe4}
\vspace{10cm}
\end{figure*}

\begin{figure*}
\epsscale{1.0}
\plottwo{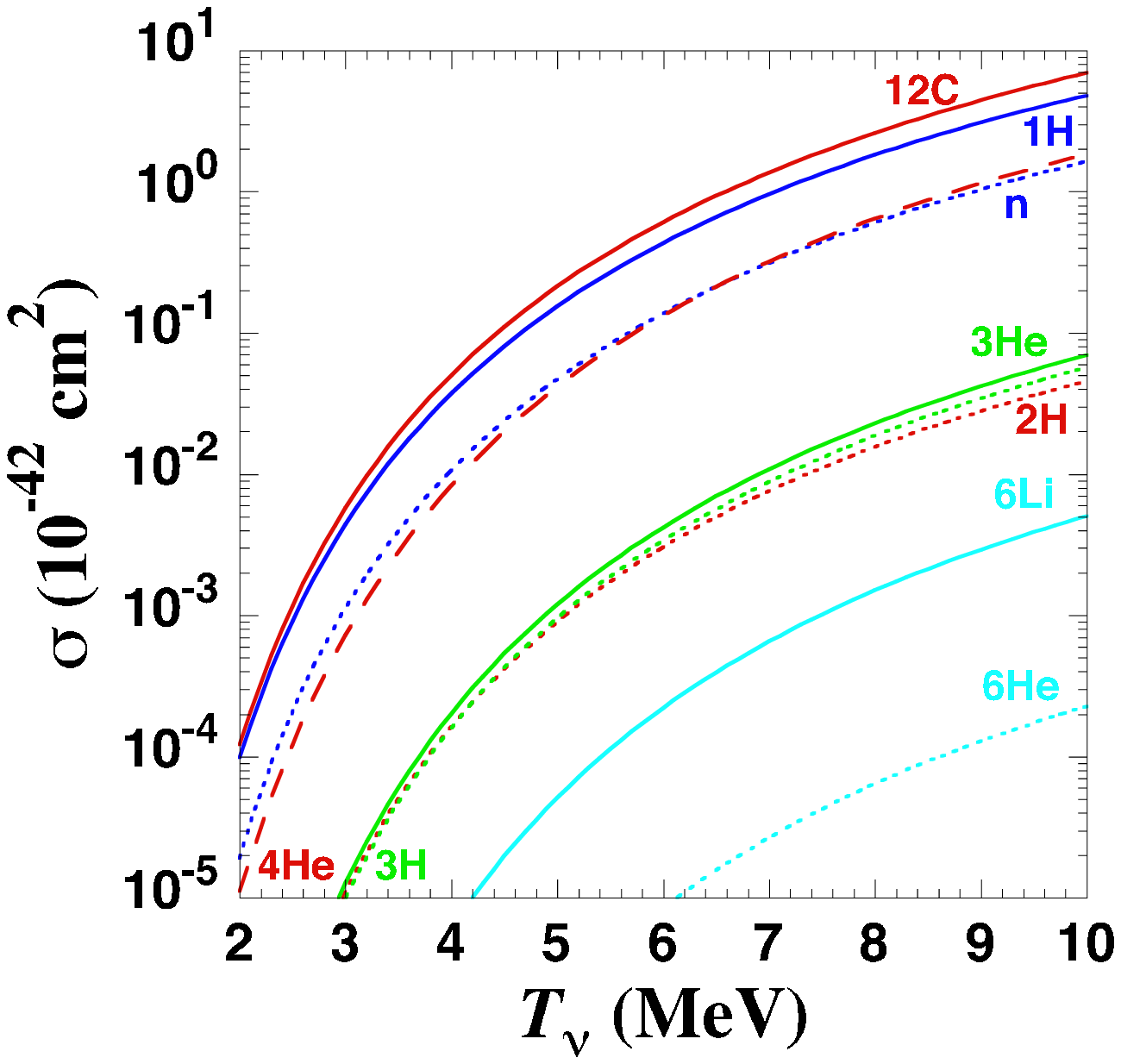}{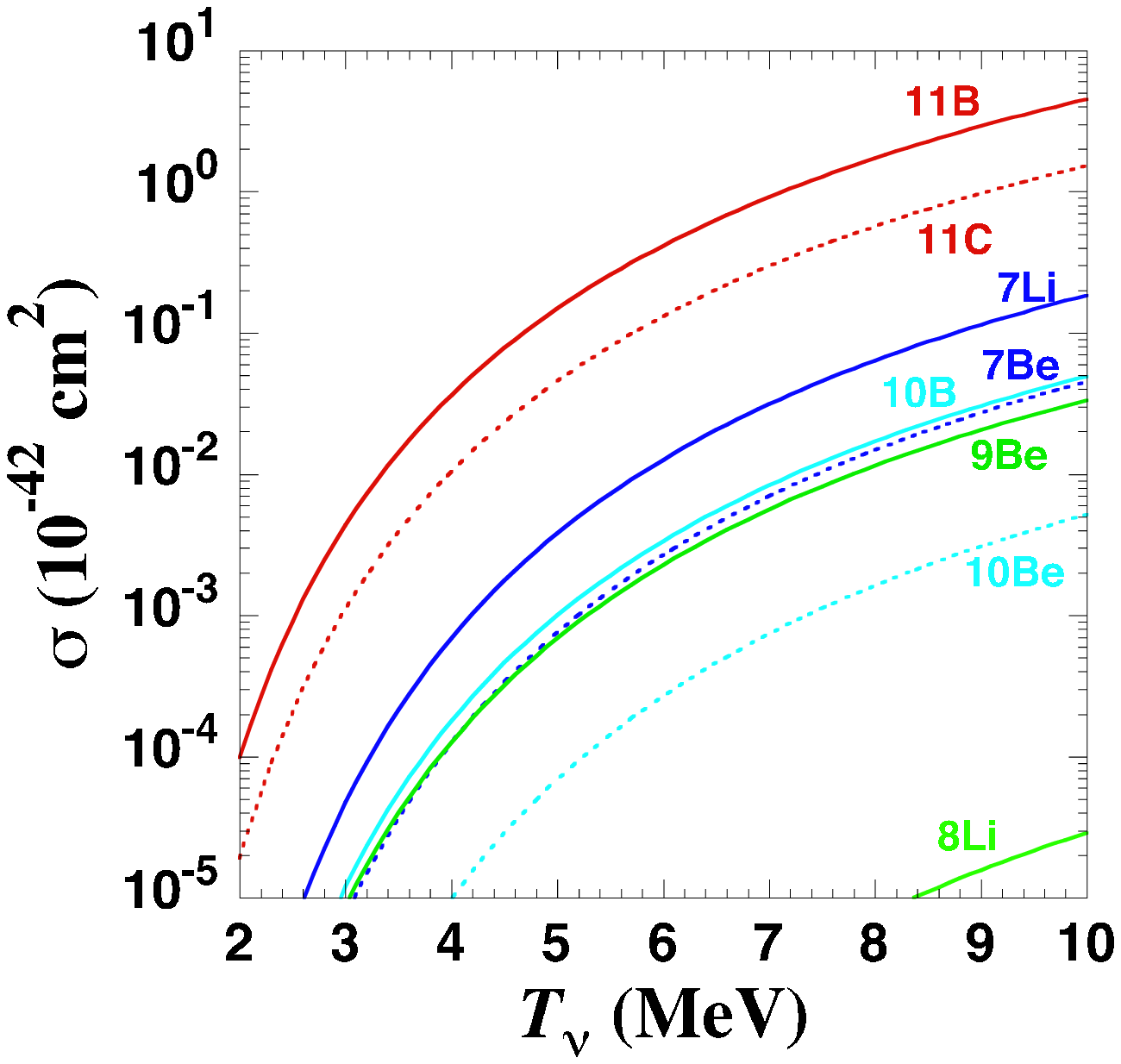}
\plottwo{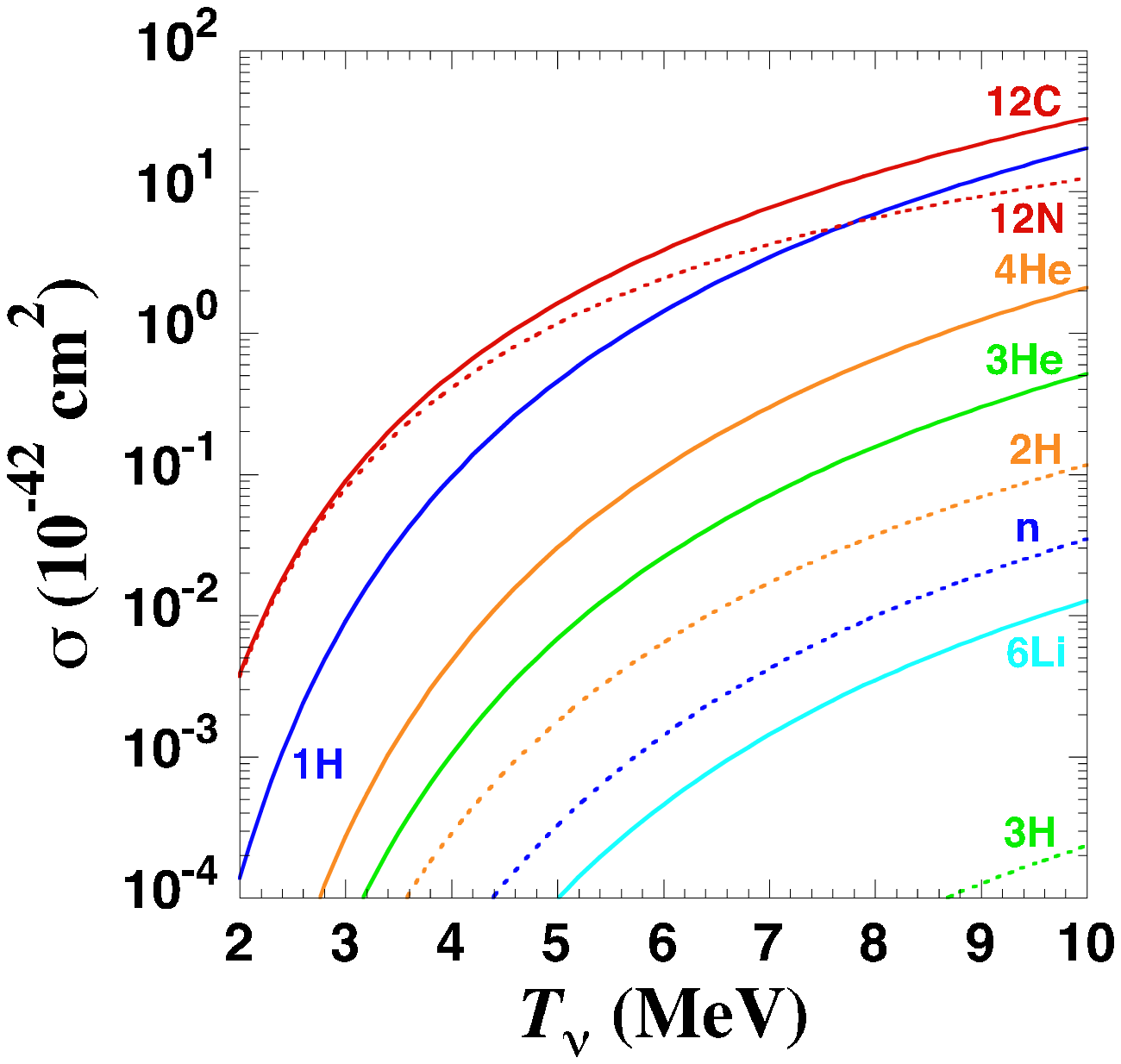}{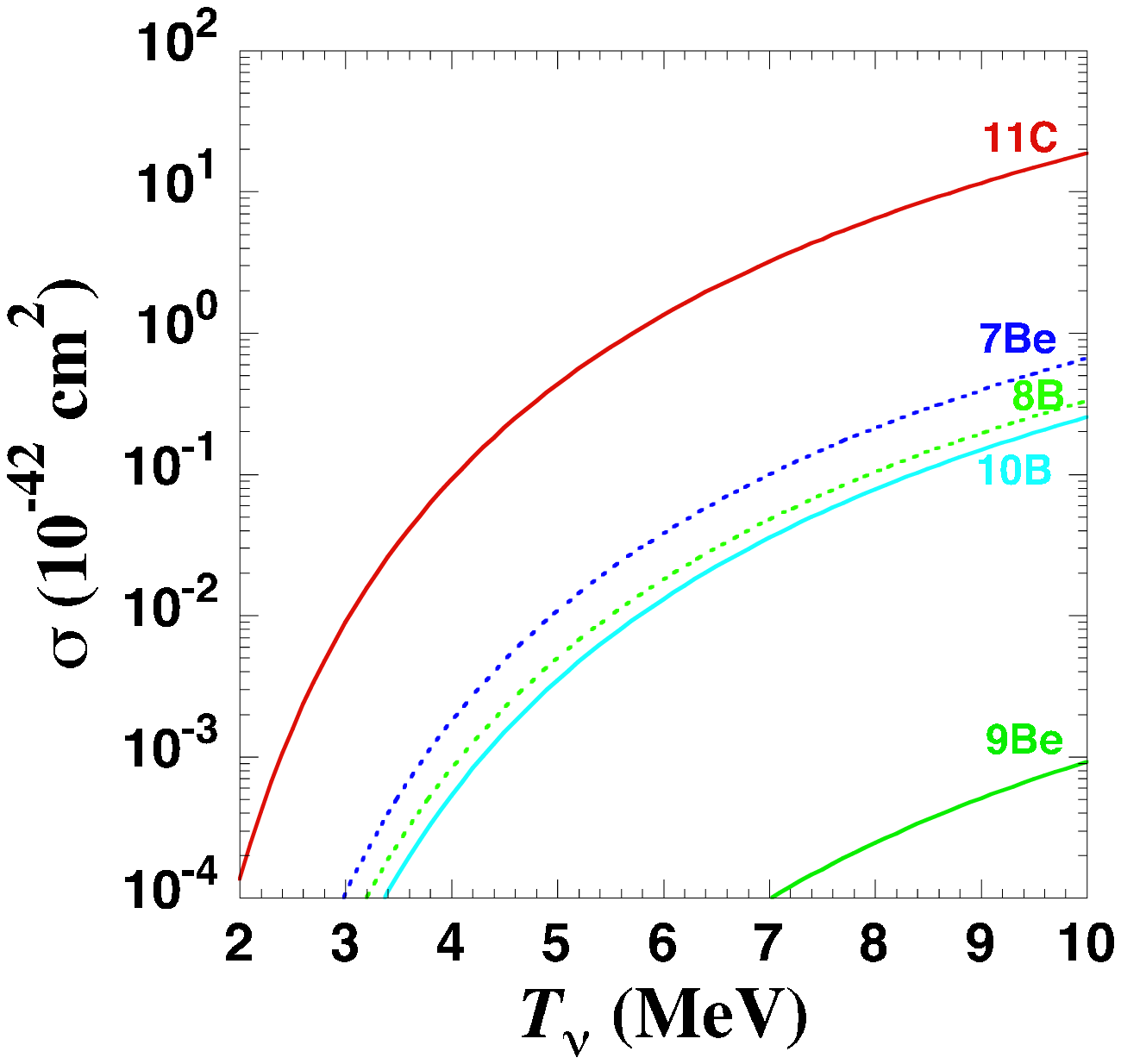}
\plottwo{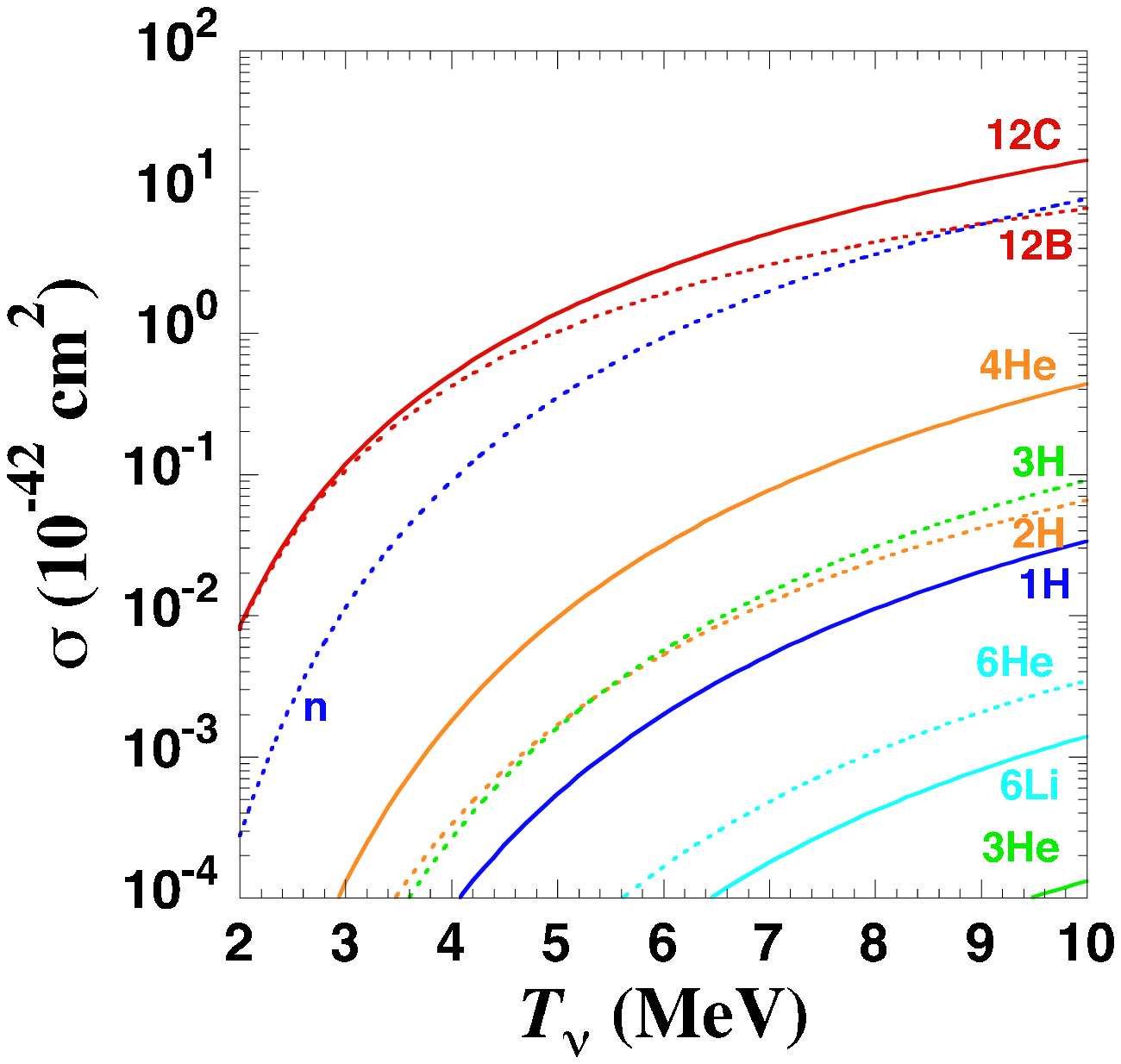}{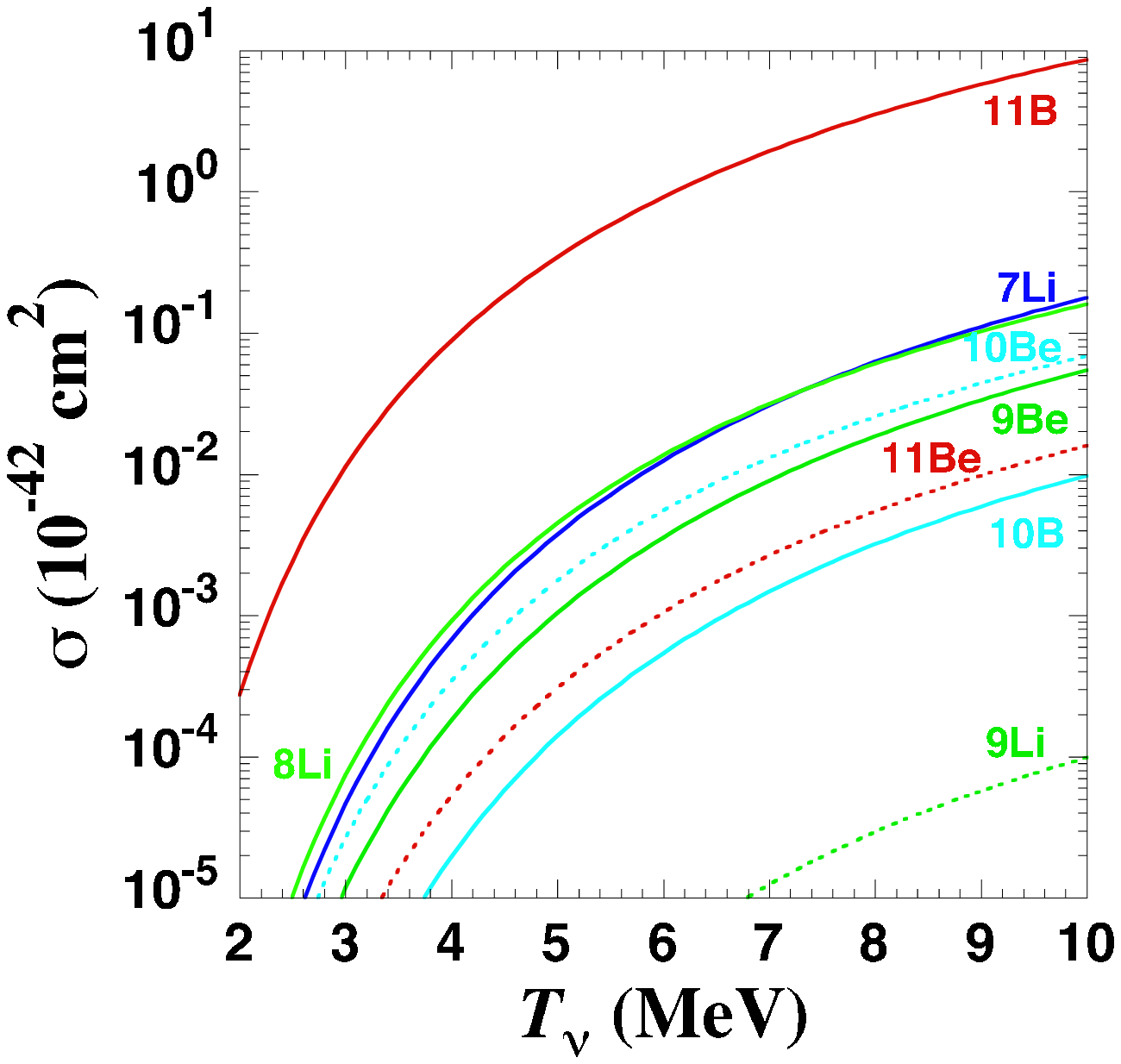}
\caption{
Averaged cross sections of $^{12}$C as a function of neutrino temperature, 
$T_\nu$, for the SFO Hamiltonian. The neutrino energy spectrum is assumed 
to follow a Fermi-Dirac distribution with zero chemical potential.
Top, meddle, and bottom panels correspond to neutral-current reactions,
charged-current reactions for $\nu_e$, and charged-current reactions for
$\bar{\nu}_e$, respectively.
The line with labeled $^{12}$C corresponds to the total decomposition rate
of $^{12}$C, $\sigma_{^{12}{\rm C},\nu}$ (see eq. [1]).
}
\label{crosssc12sfo}
\end{figure*}

\begin{figure*}
\epsscale{1.0}
\plottwo{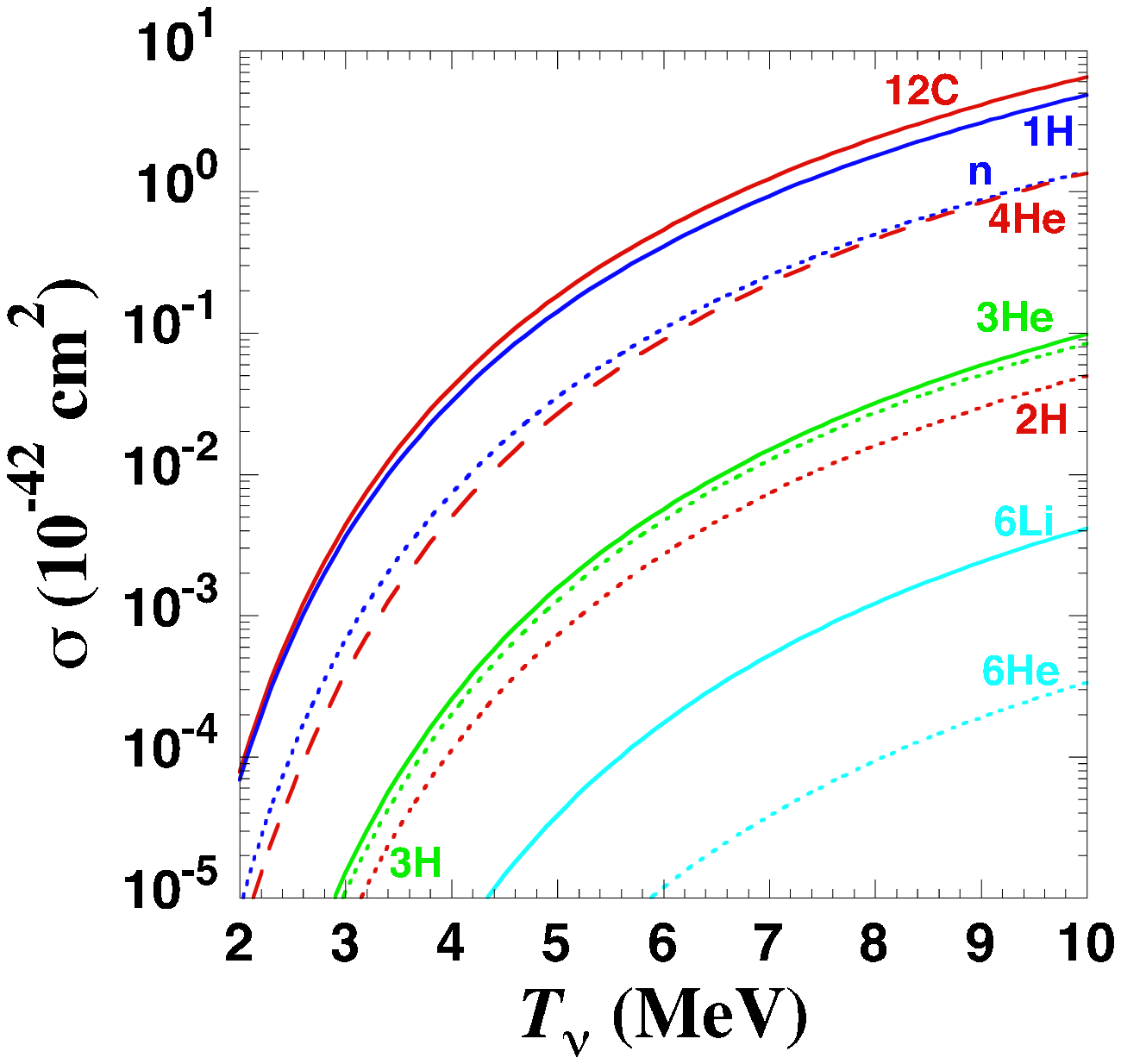}{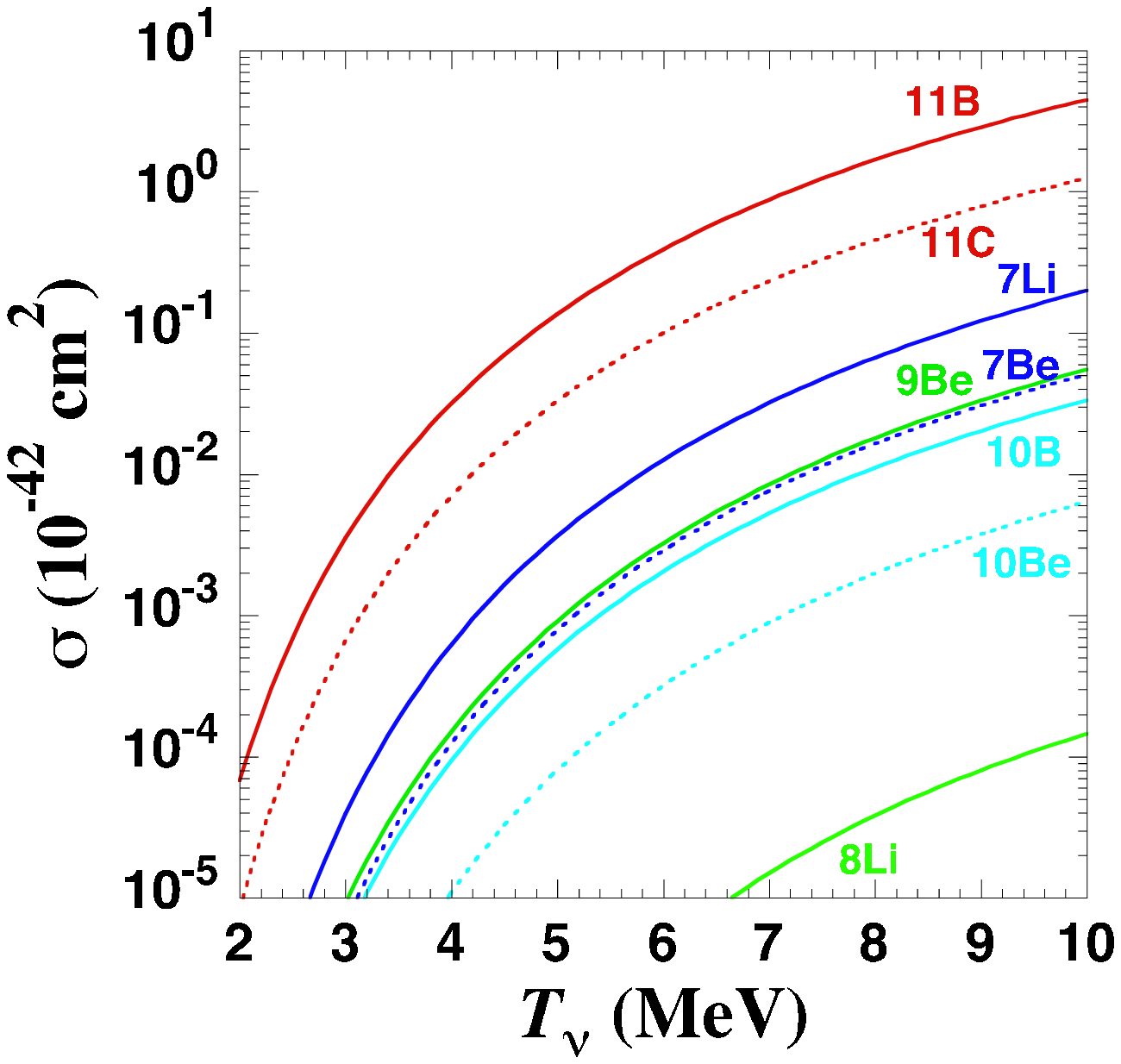}
\plottwo{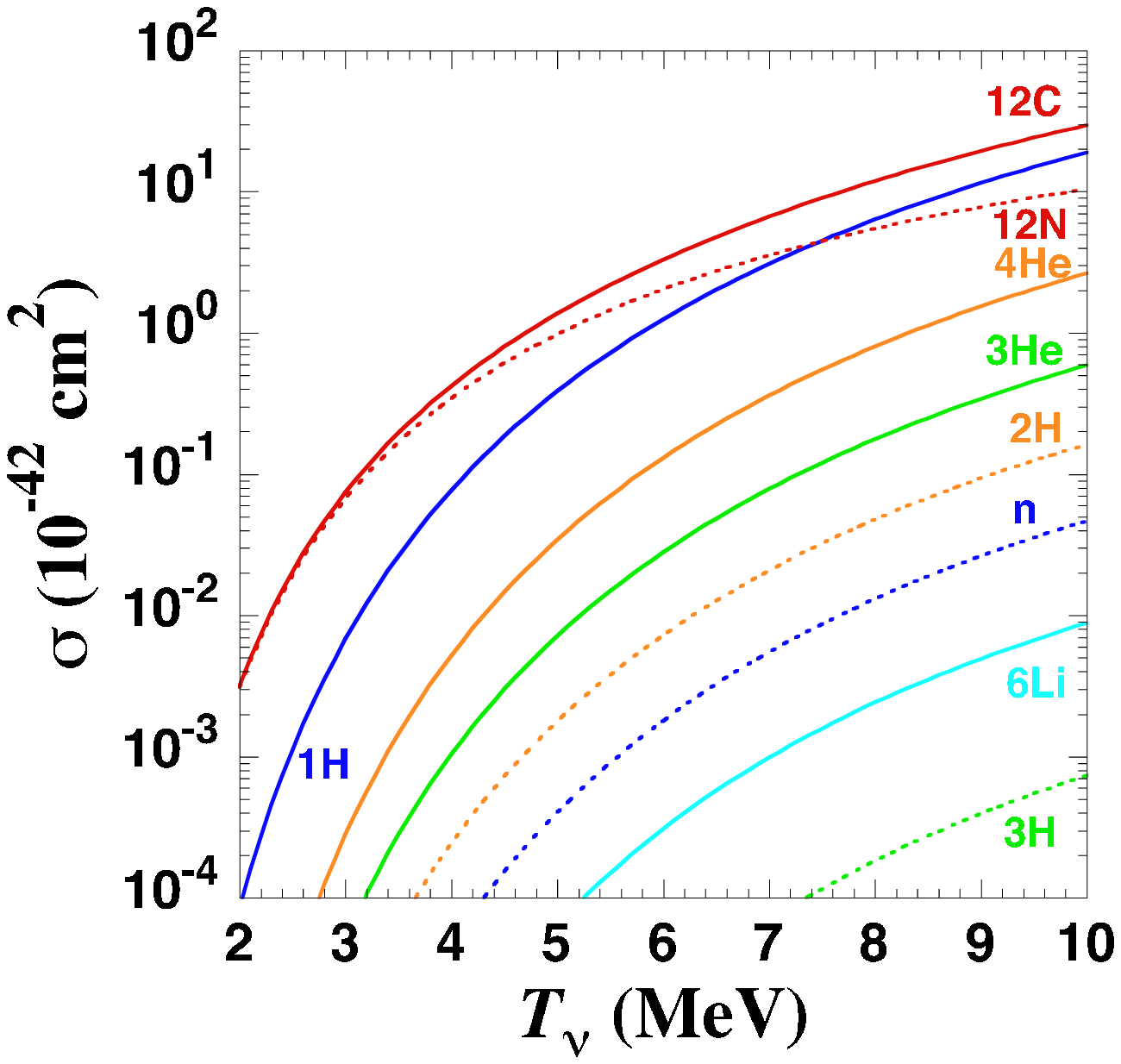}{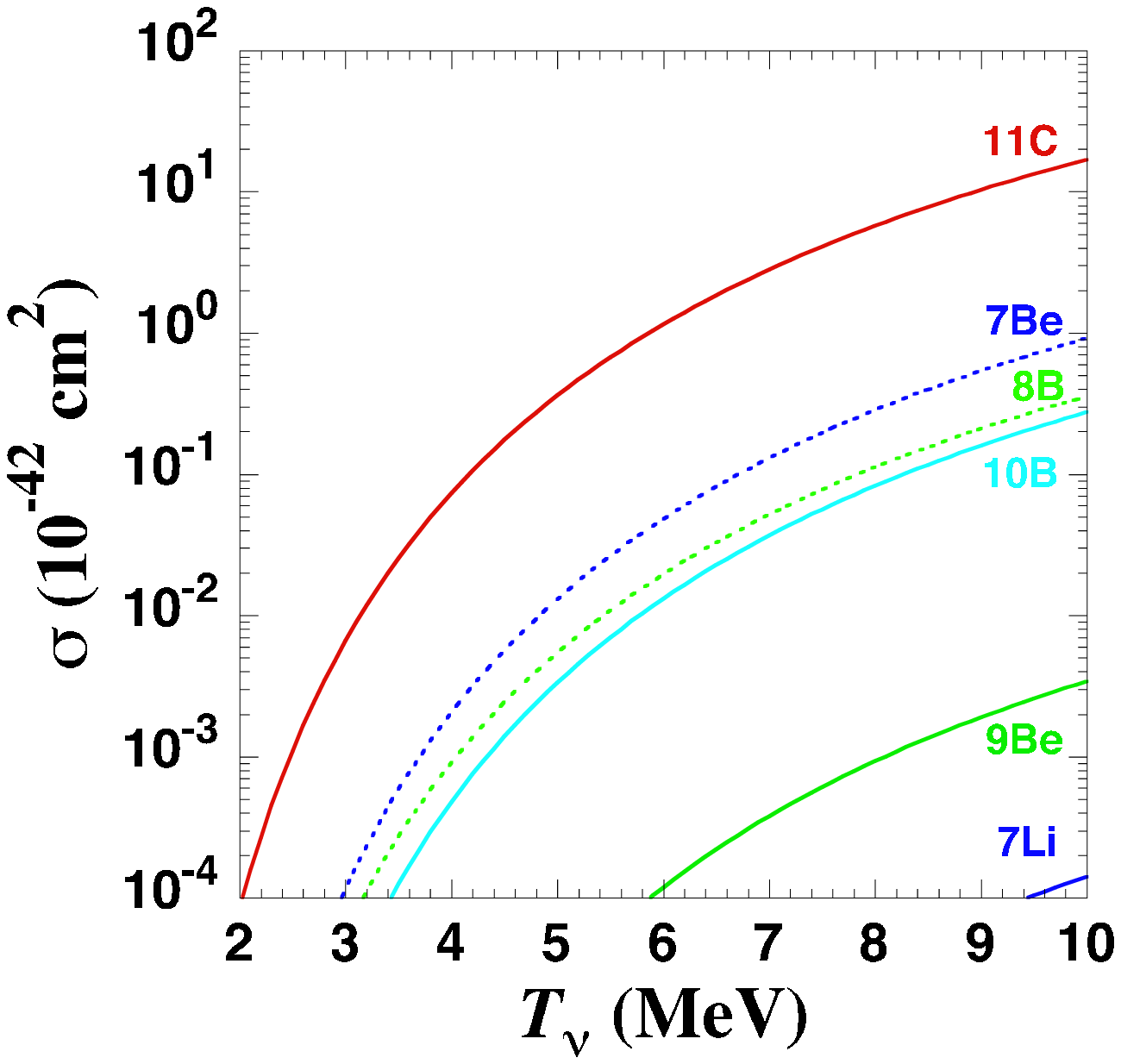}
\plottwo{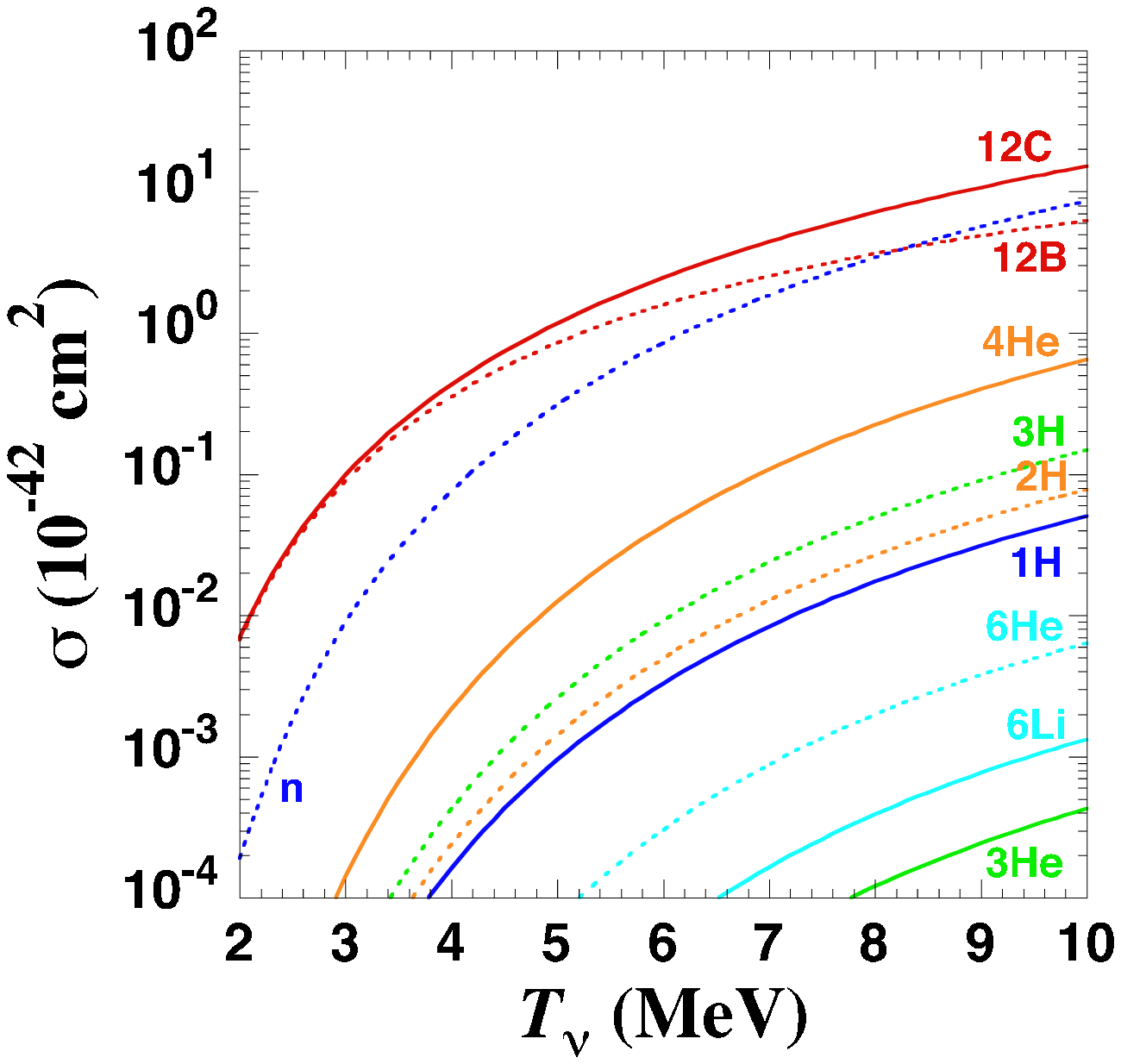}{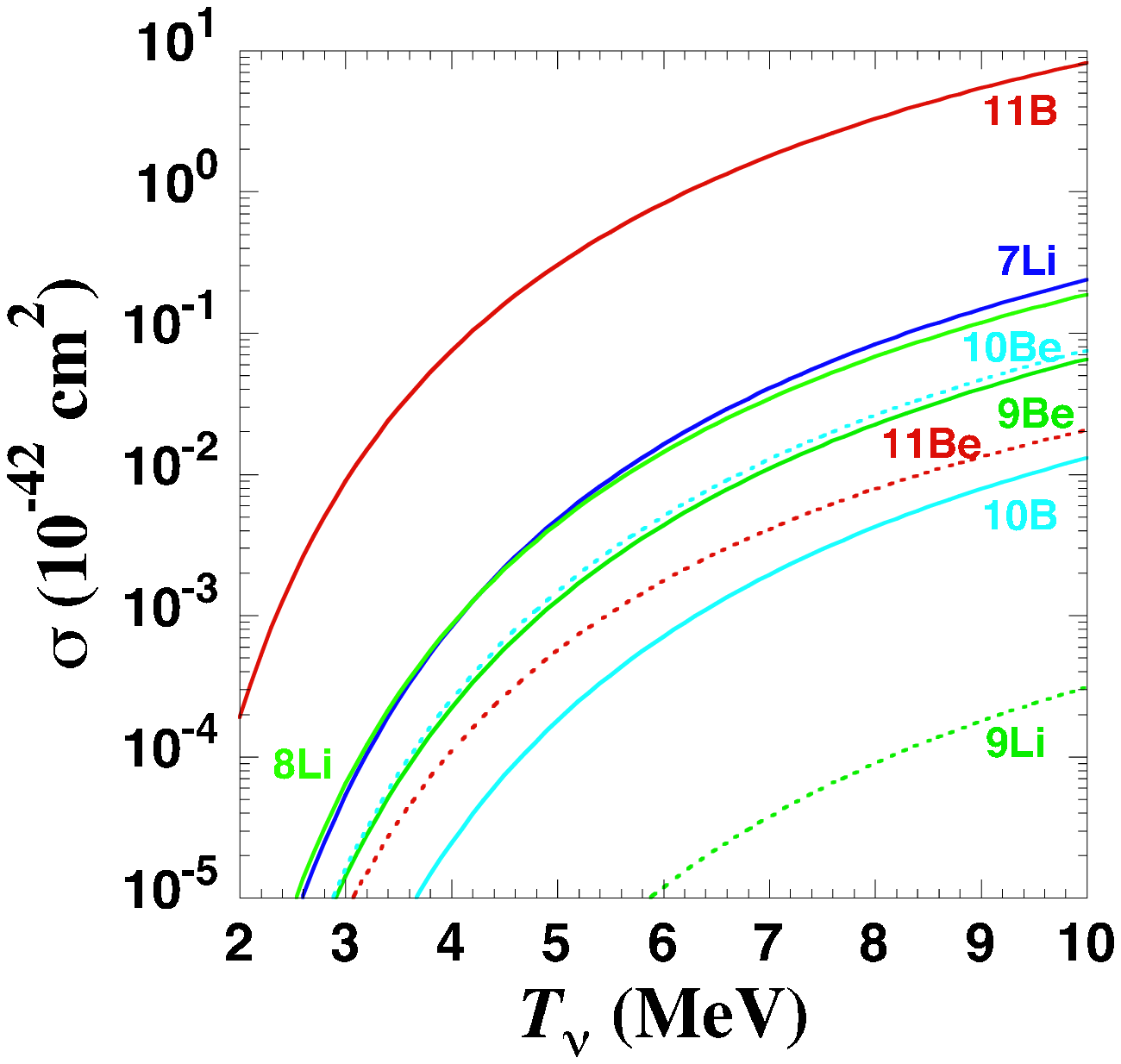}
\caption{
As in Fig.~\ref{crosssc12sfo}, but for the PSDMK2 Hamiltonian.
}
\label{crosssc12mk}
\end{figure*}

\begin{deluxetable*}{ccccccccc}
\tabletypesize{\tiny}
\tablecaption{
Neutrino-induced reaction cross sections of $^4$He in units of
$10^{-42}$ cm$^2$ with the use of the WBP Hamiltonian.
}
\tablehead{
\colhead{$E_\nu$} & & & & & & & & \\
\colhead{(MeV)} & \colhead{($\nu,\nu'p)^3$H} & \colhead{($\nu,\nu'n)^3$He} & 
\colhead{($\nu,\nu'd)^2$H} & \colhead{($\nu,\nu'nnp)^1$H} & 
\colhead{($\nu_e,e^-p)^3$He} & \colhead{($\bar{\nu}_e,e^+n)^3$H} &
\colhead{($\nu_e,e^-pp)^2$H} & \colhead{($\bar{\nu}_e,e^+nn)^2$H}
}
\startdata
10.0 & 0.000E+00 & 0.000E+00 & 0.000E+00 & 0.000E+00 & 0.000E+00 & 
0.000E+00 & 0.000E+00 & 0.000E+00 \\
20.0 & 0.000E+00 & 0.000E+00 & 0.000E+00 & 0.000E+00 & 0.000E+00 & 
0.000E+00 & 0.000E+00 & 0.000E+00 \\
30.0 & 4.018E-02 & 3.829E-02 & 2.168E-11 & 3.538E-08 & 1.604E-01 & 
1.264E-01 & 0.000E+00 & 0.000E+00 \\
40.0 & 4.609E-01 & 4.425E-01 & 3.169E-04 & 9.746E-03 & 2.094E+00 & 
1.556E+00 & 1.054E-04 & 1.587E-04 \\
50.0 & 1.802E+00 & 1.738E+00 & 7.218E-02 & 1.730E-01 & 8.957E+00 & 
5.992E+00 & 3.140E-01 & 2.211E-01 \\
60.0 & 4.777E+00 & 4.620E+00 & 3.381E-01 & 7.782E-01 & 2.564E+01 & 
1.529E+01 & 1.670E+00 & 1.053E+00 \\
70.0 & 1.017E+01 & 9.856E+00 & 8.064E-01 & 1.991E+00 & 5.842E+01 & 
3.108E+01 & 4.243E+00 & 2.409E+00 \\
80.0 & 1.874E+01 & 1.818E+01 & 1.485E+00 & 4.021E+00 & 1.145E+02 & 
5.453E+01 & 8.148E+00 & 4.167E+00
\enddata
\tablecomments{Table~\ref{tab4hewbp} is published in its entirety in the
electronic edition of the {\it Astrophysical Journal}. A portion is
shown here for guidance regarding its form and content.}
\label{tab4hewbp}
\end{deluxetable*}

\begin{deluxetable*}{ccccccccc}
\tabletypesize{\tiny}
\tablecaption{
Neutrino-induced reaction cross sections of $^4$He in units of
$10^{-42}$ cm$^2$ with the use of the SPSDMK Hamiltonian.
}
\tablehead{
\colhead{$E_\nu$} & & & & & & & & \\ 
\colhead{(MeV)} & \colhead{($\nu,\nu'p)^3$H} & \colhead{($\nu,\nu'n)^3$He} & 
\colhead{($\nu,\nu'd)^2$H} & \colhead{($\nu,\nu'nnp)^1$H} & 
\colhead{($\nu_e,e^-p)^3$He} & \colhead{($\bar{\nu}_e,e^+n)^3$H} &
\colhead{($\nu_e,e^-pp)^2$H} & \colhead{($\bar{\nu}_e,e^+nn)^2$H}
}
\startdata
10.0 & 0.000E+00 & 0.000E+00 & 0.000E+00 & 0.000E+00 & 0.000E+00 & 
0.000E+00 & 0.000E+00 & 0.000E+00 \\
20.0 & 0.000E+00 & 0.000E+00 & 0.000E+00 & 0.000E+00 & 0.000E+00 & 
0.000E+00 & 0.000E+00 & 0.000E+00 \\
30.0 & 6.992E-02 & 6.399E-02 & 0.000E+00 & 2.706E-04 & 2.045E-01 & 
1.694E-01 & 0.000E+00 & 0.000E+00 \\
40.0 & 7.360E-01 & 6.805E-01 & 3.413E-07 & 5.913E-03 & 2.709E+00 & 
1.968E+00 & 0.000E+00 & 0.000E+00 \\
50.0 & 2.879E+00 & 2.675E+00 & 6.340E-03 & 4.657E-02 & 1.211E+01 & 
7.761E+00 & 2.338E-02 & 1.717E-02 \\
60.0 & 7.633E+00 & 7.112E+00 & 5.841E-02 & 2.429E-01 & 3.557E+01 & 
2.025E+01 & 2.516E-01 & 1.628E-01 \\
70.0 & 1.616E+01 & 1.509E+01 & 1.800E-01 & 8.009E-01 & 8.209E+01 & 
4.173E+01 & 8.673E-01 & 4.997E-01 \\
80.0 & 2.944E+01 & 2.753E+01 & 3.897E-01 & 2.037E+00 & 1.614E+02 & 
7.367E+01 & 2.062E+00 & 1.063E+00
\enddata
\tablecomments{Table~\ref{tab4hemk} is published in its entirety in the
electronic edition of the {\it Astrophysical Journal}. A portion is
shown here for guidance regarding its form and content.}
\label{tab4hemk}
\end{deluxetable*}

\begin{deluxetable*}{ccccccccccc}
\tabletypesize{\tiny}
\tablecaption{
Neutrino-induced neutral-current reaction cross sections of $^{12}$C 
in units of $10^{-42}$ cm$^2$ based on the SFO Hamiltonian.
}
\tablehead{
\colhead{$E_\nu$} &  & & & & & & & & &  \\
\colhead{(MeV)} & \colhead{$^{12}$C\tablenotemark{a}} & \colhead{$n$} & \colhead{$p$} & \colhead{$d$} &
\colhead{$t$} & \colhead{$^3$He} & \colhead{$^4$He} & \colhead{$^6$He} & \colhead{$^6$Li} & \colhead{$^7$Li}
}
\startdata
10.0 & 0.000E+00 & 0.000E+00 & 0.000E+00 & 0.000E+00 & 0.000E+00 & 0.000E+00 & 0.000E+00 & 0.000E+00 & 0.000E+00 & 0.000E+00 \\
20.0 & 1.484E-03 & 1.482E-05 & 1.330E-03 & 0.000E+00 & 0.000E+00 & 0.000E+00 & 4.201E-04 & 0.000E+00 & 0.000E+00 & 0.000E+00 \\
30.0 & 4.062E-01 & 8.046E-02 & 3.147E-01 & 1.329E-04 & 9.846E-05 & 1.462E-04 & 3.296E-02 & 0.000E+00 & 1.644E-13 & 5.526E-04 \\
40.0 & 3.206E+00 & 7.104E-01 & 2.308E+00 & 1.259E-02 & 1.136E-02 & 1.458E-02 & 5.987E-01 & 2.569E-07 & 1.256E-04 & 5.501E-02 \\
50.0 & 1.123E+01 & 2.576E+00 & 7.890E+00 & 6.209E-02 & 6.915E-02 & 8.576E-02 & 2.637E+00 & 9.178E-05 & 3.726E-03 & 2.645E-01 \\
60.0 & 2.705E+01 & 6.346E+00 & 1.877E+01 & 1.765E-01 & 2.110E-01 & 2.574E-01 & 7.050E+00 & 7.484E-04 & 1.801E-02 & 7.078E-01 \\
70.0 & 5.199E+01 & 1.240E+01 & 3.578E+01 & 3.657E-01 & 4.583E-01 & 5.560E-01 & 1.439E+01 & 2.094E-03 & 4.453E-02 & 1.442E+00 \\
80.0 & 8.574E+01 & 2.068E+01 & 5.862E+01 & 6.252E-01 & 8.167E-01 & 9.890E-01 & 2.471E+01 & 3.979E-03 & 8.175E-02 & 2.486E+00 \\
\hline
\hline
\colhead{$E_\nu$} &  & & & & & & & & & \\
\colhead{(MeV)} & \colhead{$^8$Li} & \colhead{$^9$Li} & \colhead{$^7$Be} & \colhead{$^9$Be} &
\colhead{$^{10}$Be} & \colhead{$^8$B} & \colhead{$^{10}$B} & \colhead{$^{11}$B} & \colhead{$^{10}$C} & \colhead{$^{11}$C} \\
\hline
10.0 & 0.000E+00 & 0.000E+00 & 0.000E+00 & 0.000E+00 & 0.000E+00 & 0.000E+00 & 0.000E+00 & 0.000E+00 & 0.000E+00 & 0.000E+00 \\
20.0 & 0.000E+00 & 0.000E+00 & 0.000E+00 & 0.000E+00 & 0.000E+00 & 0.000E+00 & 0.000E+00 & 1.330E-03 & 0.000E+00 & 1.482E-05 \\
30.0 & 0.000E+00 & 0.000E+00 & 4.367E-05 & 1.459E-04 & 7.516E-09 & 0.000E+00 & 1.331E-04 & 3.141E-01 & 0.000E+00 & 8.042E-02 \\
40.0 & 0.000E+00 & 0.000E+00 & 9.063E-03 & 9.957E-03 & 5.222E-04 & 0.000E+00 & 1.386E-02 & 2.239E+00 & 5.356E-06 & 6.950E-01 \\
50.0 & 1.277E-06 & 1.657E-12 & 5.537E-02 & 4.668E-02 & 5.365E-03 & 1.759E-08 & 7.018E-02 & 7.520E+00 & 1.747E-04 & 2.455E+00 \\
60.0 & 6.709E-05 & 2.977E-08 & 1.657E-01 & 1.272E-01 & 1.904E-02 & 6.102E-06 & 1.920E-01 & 1.770E+01 & 9.144E-04 & 5.937E+00 \\
70.0 & 2.643E-04 & 1.879E-07 & 3.596E-01 & 2.610E-01 & 4.343E-02 & 2.902E-05 & 3.891E-01 & 3.352E+01 & 2.295E-03 & 1.147E+01 \\
80.0 & 5.593E-04 & 4.514E-07 & 6.455E-01 & 4.499E-01 & 7.859E-02 & 6.482E-05 & 6.616E-01 & 5.466E+01 & 4.216E-03 & 1.899E+01
\enddata
\tablecomments{Table~\ref{tab12c_ncsfo} is published in its entirety in the
electronic edition of the {\it Astrophysical Journal}. A portion is
shown here for guidance regarding its form and content.}
\tablenotetext{a}{Decomposition rate. See \S 2 and Eq. \ref{csc12}.}
\label{tab12c_ncsfo}
\end{deluxetable*}

\begin{deluxetable*}{cccccccccc}
\tabletypesize{\tiny}
\tablecaption{
Neutrino-induced charged-current reaction cross sections for $\nu_e$ of 
$^{12}$C in units of $10^{-42}$ cm$^2$ based on the SFO Hamiltonian.
}
\tablehead{
\colhead{$E_\nu$} & & & & & & & & & \\
\colhead{(MeV)} & \colhead{$^{12}$C\tablenotemark{a}} & \colhead{$n$} & \colhead{$p$} & \colhead{$d$} &
\colhead{$t$} & \colhead{$^3$He} & \colhead{$^4$He} & \colhead{$^6$Li} & \colhead{$^7$Li}
}
\startdata
10.0 & 0.000E+00 & 0.000E+00 & 0.000E+00 & 0.000E+00 & 0.000E+00 & 0.000E+00 & 0.000E+00 & 0.000E+00 & 0.000E+00 \\
20.0 & 2.796E-01 & 0.000E+00 & 4.275E-05 & 0.000E+00 & 0.000E+00 & 0.000E+00 & 0.000E+00 & 0.000E+00 & 0.000E+00 \\
30.0 & 5.516E+00 & 0.000E+00 & 5.781E-01 & 4.584E-07 & 0.000E+00 & 1.691E-05 & 9.536E-04 & 0.000E+00 & 0.000E+00 \\
40.0 & 2.150E+01 & 1.330E-03 & 6.374E+00 & 2.059E-02 & 0.000E+00 & 6.608E-02 & 3.252E-01 & 1.396E-04 & 0.000E+00 \\
50.0 & 5.655E+01 & 2.281E-02 & 2.658E+01 & 1.258E-01 & 2.127E-05 & 4.985E-01 & 2.159E+00 & 5.933E-03 & 6.552E-08 \\
60.0 & 1.204E+02 & 1.056E-01 & 7.247E+01 & 4.043E-01 & 5.022E-04 & 1.685E+00 & 7.021E+00 & 3.680E-02 & 1.010E-04 \\
70.0 & 2.209E+02 & 2.807E-01 & 1.537E+02 & 9.218E-01 & 1.981E-03 & 3.980E+00 & 1.630E+01 & 1.044E-01 & 6.282E-04 \\
80.0 & 3.611E+02 & 5.595E-01 & 2.751E+02 & 1.703E+00 & 4.387E-03 & 7.625E+00 & 3.092E+01 & 2.108E-01 & 1.567E-03 \\
\hline
\hline
\colhead{$E_\nu$} & & & & & & & & & \\
\colhead{(MeV)} & \colhead{$^8$Li} & \colhead{$^7$Be} & \colhead{$^9$Be} & \colhead{$^8$B} &
\colhead{$^{10}$B} & \colhead{$^9$C} & \colhead{$^{10}$C} & \colhead{$^{11}$C} & \colhead{$^{12}$N} \\
\hline
  10.0 & 0.000E+00 & 0.000E+00 & 0.000E+00 & 0.000E+00 & 0.000E+00 & 0.000E+00 & 0.000E+00 & 0.000E+00 & 0.000E+00 \\
  20.0 & 0.000E+00 & 0.000E+00 & 0.000E+00 & 0.000E+00 & 0.000E+00 & 0.000E+00 & 0.000E+00 & 4.275E-05 & 2.795E-01 \\
  30.0 & 0.000E+00 & 1.767E-04 & 0.000E+00 & 7.430E-04 & 5.340E-07 & 0.000E+00 & 4.584E-07 & 5.779E-01 & 4.937E+00 \\
  40.0 & 0.000E+00 & 1.345E-01 & 6.658E-07 & 5.803E-02 & 1.235E-02 & 0.000E+00 & 2.174E-02 & 6.147E+00 & 1.506E+01 \\
  50.0 & 0.000E+00 & 7.781E-01 & 2.669E-04 & 3.465E-01 & 1.246E-01 & 2.064E-05 & 1.328E-01 & 2.499E+01 & 2.966E+01 \\
  60.0 & 2.761E-10 & 2.296E+00 & 2.518E-03 & 1.102E+00 & 4.533E-01 & 3.532E-04 & 3.996E-01 & 6.718E+01 & 4.714E+01 \\
  70.0 & 3.717E-09 & 5.054E+00 & 7.803E-03 & 2.556E+00 & 1.099E+00 & 1.161E-03 & 8.766E-01 & 1.413E+02 & 6.558E+01 \\
  80.0 & 1.062E-08 & 9.328E+00 & 1.606E-02 & 4.855E+00 & 2.136E+00 & 2.402E-03 & 1.597E+00 & 2.515E+02 & 8.314E+01
\enddata
\tablecomments{Table~\ref{tab12c_ccesfo} is published in its entirety in the
electronic edition of the {\it Astrophysical Journal}. A portion is
shown here for guidance regarding its form and content.}
\tablenotetext{a}{Decomposition rate. See \S 2 and Eq. \ref{csc12}.}
\label{tab12c_ccesfo}
\end{deluxetable*}

\begin{deluxetable*}{ccccccccccc}
\tabletypesize{\tiny}
\tablecaption{
Neutrino-induced charged-current reaction cross sections for $\bar{\nu}_e$ of 
$^{12}$C in units of $10^{-42}$ cm$^2$ 
based the SFO Hamiltonian.
}
\tablehead{
\colhead{$E_\nu$} & & & & & & & & & & \\
\colhead{(MeV)} & \colhead{$^{12}$C\tablenotemark{a}} & \colhead{$n$} & \colhead{$p$} & \colhead{$d$} &
\colhead{$t$} & \colhead{$^3$He} & \colhead{$^4$He} & \colhead{$^6$He} & \colhead{$^6$Li} & \colhead{$^7$Li}
}
\startdata
10.0 & 0.000E+00 & 0.000E+00 & 0.000E+00 & 0.000E+00 & 0.000E+00 & 0.000E+00 & 0.000E+00 & 0.000E+00 & 0.000E+00 & 0.000E+00 \\
20.0 & 7.004E-01 & 5.771E-03 & 0.000E+00 & 0.000E+00 & 0.000E+00 & 0.000E+00 & 0.000E+00 & 0.000E+00 & 0.000E+00 & 0.000E+00 \\
30.0 & 5.252E+00 & 8.317E-01 & 1.149E-05 & 7.809E-04 & 1.223E-04 & 0.000E+00 & 3.422E-03 & 0.000E+00 & 0.000E+00 & 8.886E-04 \\
40.0 & 1.592E+01 & 5.370E+00 & 5.521E-03 & 2.646E-02 & 1.832E-02 & 0.000E+00 & 1.375E-01 & 6.683E-05 & 3.032E-06 & 5.223E-02 \\
50.0 & 3.482E+01 & 1.664E+01 & 4.072E-02 & 1.090E-01 & 1.185E-01 & 5.077E-05 & 6.556E-01 & 2.976E-03 & 7.866E-04 & 2.640E-01 \\
60.0 & 6.290E+01 & 3.647E+01 & 1.325E-01 & 2.743E-01 & 3.579E-01 & 4.628E-04 & 1.759E+00 & 1.404E-02 & 5.310E-03 & 7.095E-01 \\
70.0 & 9.941E+01 & 6.470E+01 & 2.854E-01 & 5.194E-01 & 7.521E-01 & 1.296E-03 & 3.494E+00 & 3.240E-02 & 1.341E-02 & 1.419E+00 \\
80.0 & 1.422E+02 & 9.948E+01 & 4.916E-01 & 8.266E-01 & 1.290E+00 & 2.367E-03 & 5.803E+00 & 5.553E-02 & 2.362E-02 & 2.379E+00 \\
\hline
\hline
\colhead{$E_\nu$} & & & & & & & & & & \\
\colhead{(MeV)} & \colhead{$^8$Li} & \colhead{$^9$Li} & \colhead{$^7$Be} & \colhead{$^9$Be} &
\colhead{$^{10}$Be} & \colhead{$^{11}$Be} & \colhead{$^{10}$B} & \colhead{$^{11}$B} & \colhead{$^{12}$B} & \\
\hline
10.0 & 0.000E+00 & 0.000E+00 & 0.000E+00 & 0.000E+00 & 0.000E+00 & 0.000E+00 & 0.000E+00 & 0.000E+00 & 0.000E+00 \\
20.0 & 0.000E+00 & 0.000E+00 & 0.000E+00 & 0.000E+00 & 0.000E+00 & 0.000E+00 & 0.000E+00 & 5.771E-03 & 6.946E-01 \\
30.0 & 2.533E-03 & 0.000E+00 & 0.000E+00 & 1.223E-04 & 7.809E-04 & 1.149E-05 & 1.220E-16 & 8.309E-01 & 4.417E+00 \\
40.0 & 7.545E-02 & 0.000E+00 & 0.000E+00 & 1.361E-02 & 2.765E-02 & 4.090E-03 & 8.826E-04 & 5.310E+00 & 1.044E+01 \\
50.0 & 2.821E-01 & 5.011E-05 & 5.056E-12 & 7.359E-02 & 1.164E-01 & 2.238E-02 & 1.154E-02 & 1.627E+01 & 1.772E+01 \\
60.0 & 6.551E-01 & 3.794E-04 & 3.581E-07 & 2.125E-01 & 2.872E-01 & 6.241E-02 & 4.018E-02 & 3.537E+01 & 2.536E+01 \\
70.0 & 1.206E+00 & 9.657E-04 & 1.865E-06 & 4.418E-01 & 5.389E-01 & 1.282E-01 & 8.540E-02 & 6.241E+01 & 3.273E+01 \\
80.0 & 1.922E+00 & 1.693E-03 & 4.074E-06 & 7.559E-01 & 8.586E-01 & 2.189E-01 & 1.438E-01 & 9.560E+01 & 3.950E+01
\enddata
\tablecomments{Table~\ref{tab12c_ccbsfo} is published in its entirety in the
electronic edition of the {\it Astrophysical Journal}. A portion is
shown here for guidance regarding its form and content.}
\tablenotetext{a}{Decomposition rate. See \S 2 and Eq. \ref{csc12}.}
\label{tab12c_ccbsfo}
\end{deluxetable*}

\begin{deluxetable*}{ccccccccccc}
\tabletypesize{\tiny}
\tablecaption{
Neutrino-induced neutral-current reaction cross sections of $^{12}$C 
in units of $10^{-42}$ cm$^2$ 
based on the PSDMK2 Hamiltonian.
}
\tablehead{
\colhead{$E_\nu$} & & & & & & & & & & \\
\colhead{(MeV)} & \colhead{$^{12}$C\tablenotemark{a}} & \colhead{$n$} & \colhead{$p$} & \colhead{$d$} &
\colhead{$t$} & \colhead{$^3$He} & \colhead{$^4$He} & \colhead{$^6$He} & \colhead{$^6$Li} & \colhead{$^7$Li}
}
\startdata
10.0 & 0.000E+00 & 0.000E+00 & 0.000E+00 & 0.000E+00 & 0.000E+00 & 0.000E+00 & 0.000E+00 & 0.000E+00 & 0.000E+00 & 0.000E+00 \\
20.0 & 3.289E-04 & 2.518E-07 & 2.648E-04 & 0.000E+00 & 0.000E+00 & 0.000E+00 & 1.914E-04 & 0.000E+00 & 0.000E+00 & 0.000E+00 \\
30.0 & 2.976E-01 & 4.222E-02 & 2.504E-01 & 4.759E-06 & 3.311E-05 & 1.551E-04 & 1.481E-02 & 0.000E+00 & 1.826E-24 & 3.170E-04 \\
40.0 & 2.730E+00 & 5.182E-01 & 2.114E+00 & 7.058E-03 & 1.314E-02 & 1.741E-02 & 3.537E-01 & 1.316E-07 & 3.610E-05 & 4.747E-02 \\
50.0 & 1.016E+01 & 2.077E+00 & 7.658E+00 & 5.211E-02 & 9.507E-02 & 1.138E-01 & 1.778E+00 & 9.817E-05 & 2.592E-03 & 2.633E-01 \\
60.0 & 2.526E+01 & 5.382E+00 & 1.881E+01 & 1.827E-01 & 3.082E-01 & 3.575E-01 & 5.089E+00 & 1.037E-03 & 1.452E-02 & 7.470E-01 \\
70.0 & 4.950E+01 & 1.081E+01 & 3.657E+01 & 4.194E-01 & 6.883E-01 & 7.909E-01 & 1.079E+01 & 3.135E-03 & 3.761E-02 & 1.571E+00 \\
80.0 & 8.259E+01 & 1.833E+01 & 6.067E+01 & 7.514E-01 & 1.242E+00 & 1.423E+00 & 1.894E+01 & 6.070E-03 & 6.935E-02 & 2.765E+00 \\
\hline
\hline
\colhead{$E_\nu$} & & & & & & & & & & \\
\colhead{(MeV)} & \colhead{$^8$Li} & \colhead{$^9$Li} & \colhead{$^7$Be} & \colhead{$^9$Be} &
\colhead{$^{10}$Be} & \colhead{$^8$B} & \colhead{$^{10}$B} & \colhead{$^{11}$B} & \colhead{$^{10}$C} & \colhead{$^{11}$C} \\
\hline
10.0 & 0.000E+00 & 0.000E+00 & 0.000E+00 & 0.000E+00 & 0.000E+00 & 0.000E+00 & 0.000E+00 & 0.000E+00 & 0.000E+00 & 0.000E+00 \\
20.0 & 0.000E+00 & 0.000E+00 & 0.000E+00 & 0.000E+00 & 0.000E+00 & 0.000E+00 & 0.000E+00 & 2.648E-04 & 0.000E+00 & 2.518E-07 \\
30.0 & 0.000E+00 & 0.000E+00 & 3.331E-06 & 1.525E-04 & 3.542E-09 & 0.000E+00 & 4.759E-06 & 2.501E-01 & 0.000E+00 & 4.221E-02 \\
40.0 & 0.000E+00 & 0.000E+00 & 8.400E-03 & 1.041E-02 & 4.961E-04 & 0.000E+00 & 6.689E-03 & 2.050E+00 & 4.332E-06 & 5.005E-01 \\
50.0 & 1.008E-05 & 2.562E-13 & 5.944E-02 & 6.500E-02 & 6.162E-03 & 1.030E-08 & 4.242E-02 & 7.248E+00 & 2.332E-04 & 1.929E+00 \\
60.0 & 3.630E-04 & 6.660E-08 & 1.857E-01 & 2.028E-01 & 2.324E-02 & 1.494E-05 & 1.278E-01 & 1.753E+01 & 1.294E-03 & 4.860E+00 \\
70.0 & 1.361E-03 & 5.621E-07 & 4.095E-01 & 4.464E-01 & 5.411E-02 & 7.867E-05 & 2.698E-01 & 3.374E+01 & 3.327E-03 & 9.602E+00 \\
80.0 & 2.812E-03 & 1.436E-06 & 7.399E-01 & 7.994E-01 & 9.837E-02 & 1.795E-04 & 4.658E-01 & 5.561E+01 & 6.111E-03 & 1.612E+01
\enddata
\tablecomments{Table~\ref{tab12c_ncmk2} is published in its entirety in the
electronic edition of the {\it Astrophysical Journal}. A portion is
shown here for guidance regarding its form and content.}
\tablenotetext{a}{Decomposition rate. See \S 2 and Eq. \ref{csc12}.}
\label{tab12c_ncmk2}
\end{deluxetable*}

\begin{deluxetable*}{cccccccccc}
\tabletypesize{\tiny}
\tablecaption{
Neutrino-induced charged-current reaction cross sections for $\nu_e$ of 
$^{12}$C in units of $10^{-42}$ cm$^2$ 
based on the PSDMK2 Hamiltonian.
}
\tablehead{
\colhead{$E_\nu$} & & & & & & & & & \\
\colhead{(MeV)} & \colhead{$^{12}$C\tablenotemark{a}} & \colhead{$n$} & \colhead{$p$} & \colhead{$d$} &
\colhead{$t$} & \colhead{$^3$He} & \colhead{$^4$He} & \colhead{$^6$Li} & \colhead{$^7$Li}
}
\startdata
10.0 & 0.000E+00 & 0.000E+00 & 0.000E+00 & 0.000E+00 & 0.000E+00 & 0.000E+00 & 0.000E+00 & 0.000E+00 & 0.000E+00 \\
20.0 & 2.363E-01 & 0.000E+00 & 2.663E-06 & 0.000E+00 & 0.000E+00 & 0.000E+00 & 0.000E+00 & 0.000E+00 & 0.000E+00 \\
30.0 & 4.557E+00 & 0.000E+00 & 3.901E-01 & 1.076E-06 & 0.000E+00 & 6.638E-05 & 1.909E-03 & 0.000E+00 & 0.000E+00 \\
40.0 & 1.802E+01 & 1.183E-03 & 5.282E+00 & 1.233E-02 & 0.000E+00 & 5.975E-02 & 3.247E-01 & 3.825E-05 & 0.000E+00 \\
50.0 & 4.884E+01 & 2.703E-02 & 2.365E+01 & 1.266E-01 & 3.300E-05 & 5.269E-01 & 2.478E+00 & 3.558E-03 & 1.215E-07 \\
60.0 & 1.071E+02 & 1.370E-01 & 6.690E+01 & 5.133E-01 & 1.394E-03 & 1.897E+00 & 8.601E+00 & 2.519E-02 & 1.707E-04 \\
70.0 & 2.011E+02 & 3.824E-01 & 1.450E+02 & 1.313E+00 & 6.217E-03 & 4.621E+00 & 2.065E+01 & 7.503E-02 & 1.126E-03 \\
80.0 & 3.346E+02 & 7.755E-01 & 2.632E+02 & 2.567E+00 & 1.418E-02 & 8.990E+00 & 3.981E+01 & 1.523E-01 & 2.829E-03 \\
\hline
\hline
\colhead{$E_\nu$} & & & & & & & & & \\
\colhead{(MeV)} & \colhead{$^8$Li} & \colhead{$^7$Be} & \colhead{$^9$Be} & \colhead{$^8$B} &
\colhead{$^{10}$B} & \colhead{$^9$C} & \colhead{$^{10}$C} & \colhead{$^{11}$C} & \colhead{$^{12}$N} \\
\hline
10.0 & 0.000E+00 & 0.000E+00 & 0.000E+00 & 0.000E+00 & 0.000E+00 & 0.000E+00 & 0.000E+00 & 0.000E+00 & 0.000E+00 \\
20.0 & 0.000E+00 & 0.000E+00 & 0.000E+00 & 0.000E+00 & 0.000E+00 & 0.000E+00 & 0.000E+00 & 2.663E-06 & 2.363E-01 \\
30.0 & 0.000E+00 & 7.641E-04 & 0.000E+00 & 1.012E-03 & 6.827E-06 & 0.000E+00 & 1.076E-06 & 3.892E-01 & 4.166E+00 \\
40.0 & 0.000E+00 & 1.398E-01 & 1.573E-05 & 6.329E-02 & 1.358E-02 & 0.000E+00 & 1.245E-02 & 5.052E+00 & 1.268E+01 \\
50.0 & 0.000E+00 & 9.564E-01 & 1.355E-03 & 3.774E-01 & 1.405E-01 & 3.162E-05 & 1.102E-01 & 2.176E+01 & 2.493E+01 \\
60.0 & 4.085E-09 & 3.061E+00 & 9.674E-03 & 1.201E+00 & 5.190E-01 & 8.437E-04 & 3.828E-01 & 6.016E+01 & 3.953E+01 \\
70.0 & 9.469E-08 & 7.057E+00 & 2.881E-02 & 2.776E+00 & 1.262E+00 & 3.024E-03 & 9.070E-01 & 1.287E+02 & 5.488E+01 \\
80.0 & 2.948E-07 & 1.339E+01 & 5.859E-02 & 5.242E+00 & 2.442E+00 & 6.394E-03 & 1.733E+00 & 2.315E+02 & 6.945E+01
\enddata
\tablecomments{Table~\ref{tab12c_ccemk2} is published in its entirety in the
electronic edition of the {\it Astrophysical Journal}. A portion is
shown here for guidance regarding its form and content.}
\tablenotetext{a}{Decomposition rate. See \S 2 and Eq. \ref{csc12}.}
\label{tab12c_ccemk2}
\end{deluxetable*}

\begin{deluxetable*}{ccccccccccc}
\tabletypesize{\tiny}
\tablecaption{
Neutrino-induced charged-current reaction cross sections for $\bar{\nu}_e$ of 
$^{12}$C in units of $10^{-42}$ cm$^2$ 
based on the PSDMK2 Hamiltonian.
}
\tablehead{
\colhead{$E_\nu$} & & & & & & & & & & \\
\colhead{(MeV)} & \colhead{$^{12}$C\tablenotemark{a}} & \colhead{$n$} & \colhead{$p$} & \colhead{$d$} &
\colhead{$t$} & \colhead{$^3$He} & \colhead{$^4$He} & \colhead{$^6$He} & \colhead{$^6$Li} & \colhead{$^7$Li}
}
\startdata
10.0 & 0.000E+00 & 0.000E+00 & 0.000E+00 & 0.000E+00 & 0.000E+00 & 0.000E+00 & 0.000E+00 & 0.000E+00 & 0.000E+00 & 0.000E+00 \\
20.0 & 5.917E-01 & 2.368E-03 & 0.000E+00 & 0.000E+00 & 0.000E+00 & 0.000E+00 & 0.000E+00 & 0.000E+00 & 0.000E+00 & 0.000E+00 \\
30.0 & 4.382E+00 & 6.548E-01 & 2.563E-05 & 5.835E-05 & 6.821E-05 & 0.000E+00 & 1.867E-03 & 0.000E+00 & 0.000E+00 & 1.566E-04 \\
40.0 & 1.363E+01 & 4.810E+00 & 1.195E-02 & 1.714E-02 & 3.012E-02 & 0.000E+00 & 1.608E-01 & 1.324E-04 & 1.989E-06 & 6.479E-02 \\
50.0 & 3.079E+01 & 1.571E+01 & 6.777E-02 & 1.026E-01 & 1.932E-01 & 1.069E-04 & 8.939E-01 & 5.357E-03 & 5.897E-04 & 3.480E-01 \\
60.0 & 5.724E+01 & 3.543E+01 & 2.043E-01 & 3.160E-01 & 5.845E-01 & 1.392E-03 & 2.582E+00 & 2.564E-02 & 4.866E-03 & 9.508E-01 \\
70.0 & 9.248E+01 & 6.393E+01 & 4.246E-01 & 6.618E-01 & 1.227E+00 & 4.247E-03 & 5.325E+00 & 6.002E-02 & 1.306E-02 & 1.914E+00 \\
80.0 & 1.344E+02 & 9.928E+01 & 7.099E-01 & 1.109E+00 & 2.095E+00 & 7.954E-03 & 8.998E+00 & 1.028E-01 & 2.338E-02 & 3.218E+00 \\
\hline
\hline
\colhead{$E_\nu$} & & & & & & & & & & \\
\colhead{(MeV)} & \colhead{$^8$Li} & \colhead{$^9$Li} & \colhead{$^7$Be} & \colhead{$^9$Be} & 
\colhead{$^{10}$Be} & \colhead{$^{11}$Be} & \colhead{$^{10}$B} & \colhead{$^{11}$B} & \colhead{$^{12}$B} & \\
\hline
10.0 & 0.000E+00 & 0.000E+00 & 0.000E+00 & 0.000E+00 & 0.000E+00 & 0.000E+00 & 0.000E+00 & 0.000E+00 & 0.000E+00 \\
20.0 & 0.000E+00 & 0.000E+00 & 0.000E+00 & 0.000E+00 & 0.000E+00 & 0.000E+00 & 0.000E+00 & 2.368E-03 & 5.893E-01 \\
30.0 & 1.687E-03 & 0.000E+00 & 0.000E+00 & 5.622E-05 & 5.844E-05 & 2.555E-05 & 1.521E-08 & 6.546E-01 & 3.726E+00 \\
40.0 & 6.814E-02 & 0.000E+00 & 0.000E+00 & 1.674E-02 & 1.892E-02 & 9.553E-03 & 9.832E-04 & 4.727E+00 & 8.713E+00 \\
50.0 & 2.989E-01 & 1.056E-04 & 5.199E-12 & 9.082E-02 & 1.084E-01 & 3.839E-02 & 1.462E-02 & 1.516E+01 & 1.461E+01 \\
60.0 & 7.605E-01 & 1.117E-03 & 1.744E-06 & 2.612E-01 & 3.058E-01 & 8.812E-02 & 5.364E-02 & 3.373E+01 & 2.064E+01 \\
70.0 & 1.478E+00 & 3.086E-03 & 1.104E-05 & 5.345E-01 & 6.164E-01 & 1.578E-01 & 1.161E-01 & 6.031E+01 & 2.635E+01 \\
80.0 & 2.424E+00 & 5.555E-03 & 2.528E-05 & 8.968E-01 & 1.027E+00 & 2.440E-01 & 1.955E-01 & 9.311E+01 & 3.154E+01 \\
\enddata
\tablecomments{Table~\ref{tab12c_ccbmk2} is published in its entirety in the
electronic edition of the {\it Astrophysical Journal}. A portion is
shown here for guidance regarding its form and content.}
\tablenotetext{a}{Decomposition rate. See \S 2 and Eq. \ref{csc12}.}
\label{tab12c_ccbmk2}
\end{deluxetable*}

\section{Neutrino-Nucleus Reaction Cross Sections of $^4$He and $^{12}$C}

New neutrino-induced reaction cross sections on $^{12}$C have been obtained 
by shell model calculations with the SFO Hamiltonian 
\citep[hereafter abbreviated by SC06]{sc06}.
The SFO Hamiltonian describes spin properties of $p$-shell nuclei,
such as Gamow-Teller (GT) transitions, better than conventional 
shell-model Hamiltonians, such as PSDMK2 \citep{mk75,oxb86}.
Systematic improvements in the agreement between calculated and observed
magnetic moments of $p$-shell nuclei supports the use of the SFO Hamiltonian 
\citep{sfo03}, which takes into account the important roles of spin-isospin
interactions, in particular tensor interaction, and is found to lead to 
proper shell evolution \citep{osfg05,sc06}.
 
While a slight modification of the axial-vector coupling constant, 
$g_{A}^{eff}/g_{A}$ =0.95, is enough to reproduce the GT transition in 
$^{12}$C, a large quenching of the coupling constant, 
$g_{A}^{eff}/g_{A}$ =0.7, was taken for other multipoles to
reproduce the inclusive charged-current reaction cross sections 
induced by the DAR neutrinos \citep{sc06}. This is consistent with 
the electron scattering data, where considerable quenching of the
spin $g$-factor, $g_{s}^{eff}/g_{s}$ =0.6$\sim$0.7, explains the M2 
form factor in $^{12}$C (2$^{-}$, $T=1, 19.40$ MeV)
at low momentum transfer \citep{drake68,yama71,gard84}. 
The final state interaction is included by multiplying the relativistic
Fermi function for the charged-current reactions.

Although large quenching of $g_{A}^{eff}/g_{A}$ =0.7 was adopted
for all multipoles other than the GT transitions in \citet{sc06},
electron scattering data indicate a smaller quenching of the spin
$g$-factor for 1$^{-}$ states, i.e., $g_{s}^{eff}/g_{s}$ 
$\approx$0.9 \citep{drake68,yama71}.   
Photo-reaction cross section data indicate that the electric dipole
transition strength is quenched by about 30$\%$ below $E_{x}$ =30 MeV
and a large fraction of the strength is pushed up to higher energy
\citep{ahr75,pyw85,mcl91,suzs03}. 
We therefore adopt separate quenching factors for $g_{A}$: 
$g_{A}^{eff}/g_{A}$ =0.95, 0.7, and 0.9 for the GT (1$^{+}$), 2$^{-}$
spin-dipole, and other multipoles, respectively.  
The Coulomb dipole form factor is also reduced by 30$\%$. 
As the dominant contributions come from the GT and the 2$^{-}$ 
spin-dipole transitions, the inclusive charged-current reaction
cross section in $^{12}$C remains to be explained by the modified
quenching factors. Effects of the change of the contributions 
from other multipoles are insignificant. The shell-model configuration 
space assumed here is the same as in \citet{sc06}, but multiple polarities 
up to $J=4$ are included, instead of just $J=3$.

To enable comparisons, the cross sections for $^{12}$C are obtained 
for the conventional PSDMK2 Hamiltonian in the same way, i.e., with  
$g_{A}^{eff}/g_{A}$ =1.0, 0.75, and 0.9 for the GT, 
2$^{-}$ spin-dipole, and other multipoles, respectively, and   
with the Coulomb dipole form factor reduced by 30$\%$.
 
Neutrino-induced reaction cross sections on $^{4}$He are obtained 
by shell-model calculations with the WBP \citep{wb92} and
SPSDMK \citep{mk75,oxb86} Hamiltonians, with the bare $g_{A}$ \citep{sc06}. 
The $0s$-$0p$-$1s0d$-$1p0f$ and $0s$-$0p$-$1s0d$ configurations are taken for
the shell-model space for the WBP and SPSDMK cases, respectively, and
$^4$He is not treated as a closed core. The axial-vector coupling constant is 
therefore taken to be the bare value, $g_A$.
The shell-model configuration space is extended up to 4 (5) $\hbar\omega$ 
excitations for positive (negative) parity transitions, instead of
just up to 2 (3) $\hbar\omega$ excitations. 
 
Branching ratios for $\gamma$ transitions and proton ($p$), neutron ($n$), 
and $\alpha$ knock-out channels have been obtained from
Hauser-Feshbach theory for $^{12}$C \citep{sc06}.
However, we extend the Hauser-Feshbach calculations by including 
knock-out of a deuteron ($d$), $^{3}$He, and $^{3}$H as well as 
multi-particle knock-out channels.  All possible particle knock-out 
and $\gamma$ transitions are included until the transitions end up
with a residual nucleus with mass number $A = 6 \sim 12$. For $^{4}$He, 
$p$, $n$, and $d$, knock-out channels are taken into account. 
Here, Hauser-Feshbach calculations are carried out for each Hamiltonian,
consistently with the respective energy spectrum.
We allowed $\alpha$-decay (1) after $\gamma$ transition from isospin $T=1$
states in $^{12}$C to $T=0$ states, or (2) directly from $T=1$ states
in $^{12}$C to $T=1$ states in $^8$Be.
We also assumed 1\% isospin non-conservation in $\gamma$ transitions,
as the experimental data for $^{12}$C indicate such a possibility.

Calculated reaction cross sections for various channels 
are shown in Figure 1 for $^{4}$He and in Figures 2 and 3 for $^{12}$C. 
For neutral current reactions, the average of ($\nu$, $\nu$') and
($\bar\nu$, $\bar\nu$') reactions are shown. 
Nuclei produced, including those knocked out, which cannot decay 
further by particle emissions are denoted in the figures.  

Neutral-current and charged-current reaction cross sections on $^{4}$He,
$^{4}$He($\nu, \nu'p$)$^{3}$H, $^{4}$He($\nu, \nu'n$)$^{3}$He, 
$^4$He($\nu, \nu'd$)$^2$H, $^4$He($\nu,\nu' nnp$)$^1$H, 
$^{4}$He($\nu_e, e^{-}p$)$^{3}$He, $^4$He($\nu_e, e^-pp$)$^2$H, 
$^{4}$He($\bar{\nu}_e, e^{+}n$)$^{3}$H, and 
$^4$He($\bar{\nu}_e, e^+nn$)$^2$H,
induced by supernova neutrinos with temperature $T_{\nu}$, 
are shown in Figure 1. For these averaged cross sections, the neutrino 
energy spectra are assumed to be Fermi-Dirac distributions with zero 
chemical potential, to enable comparisons with our earlier studies.
Results for the two shell-model Hamiltonians, WBP and SPSDMK, are shown. 
Results for $^{12}$C obtained with the SFO and the PSDMK2
Hamiltonians are shown in Figures 2 and 3, respectively.
We note that the decomposition cross section of $^{12}$C, 
$\sigma_{^{12}{\rm C},\nu}$, has the following relation to the production 
cross section, $\sigma_{^{12}{\rm C},\nu}(Z_i,A_i)$, of species $i$, 
of which charge number and mass number are $Z_i$ and $A_i$, respectively:
\begin{equation}
\sigma_{^{12}{\rm C},\nu} = 
\sum_i \frac{A_i}{12} \sigma_{^{12}{\rm C},\nu}(Z_i,A_i) .
\label{csc12}
\end{equation}

For use with non-thermal neutrino spectra, cross section values for $^4$He
as a function of the neutrino energy for the WBP and SPSDMK 
Hamiltonians are provided in Tables \ref{tab4hewbp} and \ref{tab4hemk}.
The neutrino-induced reaction cross sections of $^{12}$C for neutral-current
reactions, charged-current reactions of $\nu_e$, and those of $\bar{\nu}_e$
with the SFO (PSDMK2) Hamiltonian are listed in Tables 
\ref{tab12c_ncsfo} (\ref{tab12c_ncmk2}), \ref{tab12c_ccesfo} 
(\ref{tab12c_ccemk2}), and \ref{tab12c_ccbsfo} (\ref{tab12c_ccbmk2}),
respectively.

For $^{4}$He, the neutral current reaction cross sections 
obtained with the WBP Hamiltonian are rather close to those obtained with 
a microscopic ab-initio calculation using AV8' interaction \citep{gazit04},
although the dependence on $T_{\nu}$ is more moderate for WBP. 
We thus take the cross sections obtained by WBP and SFO as a ^^ ^^ standard 
set'' for the evaluation of the production yields of light elements
during supernova explosions.

We now briefly explain important neutrino-nucleus reactions on 
$^{12}$C, relevant for producing light elements. The qualitative 
nature of the reactions does not depend much on the chosen
Hamiltonians, but there are some quantitative differences. 
Light elements are mainly produced by neutral current reactions 
induced by $\nu_{\mu,\tau}$ and $\bar\nu_{\mu,\tau}$, which have
higher temperature than $\nu_{e}$ and $\bar\nu_{e}$.  Note that
neutral current processes involve six kinds of neutrinos.   

We find that $^{11}$B has the largest yield among the light 
elements. The branching ratio for $^{12}$C($\nu, \nu'p$)$^{11}$B 
is about 4 times larger than that for $^{12}$C($\nu, \nu'n$)$^{11}$C. 
The charged-current reaction cross section for
$^{12}$C($\bar{\nu}_e, e^{+}n$)$^{11}$B at $T_{\bar\nu_e}$ =5 MeV
is nearly the same as that for $^{12}$C($\nu, \nu'n$)$^{11}$C
at $T_{\nu}$ =6 MeV. 

$^{10}$B is produced mainly by neutral current reactions, 
$^{12}$C($\nu, \nu'pn$)$^{10}$B and $^{12}$C($\nu, \nu'd$)$^{10}$B.  
The amount of the production is about 6 (4) $\times$
10$^{-3}$ times that of $^{11}$B for the SFO (PSDMK2) Hamiltonian.

$^{9}$Be is produced by neutral current reactions, 
$^{12}$C($\nu, \nu'x$)$^{9}$Be ($x$ = $^3$He, $dp$, and $ppn$) 
and charged-current reactions, $^{12}$C($\bar{\nu}_e, e^{+}x$)$^{9}$Be 
($x$ = $^{3}$H, $dn$ and $pnn$). 
The contribution of the latter reaction at $T_{\bar\nu_e}$ =5 MeV is about 
10$\%$ of the former at $T_{\nu}$ =6 MeV. 
The production of $^{9}$Be is
about 4 (7) $\times$ 10$^{-3}$ times that of $^{11}$B for
the SFO (PSDMK2) Hamiltonian. 

For $^{10}$Be, charged-current reaction cross sections for 
$^{12}$C($\bar{\nu}_e, e^{+}pn$)$^{10}$Be and 
$^{12}$C($\bar{\nu}_e, e^{+}d$)$^{10}$Be are larger than neutral current 
reaction cross section for $^{12}$C($\nu, \nu'pp$)$^{10}$Be at 
($T_{\bar\nu_e}$, $T_{\nu}$) = (5 MeV, 6 MeV), while at
$T_{\bar\nu_e}$ =4 MeV the former is as small as one-fourth 
of the latter. Production yields of $^{10}$Be, thus, depend
on $T_{\bar\nu_e}$. The production of $^{10}$Be by neutral
current processes is about 5 (6) $\times$ 10$^{-4}$ times 
that of $^{11}$B for the SFO (PSDMK2) case. 

$^{7}$Li is mainly produced by neutral current reactions, 
$^{12}$C($\nu, \nu'p\alpha$)$^{7}$Li, etc.  The production
of $^{7}$Li through $^{7}$Be is about 20$\%$ of that from the
neutral current reactions. Contributions from charged-current 
processes are less than 10$\%$ of those from the neutral
current reactions at $T_{\bar\nu_e}$ =5 MeV. $^{6}$Li is 
produced by neutral current processes, but production through 
$^{6}$He is negligible. Light-element synthesis during supernova 
explosions based on the present reaction cross sections is
discussed in \S 4.

\section{SN Nucleosynthesis Model}

In this study, we adopt the same SN nucleosynthesis model employed
by \citet{yk06a,yk06b}, except for the new $\nu$-process reaction rates.
Here we briefly explain the SN explosion model, the SN neutrino model, 
and the nuclear reaction network.

\subsection{SN explosion model}

We consider a 16.2 $M_\odot$ pre-supernova model, corresponding to a
possible progenitor model for SN 1987A \citep{sn90}.
The explosion is proceeded by a spherically symmetric hydrodynamic
calculation using a piecewise parabolic method code \citep{cw84,sn92}.
The explosion energy is set to be 1 B = 1 Bethe = $1 \times 10^{51}$ ergs.
The Lagrangian location of the mass cut is fixed at 1.61 $M_\odot$.

For calculations of the effects of neutrino oscillations, we use the density
profile of the presupernova model. As discussed in \citet{yk06b}, 
shock propagation hardly affects the $\nu$-process 
(with neutrino oscillations). 
There is a resonance of the transition of 2-3 mass eigenstates in the O/C 
layer. 
When the shock wave arrives at this resonance region, the density gradient 
becomes large and, therefore, the resonance becomes non-adiabatic. 
If the adiabaticity is changed by the shock wave, the influence of neutrino 
oscillations could change as well. However, most of the supernova neutrinos 
have already passed this region before the shock arrives, so that 
the affected fraction of neutrinos is very small.

\subsection{SN Neutrino Model}

Here, we briefly explain models for the flux and energy spectra of the 
neutrinos emitted from the neutrino sphere. For simplicity, we assume 
that the neutrino luminosity decreases exponentially with a decay time 
of $\tau_\nu = 3$ s.
The total energy carried out by neutrinos is almost equal to the
binding energy released at the formation of a proto-neutron star.
A characteristic value of the energy is $3 \times 10^{53}$ ergs 
\citep[e.g.][]{wh90}, corresponding to the gravitational
binding energy of a 1.4 $M_{\odot}$ neutron star \citep{ly89,lp01}.
The spectra at the neutrino sphere are assumed 
to follow Fermi-Dirac distributions with zero-chemical potentials.
Note that the temperatures of neutrinos and the total neutrino energy
are somewhat uncertain, and that the $^{11}$B abundance in GCE can be 
used to constrain them.

We consider several neutrino models, parametrized by the neutrino 
temperatures, total energy released in neutrinos, and adopted cross sections.
Table~\ref{neutemp} lists seven models employed in this study.
We use model 1 as the ^^ ^^ standard model'' in this study, with 
$T_{\nu_e}$ = 3.2 MeV, $T_{\bar{\nu}_e}$ = 5.0 MeV, 
$T_{\nu_{\mu,\tau}}$ = 6.0 MeV, and 
$E_{\nu,total}$ = $3.0 \times 10^{53}$ ergs,
where $T_{\nu_e}$, $T_{\bar{\nu}_e}$, $T_{\nu_{\mu,\tau}}$, and 
$E_{\nu,total}$ are the temperatures of $e$-neutrinos, $e$-antineutrinos, 
$\mu$- and $\tau$-neutrinos and their antiparticles, and the total neutrino 
energy.
This set of the neutrino temperatures and total neutrino energy were used 
in the standard model of \citet{yt04,yk05,yk06a,yk06b}.
This model adopts the cross sections of $^{12}$C and $^4$He from SFO and
WBP Hamiltonians, respectively.

Models 1mk, 1p, and 1hw have the same set of the neutrino temperatures
and total neutrino energy as in model 1, but adopt different sets of $^{12}$C
and $^4$He cross sections.
Model 1mk adopts the cross sections of $^{12}$C and $^4$He with PSDMK2 and
SPSDMK Hamiltonians (Tables 1$c$, 1$d$, $6-8$).
Model 1p contains the cross sections of $^4$He($\nu, \nu'p$)$^3$H,
$^4$He($\nu, \nu'n$)$^3$He, $^4$He($\nu_e, e^-p$)$^3$He, and
$^4$He($\bar{\nu}_e, e^+n$)$^3$H for $^4$He and
$^{12}$C($\nu, \nu'p$)$^{11}$B, $^{12}$C($\nu, \nu'n$)$^{11}$C, 
$^{12}$C($\nu_e, e^-p$)$^{11}$C, and $^{12}$C($\bar{\nu}_e, e^+n$)$^{11}$B 
for $^{12}$C evaluated in \citet{sc06}.
Since the reaction rates of $^{12}$C($\nu, \nu'np$)$^{10}$B,
$^{12}$C($\nu, \nu'^3$He)$^9$Be, $^{12}$C($\nu, \nu'p\alpha$)$^7$Li,
and $^{12}$C ($\nu, \nu'n\alpha$) $^7$Be were not evaluated in \citet{sc06},
the rates of these reactions are adopted from HW92.
Model 1hw is equivalent to the standard model in \citet{yk06b}.
It adopts the cross sections of $^4$He and $^{12}$C from HW92.
Model 2 represents the same neutrino energy spectra as used in
\citet{rh02,hk05,yu05,yo07,yu08}.

\begin{deluxetable*}{lccccc}
\tabletypesize{\small}
\tablecaption{
Model parameter: Neutrino temperatures $T_{\nu_e}$, $T_{\bar{\nu}_e}$,
$T_{\nu_{\mu,\tau}}$, total released neutrino energy $E_{\nu,total}$, and
adopted neutrino-nucleus cross sections for $^{12}$C and $^4$He.}
\tablehead{
\colhead{Model} & \colhead{$T_{\nu_e}$} &
\colhead{$T_{\bar{\nu}_e}$} & \colhead{$T_{\nu_{\mu,\tau}}$} &
\colhead{$E_{\nu,total}$} &
\colhead{References for $^{12}$C and $^4$He $\nu$-process} \\
\colhead{} & \colhead{(MeV)} &
\colhead{(MeV)} & \colhead{(MeV)} &
\colhead{($\times 10^{53}$ ergs)} & 
\colhead{}
}
\startdata
model 1    & 3.2 & 5.0 & 6.0 & 3.0  & SFO, WBP \\
model 1mk  & 3.2 & 5.0 & 6.0 & 3.0  & PSDMK2, SPSDMK \\
model 1p   & 3.2 & 5.0 & 6.0 & 3.0  & 
SFO\tablenotemark{a}, WBP\tablenotemark{a}, HW92 \\
model 1hw  & 3.2 & 5.0 & 6.0 & 3.0  & HW92 \\
model 2    & 4.0 & 4.0 & 6.0 & 3.0  & SFO, WBP \\
model LT   & 3.2 & 5.0 & 6.5 & 2.35 & SFO, WBP \\
model ST   & 3.2 & 4.2 & 5.0 & 3.53 & SFO, WBP \\
\enddata
\tablecomments{Neutrino temperatures $T_{\nu_e}$, $T_{\bar{\nu}_e}$,
$T_{\nu_{\mu,\tau}}$, total released neutrino energy $E_{\nu,total}$,
and adopted neutrino-nucleus cross sections for $^{12}$C and $^4$He.}
\tablenotetext{a}{\citet{sc06}.}
\label{neutemp}
\end{deluxetable*}

When we investigate the effects of neutrino oscillations on SN nucleosynthesis,
we consider two additional sets, to take into account uncertainties 
in neutrino temperatures.
Model LT corresponds to the largest $T_{\nu_{\mu,\tau}}$ value and indicates
the $^{11}$B yield close to the upper limit still satisfying GCE constraints 
\citep{fo00,rl00,rs00,al02}.
Model ST corresponds to the smallest $T_{\nu_{\mu,\tau}}$ value and indicates
the value close to the lower limit deduced from GCE models.
We note that the temperature of $\bar{\nu}_e$ in model ST 
is changed to keep $T_{\nu_{\mu,\tau}}/T_{\bar{\nu}_e} \sim 1.2$,
which is the same as model 1.

When neutrino oscillations are taken into account, the neutrinos 
emitted from the neutrino sphere change flavor in passing through the
stellar interior.
The flavor change depends strongly on neutrino oscillation parameters.
We use the following values for these parameters.
The squared-mass differences of the mass eigenstates
$\Delta m^2_{ij} = m^2_i - m^2_j$ are set to be
\begin{eqnarray}
\Delta m^2_{21} &=& 7.9 \times 10^{-5} \quad {\rm eV^2} \nonumber \\
\quad |\Delta m^2_{31}| &=& 2.4 \times 10^{-3} \quad {\rm eV^2 .}
\end{eqnarray}
The values of the mixing angles $\theta_{12}$ and $\theta_{23}$ are
fixed to be
\begin{equation}
\sin ^2 2\theta_{12} = 0.816 \quad {\rm and} \quad 
\sin ^2 2\theta_{23} = 1.
\end{equation}
These parameter values correspond to the family of the so-called large 
mixing angle (LMA) solutions, determined with Super-Kamiokande \citep{SK04}, 
SNO 
\citep[Sudbury Neutrino Observatory;][]{SNO04}, and KamLAND \citep{KL05}.
In the case of $\Delta m^2_{31}$, only the absolute value has been determined.
The positive and negative values correspond to normal and inverted mass
hierarchies, respectively.
For the mixing angle $\theta_{13}$, the upper limit of $\sin ^22\theta_{13}$
has been determined to be $\sin^22\theta_{13} \sim 0.1$ from the CHOOZ
experiment \citep{ap03}.
In this study, we use values of $\sin^22\theta_{13}$ 
between $1 \times 10^{-6}$ and 0.1.

There are two resonances in the transitions between two mass eigenstates 
in the stellar interior of the pre-supernova. The resonance density is 
obtained from 
\begin{eqnarray}
\rho_{res} Y_e &=&
\frac{m_u \Delta m^2_{ji} c^4 \cos 2\theta_{ij}}
{2 \sqrt{2} G_F (\hbar c)^3 \varepsilon_\nu} \\
&=& 6.55 \times 10^6 
 \left(\frac{\Delta m^2_{ji}}{1 {\rm eV^2}}\right)
 \left(\frac{1 {\rm MeV}}{\varepsilon_\nu}\right) \cos 2\theta_{ij}
\quad {\rm g cm^{-3}}, \nonumber
\end{eqnarray}
where $\rho_{res}$ is the resonance density, $Y_e$ is the electron fraction,
$m_u$ is the atomic mass unit, $c$ is the speed of light, $G_F$ is the Fermi
constant, $\hbar$ is th Planck constant divided by $2 \pi$, and $\varepsilon_\nu$
is the neutrino energy.

One resonance is related to the transition between the 2-3 mass eigenstates.
We refer to this resonance as the ^^ ^^ H resonance''.
The density range of the H resonance is $\rho_{res} \sim 300-3000$ g cm$^{-3}$ 
with the energy range of $\varepsilon_\nu \sim 10 - 100$ MeV.
This density range corresponds to the C/O layer and the inner region of
the He layer. Adiabaticity of the H resonance depends on the value of 
the oscillation parameter $\sin^22\theta_{13}$.

The other resonance is due to the transition between the 1-2 mass eigenstates.
We refer to this resonance as the ^^ ^^ L resonance''.
The density range of the L resonance is $\rho_{res} \sim 4-40$ g cm$^{-3}$.
The location of the L resonance is in the He layer.
The L resonance is an adiabatic resonance in the range of neutrino
oscillation parameters considered in this study.
Details of neutrino oscillations in this supernova model are provided
in \citet{yk06b}.

\subsection{Nucleosynthesis Model}

We calculate the nucleosynthesis of the supernova explosion using a
nuclear reaction network consisting of 291 nuclear species as used
in \citet{yt04,yk05,yk06a,yk06b}, and tabulated in Table 1 in \citet{yt04}.
The difference from previous studies \citep{yk06a,yk06b} is that here
new cross sections for neutrino-$^{12}$C and neutrino-$^4$He reactions
are used (see \S 2). Reaction rates are calculated using these new cross 
sections and the neutrino energy spectra discussed above.
When we take neutrino oscillations into account, the formulation of 
the rates of the charged-current $\nu$-process reactions is given by
equation (8) in \citet{yk06b}.
The range of the neutrino energy for integration is capped at 160 MeV.
The reaction rates of the other $\nu$-process reactions are adopted
from HW92, and the effects of neutrino oscillations are not included 
for those reactions.

\section{Light Element Yields}

\subsection{Production of Light Elements in SNe}

Below we discuss the production processes of light elements in
the SN model with the new cross sections for $^4$He and $^{12}$C.
The mass fraction distribution of the light elements at 1000 s 
after core bounce is shown in Figure~\ref{massfrac}.
The mass fractions of $^7$Li and $^{11}$B are larger than those of
other light elements. The third abundant species is $^{10}$B.
The abundances of $^6$Li and $^9$Be are smaller than that of $^{10}$B
by more than an order of magnitude. 
The radioactive isotope $^{10}$Be is produced at 
a level similar to those of $^6$Li and $^9$Be.

\subsubsection{$^7$Li}

Most of $^7$Li is produced in the He/C layer.
The $^7$Li is originally produced as $^7$Li and its isobar $^7$Be.
In an inner region of the He/C layer ($M_r \la 4.8 M_\odot$), the production 
of $^7$Be dominates.
Most of $^7$Be is produced through $^4$He($\nu,\nu'n)^3$He and
$^3$He($\alpha,\gamma)^7$Be. The charged-current reaction 
$^4$He($\nu_e,e^-p)^3$He contributes less to the $^7$Be production, 
but is important when neutrino oscillations are 
considered (see \S 5). In the O/Ne layer, the mass fraction of $^7$Be is small.
Most of $^7$Be is produced through the $\nu$-process of $^{12}$C.
When the shock wave arrives, almost all the $^7$Be is photo-disintegrated by 
$^7$Be($\gamma,\alpha)^3$He. During the expansion stage, $^7$Be is again produced 
from $^{12}$C.

In the outer region of the He/C layer, most of $^7$Li is produced via the
reaction sequence $^4$He($\nu,\nu'p)^3$H($\alpha,\gamma)^7$Li.
Some of $^7$Li is contributed by the charged-current reaction 
$^4$He($\bar{\nu}_e,e^+n)^3$H.
In this region, all $^3$H produced through the $\nu$-process is consumed 
by $\alpha$-capture to $^7$Li during the explosion.
On the other hand, a much smaller amount of $^7$Be is produced, because the 
shock temperature is too low to effectively enable $^3$He($\alpha,\gamma)^7$Be 
during the explosion.
In the inner region of the He/C layer, the produced $^7$Li experiences
$\alpha$-capture to yield $^{11}$B during the explosion.
In the O-rich layer, $^7$Li is produced through the $\nu$-process of $^{12}$C, 
and further $\alpha$-capture produces $^{11}$B.

\subsubsection{$^{11}$B}

\begin{figure}
\epsscale{1.1}
\plotone{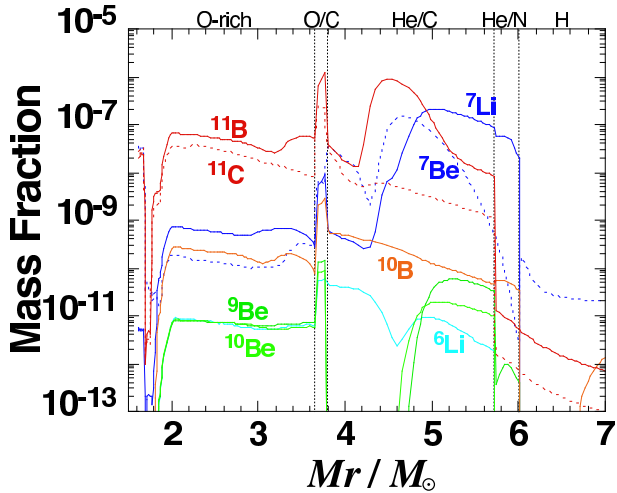}
\caption{
Mass fraction distributions for model 1.
Blue solid and dashed lines correspond to the mass fractions of $^7$Li and 
$^7$Be, respectively.
Red solid and dashed lines are the mass fractions of $^{11}$B and $^{11}$C.
}
\label{massfrac}
\end{figure}

About 60\% of $^{11}$B is produced in the He/C layer.
Most of the $^{11}$B is produced through $^7$Li($\alpha,\gamma)^{11}$B
in the mass coordinate range $4.2 M_\odot \la M_r \la 4.9 M_\odot$.
In this region, the mass fraction of $^7$Li becomes very small owing to
this reaction (see Figure~\ref{massfrac}). A very small amount of the 
isobar $^{11}$C is co-produced through $^{12}$C($\nu,\nu'n)^{11}$C.

In the O-rich layer (O/Ne and O/C layers), both $^{11}$B and $^{11}$C
are produced through the $\nu$-process of $^{12}$C. The sum of their 
mass fractions is about $10^{-6}$ in the O/C layer, because the mass 
fraction of $^{12}$C is also large. The mass fractions of $^{11}$B and 
$^{11}$C are similar about 10 s after the explosion.
Some $^{11}$B is destroyed by $^{11}$B($\alpha,p)^{14}$C.
In the O/Ne layer, $^{11}$B is produced as $^{11}$C.
The branching ratio of $^{11}$B is larger than that of $^{11}$C in the
$\nu$-process of $^{12}$C.
Therefore, the amount of $^{11}$B produced through the $\nu$-process is
larger than the amount of $^{11}$C.
However, more than 90\% of $^{11}$B is lost due to the reaction
$^{11}$B($\alpha,p)^{14}$C at shock arrival.
The $\nu$-process that continues after the explosion increases the $^{11}$B
abundance again, but to a lesser extent than $^{11}$C.
The main destruction reaction of $^{11}$C is $^{11}$C($\alpha,p)^{14}$N,
but it is practically negligible.
The produced $^{11}$C decays to $^{11}$B by $\beta^+$-decay and electron
capture with a half-life of 20.39 minutes.
Thus, the mass fraction of $^{11}$B in the O-rich layer is larger than
that of $^{11}$C in Figure~\ref{massfrac}.

\subsubsection{$^6$Li and $^9$Be}

The production processes of $^6$Li and $^9$Be are connected.
About 65\% and 60\% of $^6$Li and $^9$Be, respectively, are produced 
in the He/C layer.
Most of $^9$Be is produced through the $\nu$-process reaction
$^{12}$C($\nu,\nu'x)^9$Be.
However, it is decomposed after shock arrival at $M_r \la 5.0 M_\odot$.
The main destructive reaction is $^9$Be($\alpha,n)^{12}$C.
About a half of $^6$Li is produced through 
$^4$He($\nu, \nu'd)^2$H($\alpha, \gamma)^6$Li in the region
$3.8 M_\odot \la M_r \la 4.6 M_\odot$ in the He/C layer.
Additional $^6$Li is synthesized through $^{12}$C($\nu, \nu'x)^6$Li
before shock arrival and through $^9$Be($p,\alpha)^6$Li after
shock arrival.
In the inner region of the He/C layer $^6$Li is destroyed through
$^6$Li($p,\alpha)^3$He.

In the O-rich layer, $^9$Be is mainly produced through 
$^{12}$C($\nu,\nu'x)^9$Be.
Almost all of the $^9$Be produced before shock arrival is destroyed
completely through $^9$Be($\alpha,n)^{12}$C by the shock.
It is supplied again through $^{12}$C($\nu,\nu'x)^9$Be during 
the expansion stage. It is partly produced through 
$^{12}$C($\bar{\nu}_e,e^+)^{12}$B($p,\alpha)^9$Be. In this layer, the 
main production process for $^6$Li is $^9$Be($p,\alpha)^6$Li.
This reaction is effective even before shock arrival.
The temperature increase from the shock reduces the $^6$Li abundance
through $^6$Li($p,\alpha)^3$He.
However, $^6$Li is supplied again via $^9$Be($p,\alpha)^6$Li during
post-shock expansion. The contribution from $^{12}$C($\nu,\nu'x)^6$Li is 
small.

\subsubsection{$^{10}$Be}

Radioactive $^{10}$Be is produced in both the O-rich and He/C layers.
It is mainly produced through the charged-current reaction
$^{12}$C($\bar{\nu}_e, e^+x)^{10}$Be and the neutral-current reaction
$^{12}$C($\nu,\nu'x)^{10}$Be.
The contribution from the charged-current reaction is larger than
that from the neutral-current reaction in this model.
The produced $^{10}$Be is destroyed by $^{10}$Be($\alpha,n)^{13}$C
at shock arrival at $M_r \la 4.8 M_\odot$.
In the O-rich layer, however, the $\nu$-process reaction still increases
the $^{10}$Be amount during the expansion stage.
We note that the $\nu$-process reactions producing $^{10}$Be directly
are included in this study for the first time.
When we do not include these reactions, $^{10}$Be is produced through
$^{12}$C($\bar{\nu}_e,e^+p)^{11}$Be($\gamma,n)^{10}$Be.

\subsubsection{$^{10}$B}

About 80\% of the $^{10}$B amount is produced through the $\nu$-process of 
$^{12}$C, mainly via $^{12}$C($\nu,\nu' x)^{10}$B.
About a half of $^{10}$B is produced in the O-rich (O/Ne and O/C) layer.
In the O-rich layer, some $^{10}$B is produced by 
$^{13}$C($p,\alpha$)$^{10}$B after shock arrival.
The amount of $^{10}$B increases again in the expansion due to the supply,
through the $\nu$-process, of $^{12}$C.
A small amount of $^{10}$B is also produced through 
$^6$Li($\alpha, \gamma)^{10}$B in the He/C layer.
Destruction after shock passage is negligible in the He/C layer.

\begin{deluxetable*}{lccccc}
\tabletypesize{\small}
\tablecaption{
Yields of $^7$Li, $^{11}$B, $^6$Li, $^9$Be, $^{10}$Be, and $^{10}$B.}
\tablehead{
\colhead{Species} & \colhead{model 1} & \colhead{model 1mk} & 
\colhead{model 2} & \colhead{model 1p} & \colhead{model 1hw} \\
\colhead{} & \colhead{($M_\odot$)} & \colhead{($M_\odot$)} & 
\colhead{($M_\odot$)} & \colhead{($M_\odot$)} & \colhead{($M_\odot$)}
}
\startdata
$^7$Li    & $2.67 \times 10^{-7}$  & $4.14 \times 10^{-7}$  &
$2.54 \times 10^{-7}$  & $3.06 \times 10^{-7}$\tablenotemark{a}  & 
$2.36 \times 10^{-7}$\tablenotemark{b}  \\
$^{11}$B  & $7.14 \times 10^{-7}$  & $8.67 \times 10^{-7}$  &
$6.72 \times 10^{-7}$  & $7.51 \times 10^{-7}$\tablenotemark{a}  & 
$6.26 \times 10^{-7}$\tablenotemark{b}  \\
$^6$Li    & $4.67 \times 10^{-11}$ & $3.60 \times 10^{-11}$ & 
$4.19 \times 10^{-11}$ & $4.61 \times 10^{-12}$ & $3.46 \times 10^{-12}$ \\
$^9$Be    & $6.56 \times 10^{-11}$ & $9.65 \times 10^{-11}$ & 
$5.57 \times 10^{-11}$ & $1.69 \times 10^{-11}$ & $1.36 \times 10^{-11}$ \\
$^{10}$Be & $3.54 \times 10^{-11}$ & $3.55 \times 10^{-11}$ & 
$1.69 \times 10^{-11}$ & $4.18 \times 10^{-12}$ & $4.18 \times 10^{-12}$ \\
$^{10}$B  & $1.08 \times 10^{-9}$ & $6.10 \times 10^{-10}$ & 
$1.05 \times 10^{-9}$ & $2.45 \times 10^{-9}$  & $2.45 \times 10^{-9}$  \\
\enddata
\tablenotetext{a}{Data are adopted from Table IV of \citet{sc06}.}
\tablenotetext{b}{Data are adopted from \citet{yk06b}.}
\label{lyield}
\end{deluxetable*}

\subsection{Yields of Light Elements}

We consider light element yields resulting from different sets of the relevant 
$\nu$-process cross sections.
The yields of $^{7}$Li, $^{11}$B, $^{6}$Li, $^{9}$Be, $^{10}$Be, and $^{10}$B
are listed in Table~\ref{lyield}. 
We first compare the yields of the light elements of model 1 with
those of model 1p (see \S 3.2 and Table~\ref{neutemp}).
In model 1 and model 2, new reaction rates obtained in \S 2 with the 
WBP+SFO Hamiltonians and new branching ratios are used.  
Model 1p uses the reaction rates of $^4$He and $^{12}$C evaluated in SC06.
The branching ratios used to produce $^7$Li, $^7$Be, $^9$Be, and $^{10}$B 
from $^{12}$C were not evaluated in SC06.
Therefore, we adopted the rates of these reactions from HW92 in model 1p
(see also \S 3.2).

The yields of $^7$Li and $^{11}$B in model 1 become slightly smaller than
those in model 1p, but are not very different.
$^7$Li and $^{11}$B are the main products of the $\nu$-process
from $^4$He and $^{12}$C.
The cross sections of $^4$He($\nu,\nu'p)^3$H and $^4$He($\nu,\nu'n)^3$He
in this study are slightly smaller than those in SC06, owing to 
the consideration of the branches of $dd$ and $nnpp$.
The cross sections of $^{12}$C($\nu,\nu'x)^{11}$B and 
$^{12}$C($\nu,\nu'x)^{11}$C in this study scarcely change from those of SC06.

The $^{10}$B yield of model 1 is smaller than that of model 1p by a factor
of 2.3. This reflects the difference of the $\nu$-process reaction rates 
to produce $^{10}$B from $^{12}$C.
The total $\nu$-process reaction rate to produce $^{10}$B from $^{12}$C
in this study is smaller than that of HW92 by a factor of 3.
The $^{10}$B production through $^6$Li($\alpha, \gamma)^{10}$B in the
He/C layer slightly suppresses the decrease.

The $^9$Be yield in model 1 is larger than those in model 1p by a factor 4.
The neutrino reaction rate responsible for production of $^9$Be in this study 
is larger than that used by HW92 by a factor of 6. 
Therefore, the enhancement of the $^9$Be yield is not as large as
that of the $\nu$-process product $^9$Be.
The destruction of $^9$Be during the explosion might suppress the
yield enhancements.

The $^6$Li yield in model 1 is larger than that in model 1p by about
1 order of magnitude.
As explained in \S 4.1.3, the new branches $^4$He($\nu, \nu'd)^2$H,
$^4$He($\nu_e,e^-pp)^2$H, and $^4$He($\bar{\nu}_e, e^+nn)^2$H strongly
enhance deuteron production.
The produced deuterons are captured to produce $^6$Li through
$^2$H($\alpha, \gamma)^6$Li.
This reaction sequence enhances the $^6$Li yield by a factor of 5.
The newly evaluated branches to produce $^6$Li and the increasing
reaction rate of $^9$Be production through the $\nu$-process from
$^{12}$C also enhance the $^6$Li yield.
Thus, it is important to evaluate the rates of the $\nu$-process
branches from $^4$He and $^{12}$C when the $^6$Li yield is investigated.

We calculated the reaction cross sections of the $\nu$-process branches
to produce $^{10}$Be. These reactions should enhance the yield of $^{10}$Be.
The yield of $^{10}$Be in model 1 is larger than that in model 1p by 
a factor of 8.5. 
This is due to the additional $\nu$-process reactions.
We note that the yield of $^{10}$Be strongly depends on the $\bar{\nu}_e$
temperature because the cross section of $^{12}$C($\bar{\nu}_e,e^+x)^{10}$Be
is large. 
In the case of model 2, which uses a $\bar{\nu}_e$ temperature smaller
than that in model 1, the $^{10}$Be yield is smaller than in model 1 
by a factor of 2. 
This decrease is due to the decrease in the rate of 
$^{12}$C($\bar{\nu}_e,e^+x)^{10}$Be over that of model 1 by a factor of 5.

We compare $^7$Li and $^{11}$B yields of models 1 and 1hw 
(see Table~\ref{lyield}).
The $^7$Li and $^{11}$B yields of model 1 are larger by factors of 1.13
and 1.14 than the corresponding yields in model 1hw.
The larger yields reflect the fact that the cross sections of 
neutrino-$^4$He reactions for neutral- and charged-current used in this study
are larger than those of the corresponding values in HW92.
On the other hand, the production of $n$ and $p$ through the $\nu$-process
might suppress the enhancement of the $^7$Li and $^{11}$B production.
We note that the total cross section of neutral-current $\nu$-process
reactions on $^{12}$C in this study is slightly smaller than that in HW92.
However, the $^{11}$B yield is not smaller than the one obtained with the old
cross sections. Most of $^{11}$B is produced through $^7$Li($\alpha,\gamma)^{11}$B
and the $\nu$-process from $^{12}$C. The production through $^7$Li($\alpha,\gamma)^{11}$B 
increases the $^{11}$B yield when the new reaction rates are used.

We also compare light element yields of models 1 and 1mk.
model 1mk uses the same neutrino temperature set as model 1, and 
cross sections are evaluated using the PSDMK2 Hamiltonian for $^{12}$C and
SPSDMK for $^4$He. The yield of $^7$Li in model 1mk is larger than 
that in model 1 by a factor of 1.6.
This is because the cross sections of $^4$He with the SPSDMK Hamiltonian
are larger than the corresponding ones with the WBP Hamiltonian
for a given neutrino temperature.
The yields of other light elements have dependencies similar to the
cross sections of the $\nu$-process reactions to produce the corresponding
nuclei.
In the case of $^9$Be, the yield in model 1mk is larger than 
the corresponding yield in model 1.
The $\nu$-process production cross section exhibits the same trend.
On the other hand, the yields of $^6$Li and $^{10}$B in model 1mk are smaller 
than those in model 1. 
The SPSDMK cross section of $^4$He($\nu, \nu'd)^2$H is smaller than
the WBP one by more than a factor of 2.
The PSDMK2 cross section to produce $^{10}$B is also 
smaller than the one from the SFO Hamiltonian.
The $^{10}$Be yield is almost same between the two models.
We do not find large differences in the cross section to produce
$^{10}$Be from $^{12}$C in the neutrino temperature range in this study.

The case of $^{11}$B is an exception.
The yield of $^{11}$B in model 1mk is larger than that in model 1 
by a factor of 1.21, although the $\nu$-process cross sections to produce 
$^{11}$B and $^{11}$C evaluated using PSDMK2 Hamiltonian are smaller than 
those using SFO.
A large amount of $^{11}$B is produced via $^7$Li($\alpha,\gamma)^{11}$B
through the reaction sequence from 
$^4$He($\nu,\nu'p)^3$H($\alpha,\gamma)^7$Li.
The larger production of $^{11}$B reflects the larger cross sections
of $^4$He($\nu,\nu'p)^3$H evaluated using the SPSDMK Hamiltonian.

\begin{figure*}
\epsscale{1.1}
\plottwo{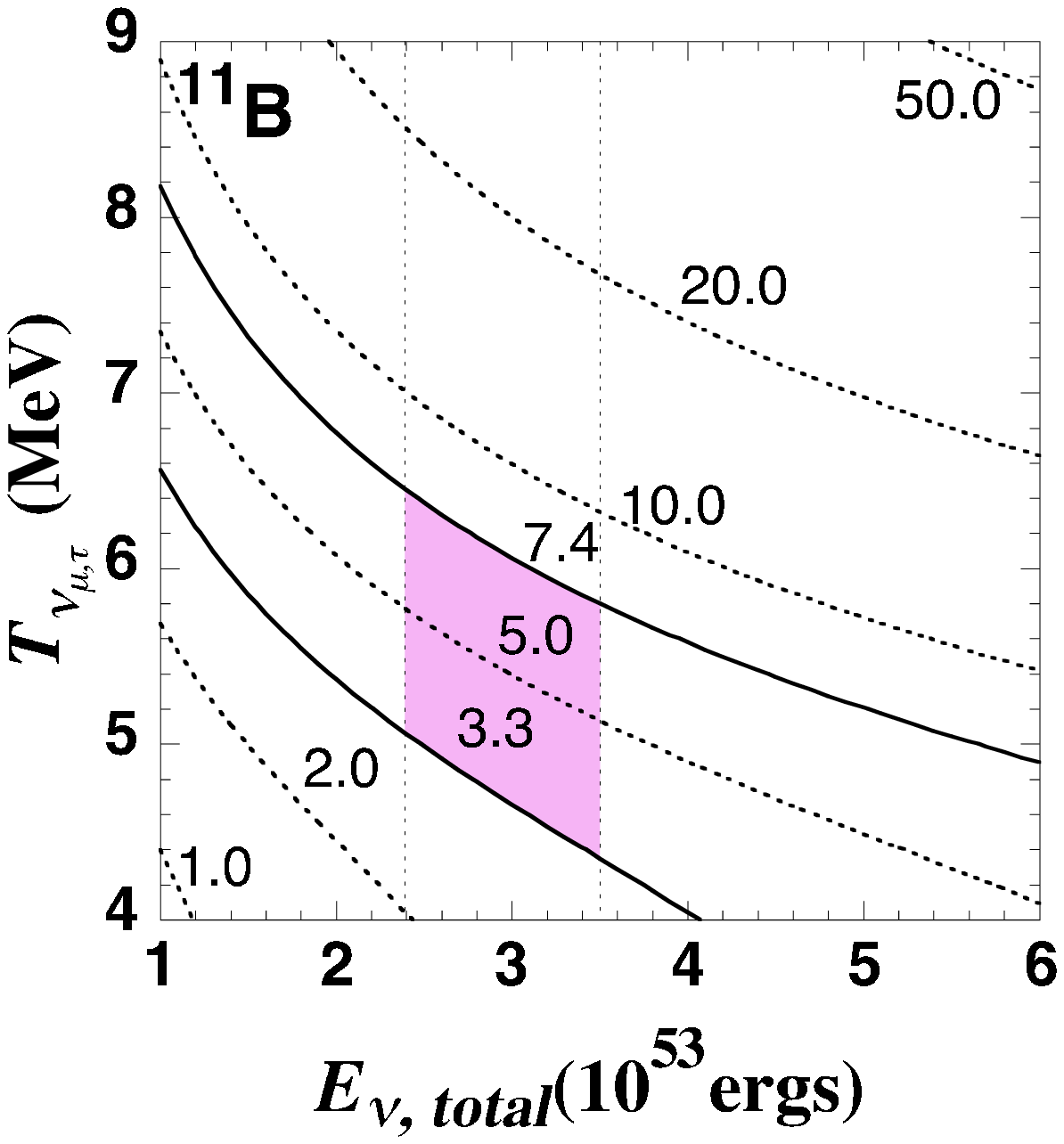}{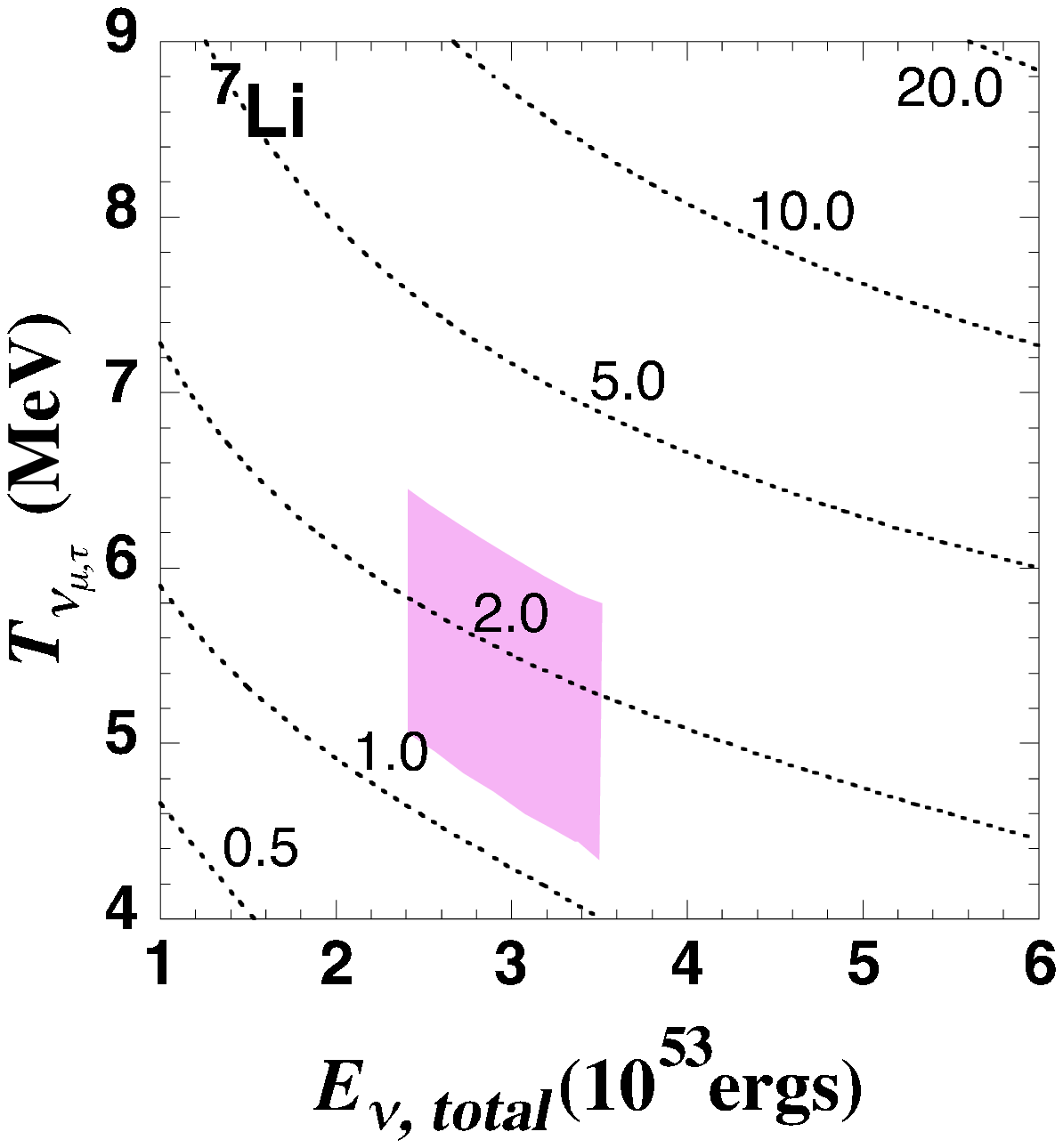}
\caption{
Contours of the yields of $^{11}$B ({\it left}) and $^7$Li ({\it right})
with SFO+WBP neutrino cross sections, as a function of total neutrino energy
and the neutrino temperature $T_{\nu_{\mu,\tau}}$.
The temperature of $\nu_e$ and $\bar{\nu}_e$ is assumed to be fixed at 4 MeV.
The number attached to each line indicates the yield in units of 
$10^{-7} M_\odot$.
The region between the two solid contour lines satisfies the SN contribution
constraint from GCE modeling.
The range between the two vertical dotted lines indicate the possible neutrino
energy range evaluated for the gravitational energy of a neutron star.
The shaded region satisfies both the $^{11}$B GCE constraint and the total
neutrino energy.
}
\label{contSFO}
\end{figure*}

\begin{deluxetable*}{lcccc}
\tabletypesize{\small}
\tablecaption{
Yield ranges of light elements constrained by the SN contribution of $^{11}$B
in GCE models.}
\tablehead{
\colhead{} & \multicolumn{2}{c}{WBP+SFO Model} & 
\multicolumn{2}{c}{SPSDMK+PSDMK2 Model} \\
\cline{2-3} \cline{4-5}
\colhead{Species} & \colhead{Minimum Yield} & \colhead{Maximum Yield} & 
\colhead{Minimum Yield} & \colhead{Maximum Yield} \\
\colhead{} & \colhead{($M_\odot$)} & \colhead{($M_\odot$)} &
\colhead{($M_\odot$)} & \colhead{($M_\odot$)}
}
\startdata
$^{11}$B  & $3.30 \times 10^{-7}$  & $7.40 \times 10^{-7}$  &
$3.30 \times 10^{-7}$  & $7.40 \times 10^{-7}$  \\
$^6$Li    & $2.21 \times 10^{-11}$ & $5.25 \times 10^{-11}$ &
$1.39 \times 10^{-11}$ & $3.11 \times 10^{-11}$ \\
$^7$Li    & $1.17 \times 10^{-7}$  & $2.82 \times 10^{-7}$  &
$1.40 \times 10^{-7}$  & $3.37 \times 10^{-7}$  \\
$^9$Be    & $2.94 \times 10^{-11}$ & $7.08 \times 10^{-11}$ &
$3.23 \times 10^{-11}$ & $7.96 \times 10^{-11}$ \\
$^{10}$Be & $2.86 \times 10^{-11}$ & $3.94 \times 10^{-11}$ &
$2.81 \times 10^{-11}$ & $3.96 \times 10^{-11}$ \\
$^{10}$B  & $4.39 \times 10^{-10}$ & $1.20 \times 10^{-9}$ &
$1.88 \times 10^{-10}$ & $5.30 \times 10^{-10}$ \\
\enddata
\label{GCEyield}
\end{deluxetable*}

\subsection{Constraints on the Neutrino Energy Spectrum}

Light elements are continuously produced by Galactic cosmic rays (GCRs),
nucleosynthesis in SNe, AGB stars, and so on. GCE models deduce the 
contributions of various production sites from the observed light-element 
abundances in stars as a function of their metallicity. The contribution 
of the $^{11}$B yield in SNe was evaluated in GCE models
\citep{fo00,rl00,rs00,al02}. The yield of $^{11}$B in representative 
SNe of progenitor mass $\sim 20 M_\odot$ is
\begin{equation}
3.3 \times 10^{-7} \, M_\odot \, \la M({\rm ^{11}B}) \, \la \, 
7.4 \times 10^{-7} \, M_\odot.
\end{equation}
\citet{yk05} evaluated the range of the temperature of $\nu_{\mu,\tau}$
and $\bar{\nu}_{\mu,\tau}$ neutrinos, $T_{\nu_{\mu,\tau}}$, to be between 
4.8 and 6.6 MeV. It was assumed that the energy spectra of SN neutrino
follow Fermi-Dirac distributions with zero chemical potentials and
($T_{\nu_e}, T_{\bar{\nu}_e}$) = (3.2 MeV, 5 MeV). They also discussed the 
effects of the degeneracy of the neutrino energy spectra. However, the range 
of allowed neutrino temperatures also depends on the cross sections of the 
$\nu$-process reactions. We re-evaluate the range of neutrino temperatures 
from GCE model constraints with the new cross sections, and also discuss the 
yields of other light elements in the temperature range given by the $^{11}$B 
constraint.

We calculated light-element nucleosynthesis in the $T_{\nu_{\mu,\tau}}$ range 
between 4 and 9 MeV on grids with steps of
0.2 MeV and in the $E_\nu$ range between $1 \times 10^{53}$ and 
$6 \times 10^{53}$ ergs on grids with steps of $1 \times 10^{53}$ ergs.
We fixed the temperatures of $\nu_e$ and $\bar{\nu}_e$ at 3.2 and 5 MeV,
respectively, for simplicity as in \citet{yk05}.
Based on the nucleosynthesis calculations, we derive contours of $^{11}$B
yield in $E_\nu$ - $T_{\nu_{\mu,\tau}}$ space.
Figure~\ref{contSFO} shows the contours of the $^{11}$B yield for the $\nu$-process 
cross sections from the WBP+SFO model. The total neutrino energy deduced from the 
gravitational binding energy of a neutron star is in the range
$2.4 \times 10^{53} {\rm ergs} \le E_\nu \le 3.5 \times 10^{53} {\rm ergs}$.
From the GCE-range of the $^{11}$B yield (SN component) and the above range for
the total neutrino energy, we constrain the neutrino temperature to be 
confined in
\begin{equation}
4.3 {\rm MeV} \la T_{\nu_{\mu,\tau}} \la 6.5 {\rm MeV}.
\end{equation}
This range is slightly smaller than the previously evaluated range;
4.8 MeV $\la$ $T_{\nu_{\mu,\tau}}$ $\la$ 6.6 MeV \citep{yk05}.
The contours of the $^7$Li yield are shown in Figure~\ref{contSFO}($b$).
The yield expected from GCE considerations is indicated by the shaded region.
We also evaluated the range of yields for other species, as allowed by
the $T_{\nu_{\mu,\tau}}$ and $E_{\nu,total}$ constraints 
(see Table~\ref{GCEyield}). Yields of model 1 are in the allowed range 
for each nuclear species considered.

We note that the lower limit of $T_{\mu,\tau}$ is smaller than the assumed
value of $T_{\bar{\nu}_e}$.
If we assume that $T_{\nu_{\mu,\tau}}/T_{\bar{\nu}_e}$ keeps a constant
ratio, the lower limit of $T_{\nu_{\mu,\tau}}$ should be larger.
When we assume $T_{\nu_{\mu,\tau}}/T_{\bar{\nu}_e} = 1.2$, which is equal
to the ratio in model 1, the lower limit of $T_{\nu_{\mu,\tau}}$ is
5.0 MeV. This corresponds to model ST (see Table~\ref{neutemp}).

\begin{figure*}
\plottwo{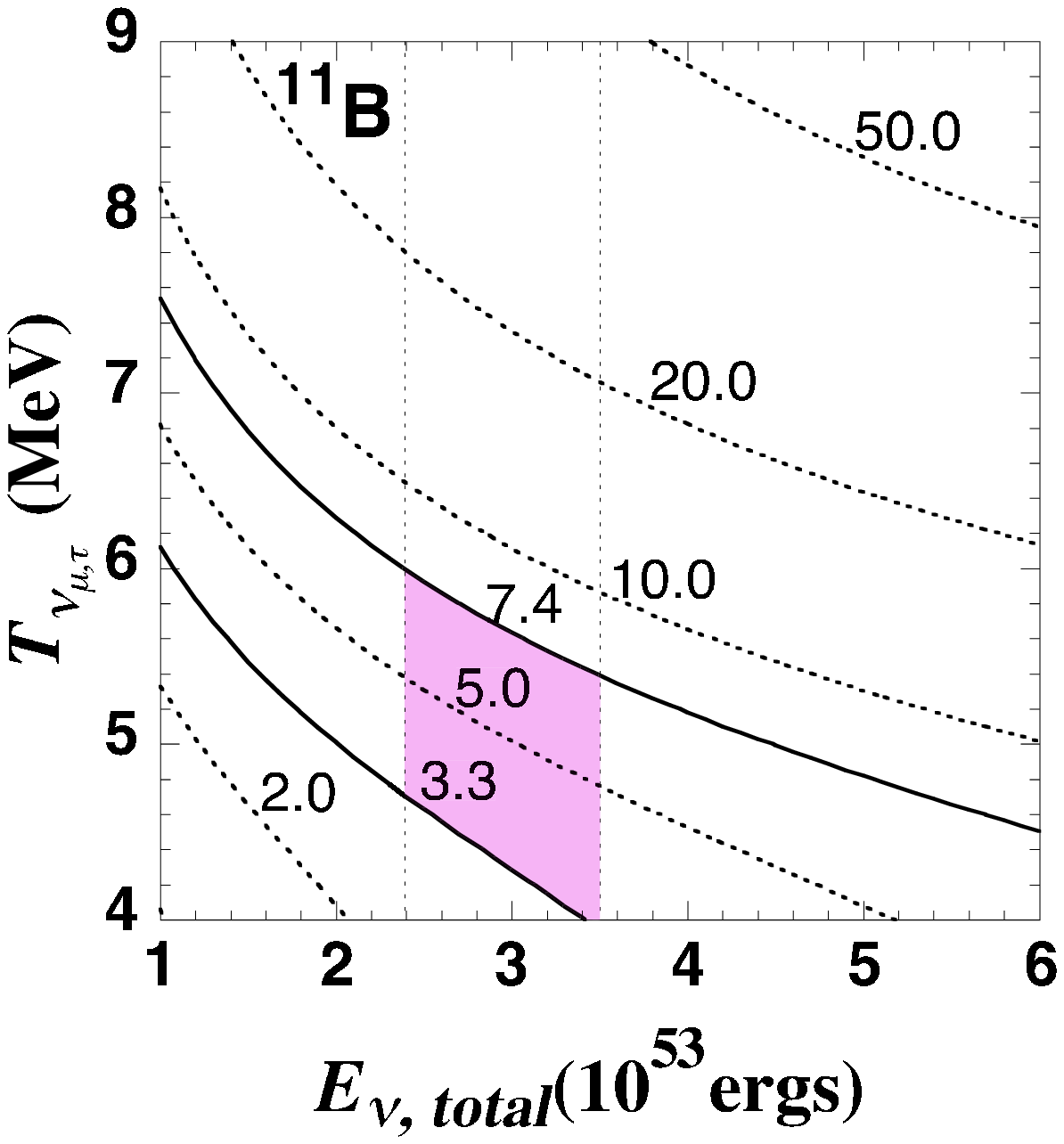}{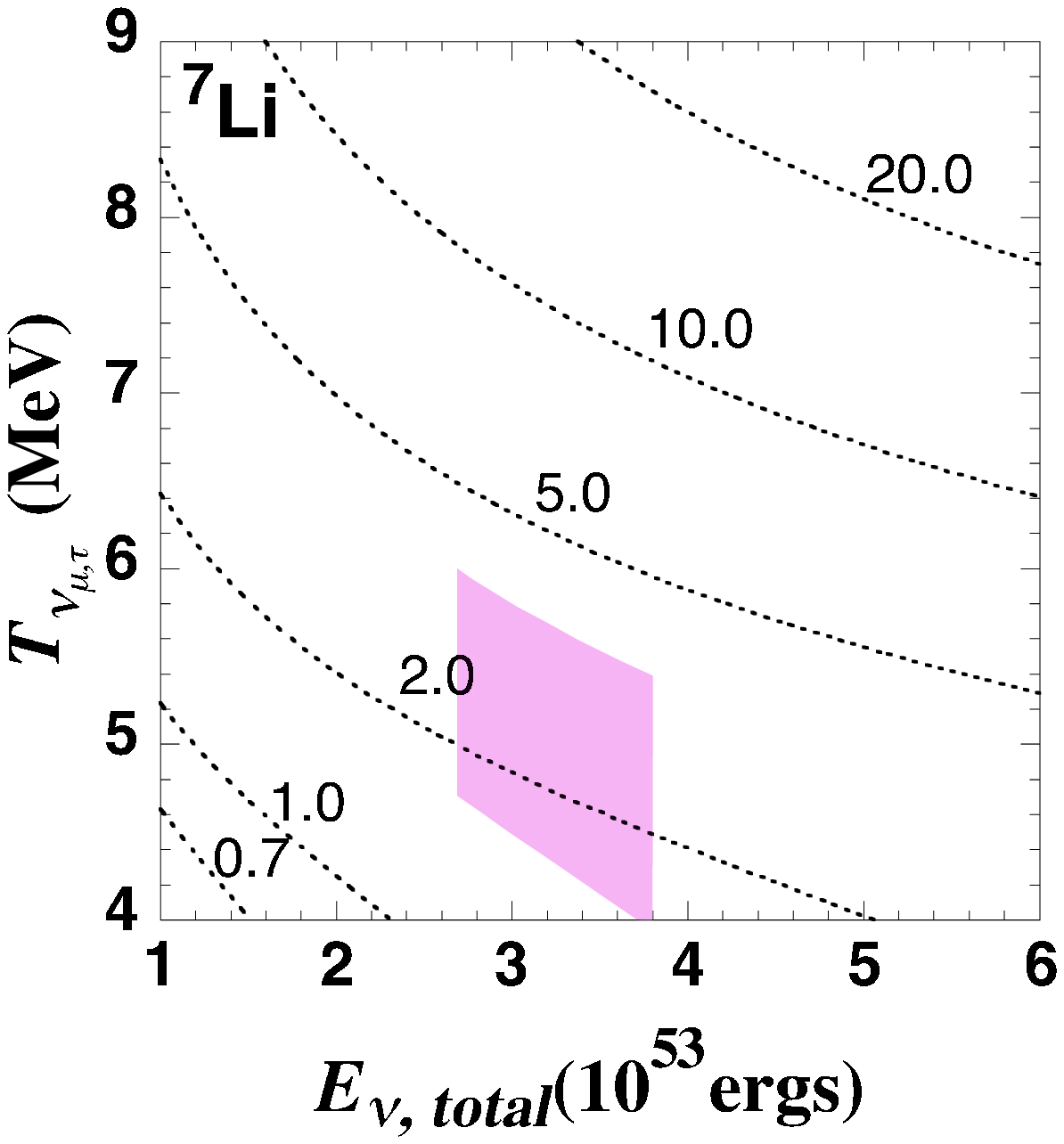}
\caption{
Same as Figure~\ref{contSFO}, but for the SN model with PSDMK2+SPDMK-based 
neutrino cross sections.
}
\label{contMK}
\end{figure*}

In order to exhibit the effects of different shell model Hamiltonians,
we evaluated the range of the neutrino temperature using 
the $\nu$-process cross sections of the SPSDMK + PSDMK2 model.
Figure~\ref{contMK}($a$) shows the corresponding contours of the 
$^{11}$B yield for the SPSDMK + PSDMK2 model.
With this cross section set, we find a larger yield of $^{11}$B
than for the WBP+SFO model, with the same neutrino radiation.
The range of the neutrino temperature consistent with the $^{11}$B 
constraint in GCE models is now
\begin{equation}
4.0 {\rm MeV} \la T_{\nu_{\mu,\tau}} \la 6.0 {\rm MeV},
\end{equation}
slightly shifted to smaller values than obtained in the WBP+SFO model.

Yields of other elements, constrained by the allowed ranges of neutrino 
temperature and total neutrino energy, are given in Table~\ref{GCEyield}.
The contours of the $^7$Li yield are shown in Figure~\ref{contMK}($b$).
The upper and lower limits of the yields are different from those of the
WBP+SFO model, but by less than 20\% for most cases.
This change is much smaller than the basic yield range for each species.
However, the yields of $^6$Li and $^{10}$B show relatively large differences 
due to the difference of the contribution from $^4$He($\nu, \nu'd)^2$H.

\subsection{Effect of Neutrino Chemical Potential}

Most of the calculations on the $\nu$-process in supernovae are performed
with the neutrino energy spectra, assuming of Fermi-Dirac distributions 
and zero-chemical potentials.
On the other hand, studies of detailed neutrino transport in core-collapse
supernovae have shown that the neutrino energy spectra are closer to 
a slightly ^^ ^^ degenerate'' distribution than those with
zero-chemical potential.
In order to investigate the detailed dependence on neutrino degeneracy,
the $\nu$-process cross sections as a function of the neutrino energy
are required. 
Here we investigate the effects of the neutrino degeneracy 
on the light-element yields.
We note that we take into account the neutrino degeneracy
only for the $\nu$-process of $^4$He and $^{12}$C.
We do not take account of the neutrino degeneracy for other $\nu$-process
reactions adopted from HW92 because their reaction rates are shown with 
definite neutrino temperatures derived with the assumption of Fermi-Dirac 
distributions with zero-chemical potential.

\begin{figure}[b]
\epsscale{1.2}
\plotone{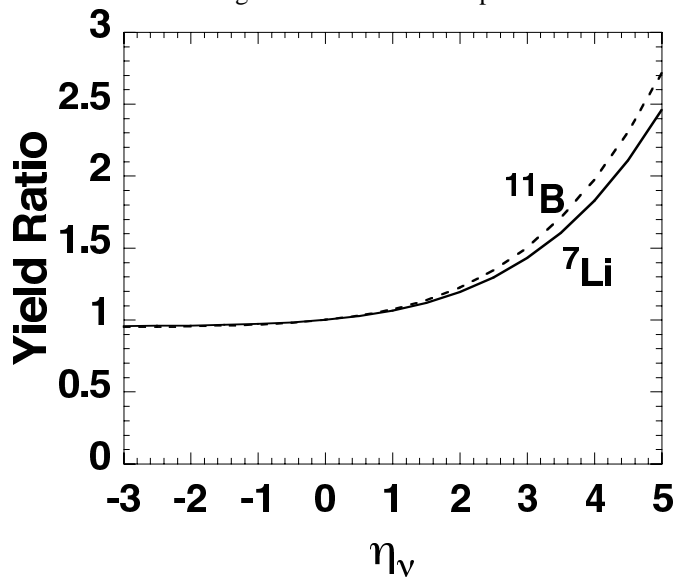}
\caption{
Relation between the yield ratios of $^7$Li ({\it solid lines}) and $^{11}$B 
({\it dotted lines}) to those in case of model 1 and the neutrino degeneracy
parameter $\eta_\nu$.
The neutrino temperatures and the total neutrino energy are fixed to the
values of ($T_{\nu_e}$, $T_{\bar{\nu}_e}$, $T_{\nu_{\mu,\tau}}$,
$E_{\nu,total}$) = (3.2 MeV, 5 MeV, 6 MeV, $3 \times 10^{53}$ ergs).
}
\label{muYield}
\end{figure}

The yield ratios of $^7$Li and $^{11}$B relating to the neutrino degeneracy
parameter $\eta_\nu = \mu_\nu/kT_\nu$ are shown in Figure~\ref{muYield}.
Here we assumed that the degeneracy parameter $\eta_\nu$ does not depend
on neutrino flavors.
The yield ratios increase with the neutrino degeneracy.
In the case of $\eta_\nu = 3$, the yield ratios of $^7$Li and $^{11}$B
are 1.4 and 1.5, respectively.
\citet{yk05} discussed the effect of the neutrino degeneracy on the 
light-element yields using an analytical approximation for the $\nu$-process
cross sections of $^4$He and $^{12}$C.
We find that the $^7$Li and $^{11}$B yields in the case of $\eta_\nu = 3$
would be increased by about 50\% compared to the yields for $\eta_\nu = 0$.
Therefore, our analytical evaluation approximates the numerical evaluation 
well. We obtained a similar dependence of the yield ratios on the neutrino
degeneracy for the other light elements; the yield ratios are between
1.4 and 1.5 in the case of $\eta_\nu = 3$.
The constraint of the neutrino temperature is smaller by about
0.1 MeV in the case of $\eta_\nu = 3$ \citep{yk05}.

\begin{figure*}
\plottwo{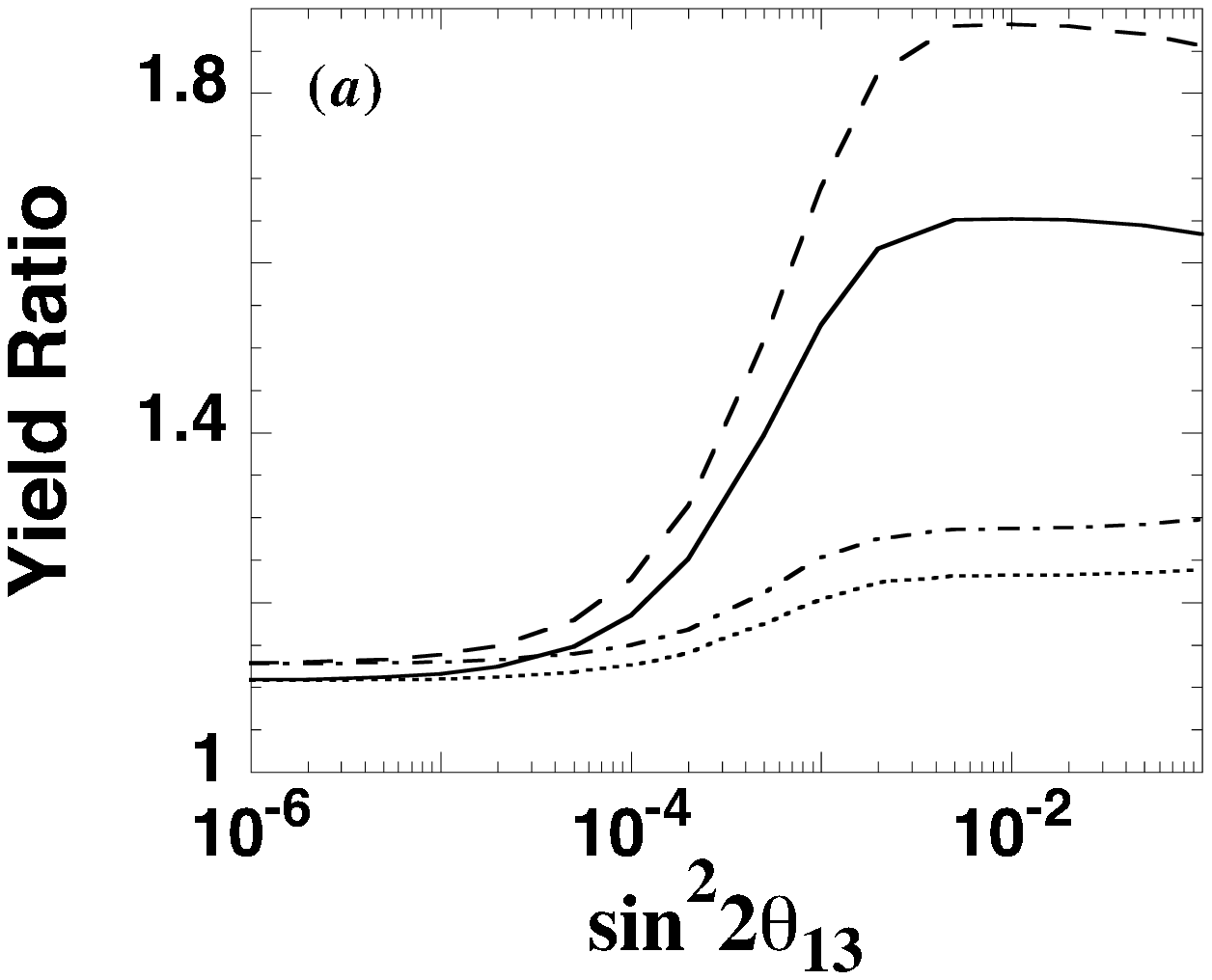}{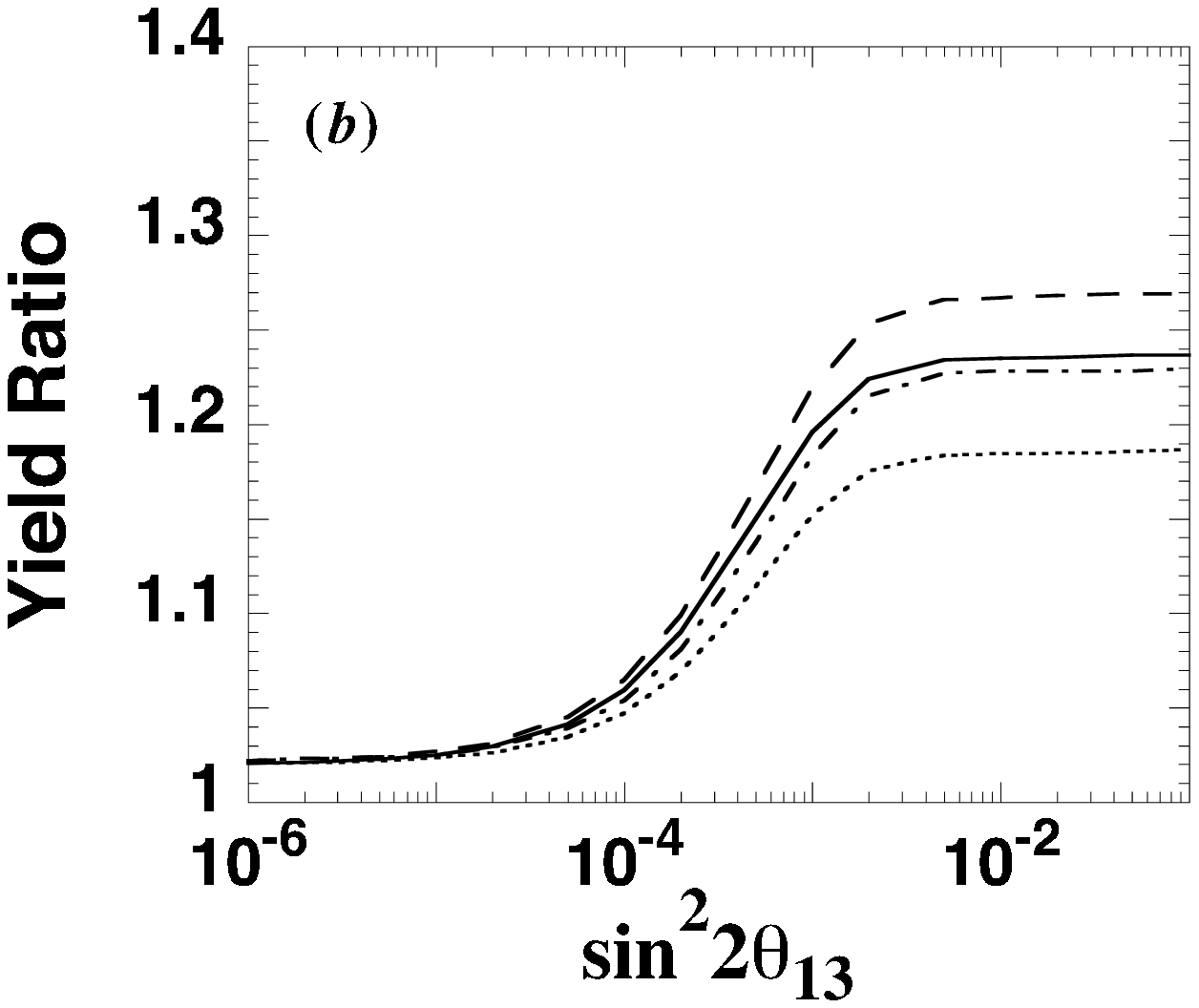}
\caption{
Yield ratios of ($a$) $^7$Li and ($b$) $^{11}$B.
Solid and dotted lines correspond to normal and inverted mass hierarchies,
respectively, in model 1.
Dashed and dot-dashed lines correspond to normal and inverted mass hierarchies
in model 1hw.
}
\label{Yratio_SFO}
\end{figure*}

\section{Changed $^7$Li and $^{11}$B Yields due to Neutrino Oscillations}

Neutrino oscillations change the flavors of the neutrinos emitted from
the neutrino sphere during their passage through the stellar interior.
The average neutrino energies of $\nu_e$ and $\bar{\nu}_e$ increase
due to the neutrino oscillations, and their enhancement depends on
neutrino oscillation parameters, i.e., mass hierarchy and the mixing
angle $\theta_{13}$.
In this study, we evaluate the flavor transition probabilities by the
neutrino oscillations using the same procedure as in \citet{yk06a,yk06b}.
We evaluate the rates of the charged-current $\nu$-process reactions of 
$^4$He and $^{12}$C using the flavor-transition probabilities and the
cross sections derived in \S 2.
Then we calculate detailed nucleosynthesis with the $\nu$-process reactions.

\subsection{Differences Due to New Neutrino-Nucleus Cross Sections}

The dependence of the $^7$Li and $^{11}$B yields on neutrino oscillation 
parameters is influenced by the cross sections of neutrino-nucleus
reactions \citep{yk06a,yk06b}. We study the effects on the $^7$Li and 
$^{11}$B yields due to neutrino oscillations by comparing models 1 and 1hw.
model 1hw is the standard model in \citet{yk06a,yk06b}.

Figure~\ref{Yratio_SFO}$a$ shows the relation between the $^7$Li yield 
ratio, i.e., the ratio of the $^7$Li yield to the one without neutrino 
oscillations, and the mixing angle $\sin^22\theta_{13}$.
We observe that the dependence on the oscillation parameters, i.e., mass 
hierarchies and mixing angle $\theta_{13}$ does not change qualitatively.
The increase in the $^7$Li yield is larger in a normal mass hierarchy
than in an inverted hierarchy, and the yield increases for the case of
$\sin^22\theta_{13} \ga 2 \times 10^{-3}$, i.e., the H resonance is adiabatic.
On the other hand, the maximum value of the yield ratio is somewhat reduced.
The maximum yield ratio is 1.65, which is smaller than the value of 1.87 
found in the previous study. 
The maximum $^7$Li yield is $4.41 \times 10^{-7} M_\odot$ in 
a normal mass hierarchy and for $\sin^22\theta_{13} = 1 \times 10^{-2}$.
The difference of the neutral-current cross section of $^4$He becomes larger 
for a smaller neutrino temperature. In this case, the contribution from the 
neutral-current reactions of $\nu_e$ and $\bar{\nu}_e$ is larger in the new 
rates than it is for the HW92 rates. The contribution from charged-current 
reactions in the new rates becomes smaller.

In the case of an inverted mass hierarchy, the maximum increase in the
$^7$Li yield is a factor of 1.24, which is slightly smaller than the value 
found in the previous study. 
The new reaction rates slightly decrease the effect of neutrino oscillations. 
The maximum yield of $^7$Li is $3.31 \times 10^{-7} M_\odot$ for 
$\sin^22\theta_{13}=0.1$.

Figure~\ref{Yratio_SFO}$b$ shows the dependence of the $^{11}$B yield ratio
on the mixing angle $\sin^22\theta_{13}$ using the new and old cross sections.
The $^{11}$B yield increases most effectively in a normal mass hierarchy 
and for the case of the adiabatic H resonance. 
The $^{11}$B yield reaches $8.83 \times 10^{-7} M_\odot$ in the case of 
a normal mass hierarchy and $\sin^22\theta_{13}=0.1$. 
The maximum yield ratio is 1.24.
The maximum increase in the $^{11}$B yield is smaller with the new cross 
sections. 
In the case of an inverted mass hierarchy, the increase in the $^{11}$B
yield is smaller than the corresponding value in a normal mass hierarchy.
The maximum yield is $8.48 \times 10^{-7} M_{\odot}$ for 
$\sin^22\theta_{13} = 0.1$. 
The maximum yield ratio is 1.19.

\begin{figure*}
\plottwo{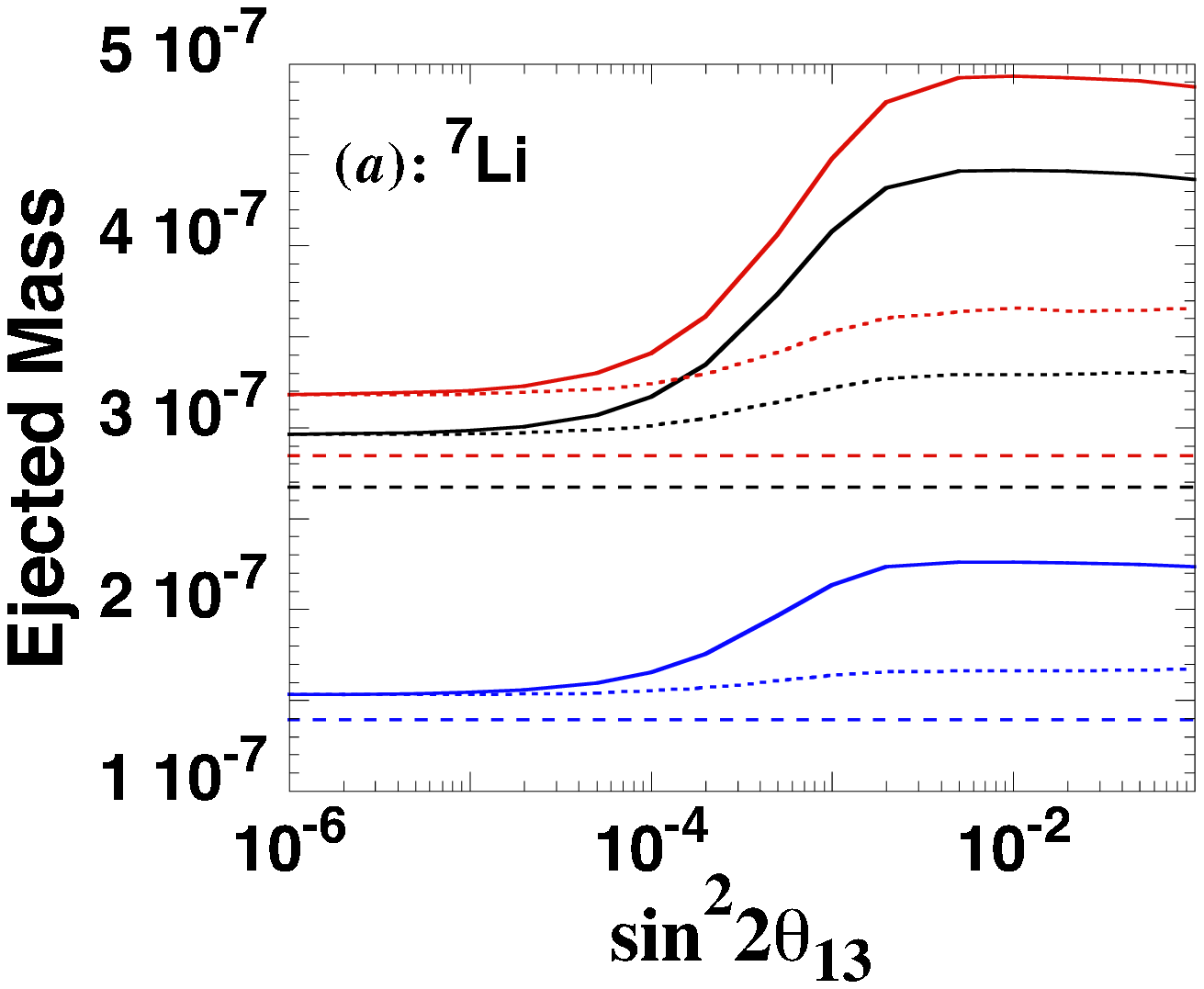}{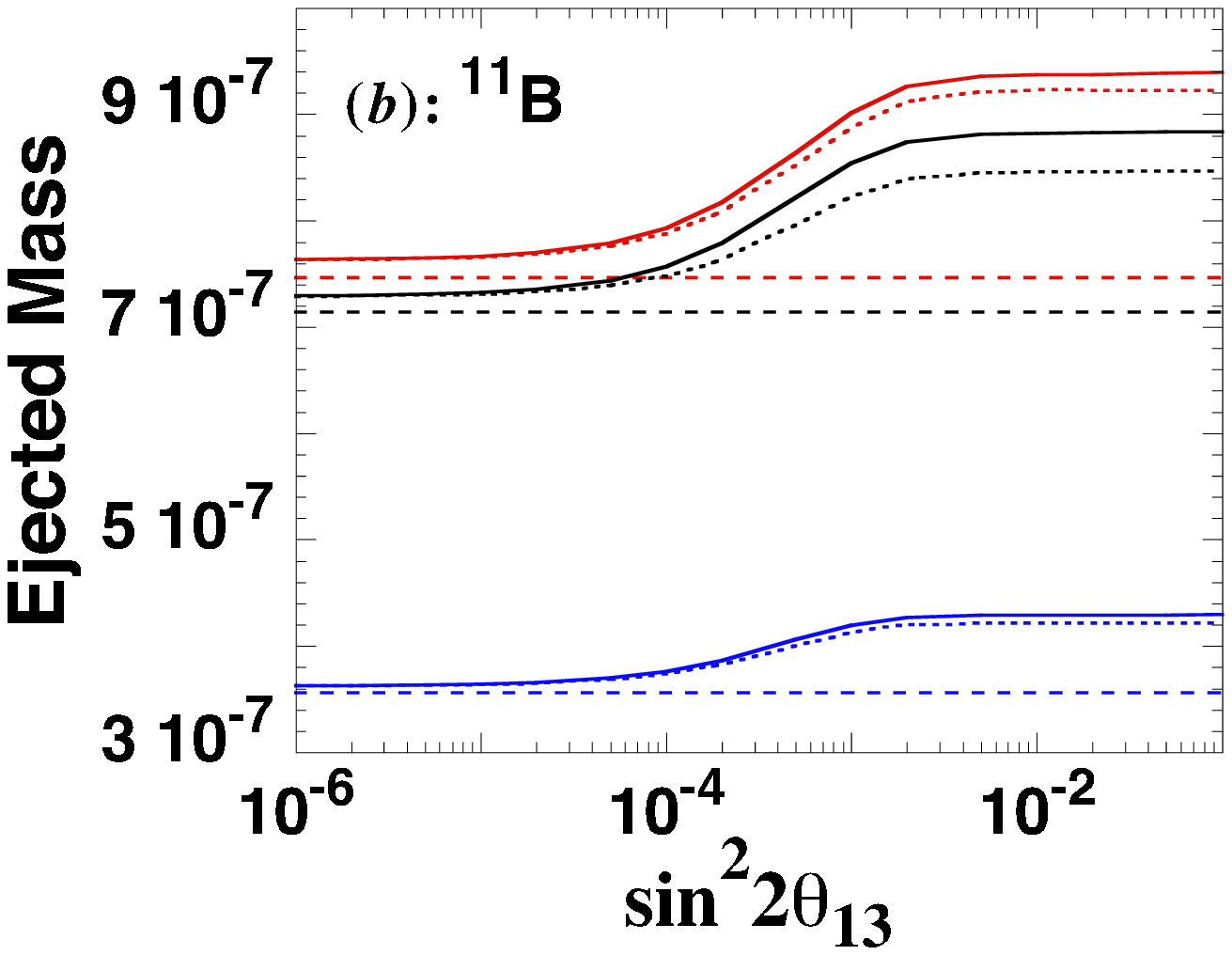}
\caption{
Dependence of the yields of ($a$) $^7$Li and ($b$) $^{11}$B
in units of $M_\odot$ on mixing angle $\sin^22\theta_{13}$ 
with different assumptions for 
the neutrino temperature $T_{\nu_{\mu,\tau}}$.
Black, blue, and red lines indicate models, 1, ST, and LT, respectively.
Solid and dotted lines correspond to the normal and inverted mass
hierarchies, respectively.
}
\label{Yields_Tx}
\end{figure*}

\subsection{Dependence on $T_{\nu_{\mu,\tau}}$}

We investigated the dependence of the $^{11}$B yield on the temperature
of $\nu_{\mu,\tau}$ and $\bar{\nu}_{\mu,\tau}$ neutrinos, and evaluated the 
temperature range satisfying the $^{11}$B abundance constraints in GCE models 
\citep{yk05}. This range also depends on the cross sections of the $\nu$-process.
Therefore, we consider models LT and ST, which present $T_{\nu_{\mu,\tau}}$ 
and $E_\nu$ values different from model 1 (see \S 3.2). 
The values of the neutrino 
temperatures and the total neutrino energy are given in Table~\ref{neutemp}.

When neutrino oscillations are not considered, the $^7$Li yield varies between 
$1.39 \times 10^{-7} M_\odot$ and $2.88 \times 10^{-7} M_\odot$ due to 
the allowed  range of the neutrino temperature $T_{\nu_{\mu,\tau}}$. 
The variation of the $^{11}$B yield is between $3.56 \times 10^{-7} M_\odot$ 
and $7.46 \times 10^{-7} M_\odot$.
Both yields thus change by about a factor of 2 over this temperature range.
Therefore, we must consider variations due to neutrino oscillations as well as 
neutrino temperature.

The dependence of the $^7$Li and $^{11}$B yields on mass hierarchies and 
the mixing angle $\theta_{13}$ in these three models is shown in 
Figure~\ref{Yields_Tx}.
We observe that the $^7$Li yield varies between $1.39 \times 10^{-7} M_\odot$ 
and $4.93 \times 10^{-7} M_\odot$, widening with increasing temperature. 
However, it is 
difficult to distinguish the effect of neutrino oscillations and temperature.
If the $^7$Li yield is smaller than $3.2 \times 10^{-7} M_\odot$, the increase
in the yield due to the neutrino oscillations cannot be distinguished from the
$^7$Li yield range deduced from the uncertainty of the neutrino temperature.
Even for larger $^7$Li yield, the constraints on the mass hierarchy
and the mixing angle $\theta_{13}$ become ambiguous due to the uncertainty
in the neutrino temperature.

Figure~\ref{Yields_Tx}$b$ shows that the variation of the $^{11}$B yield
is between $3.56 \times 10^{-7} M_\odot$ and $9.40 \times 10^{-7} M_\odot$.
However, as pointed out above, it is difficult to constrain oscillation 
parameters.
If the $^{11}$B yield is smaller than $7.6 \times 10^{-7} M_\odot$,
the uncertainty due to the neutrino temperature and the increase
in the yield due to neutrino oscillations are not distinguishable.
Even for larger yields, there are no clear differences between the yields 
in a normal mass hierarchy and in an inverted mass hierarchy.


\subsection{Constraints on Oscillation Parameters from the $^7$Li/$^{11}$B Ratio}

In \citet{yk06a,yk06b} we proposed a constraint of neutrino oscillation
parameters derived from the $^7$Li/$^{11}$B abundance ratio. 
We have shown above that both the $^7$Li and $^{11}$B yields change with 
the neutrino temperature by about a factor two. 
When the abundance ratio of $^7$Li/$^{11}$B is considered, the uncertainty 
due to the neutrino temperature cancels out. 
Then, the dependence on mass hierarchy and mixing angle $\theta_{13}$ is 
most clearly revealed. 
We found that the $^7$Li/$^{11}$B ratio is larger than 0.83 in a normal 
mass hierarchy and $\sin^22\theta_{13} \ga 2 \times 10^{-3}$. 
When we do not consider neutrino oscillations, the $^7$Li/$^{11}$B
ratio is 0.71, at most. 
However, the $^7$Li/$^{11}$B ratio does depend on the relevant 
$\nu$-process cross sections. 
We evaluate the range of the $^7$Li/$^{11}$B ratio with the 
new cross sections discussed in \S 2.

\begin{figure}[b]
\plotone{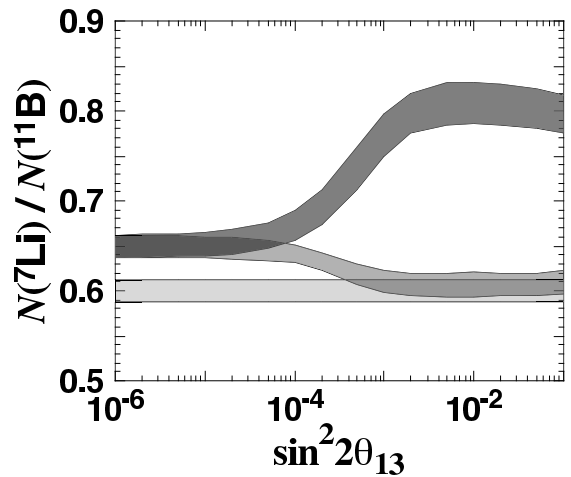}
\caption{
$^7$Li/$^{11}$B abundance ratio as a function of the mixing angle
$\sin^22\theta_{13}$. Dark and medium shaded regions correspond to 
normal and inverted mass hierarchies, respectively. The lightly shaded 
region indicates the ratio obtained without neutrino oscillations.
Each range is drawn using the results of models 1, 2, LT, and ST.
}
\label{LiBratio}
\end{figure}

We evaluate the $^7$Li/$^{11}$B range using models 1, 2, LT, and ST.
Figure~\ref{LiBratio} shows the abundance ratio of $^7$Li/$^{11}$B 
with the relation to $\sin^22\theta_{13}$ evaluated using the WBP+SFO model.
When we do not consider neutrino oscillations, the $^7$Li/$^{11}$B
ratio lies between 0.59 and 0.61. The $^7$Li/$^{11}$B ratio changes by 
about 12\% among these four neutrino temperature models.
As shown in \citet{yk06a,yk06b}, the uncertainties of the $^7$Li and 
$^{11}$B yields by neutrino temperatures are canceled out when we adopt
the $^7$Li/$^{11}$B ratio.

\begin{figure*}
\epsscale{1.0}
\plottwo{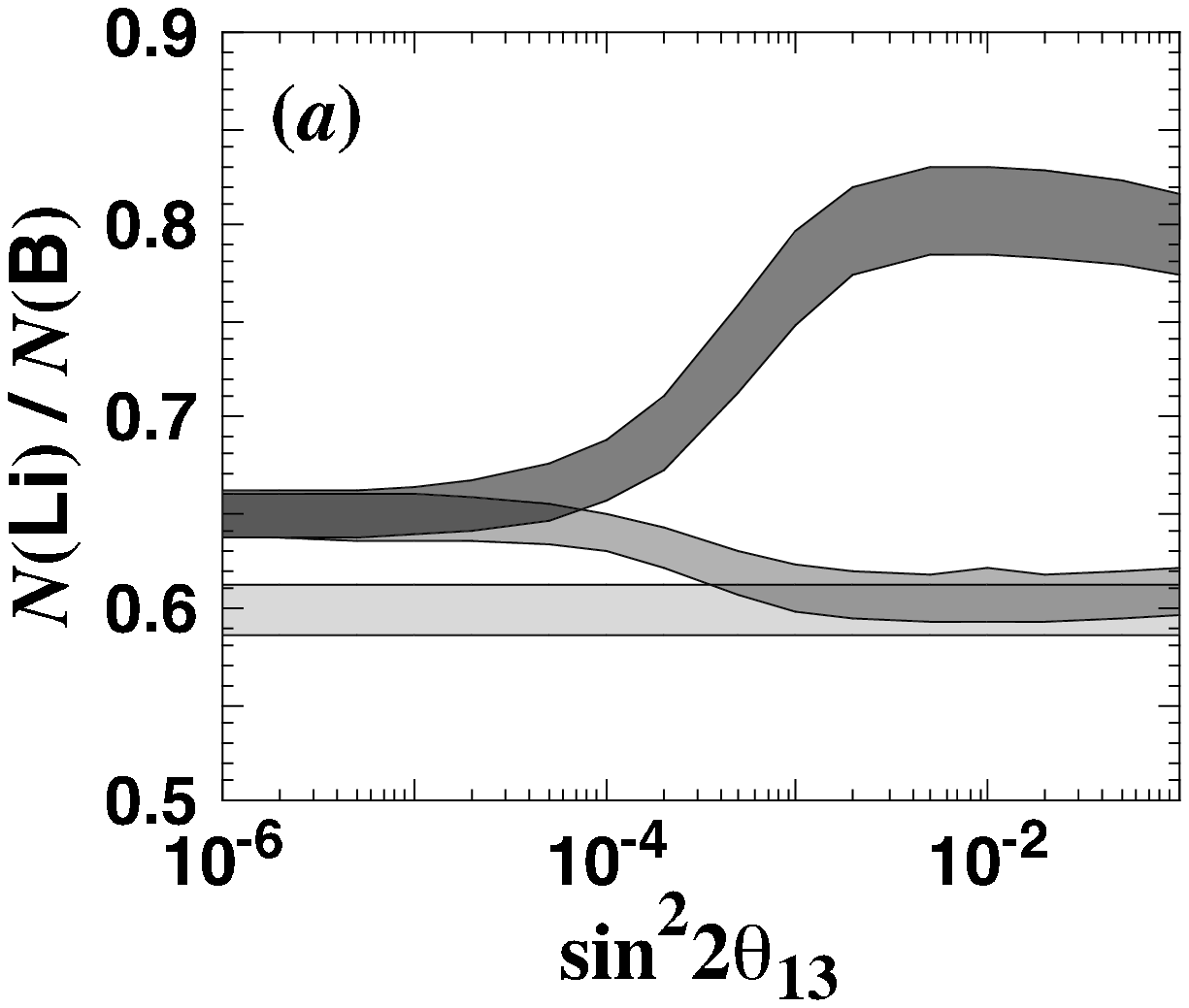}{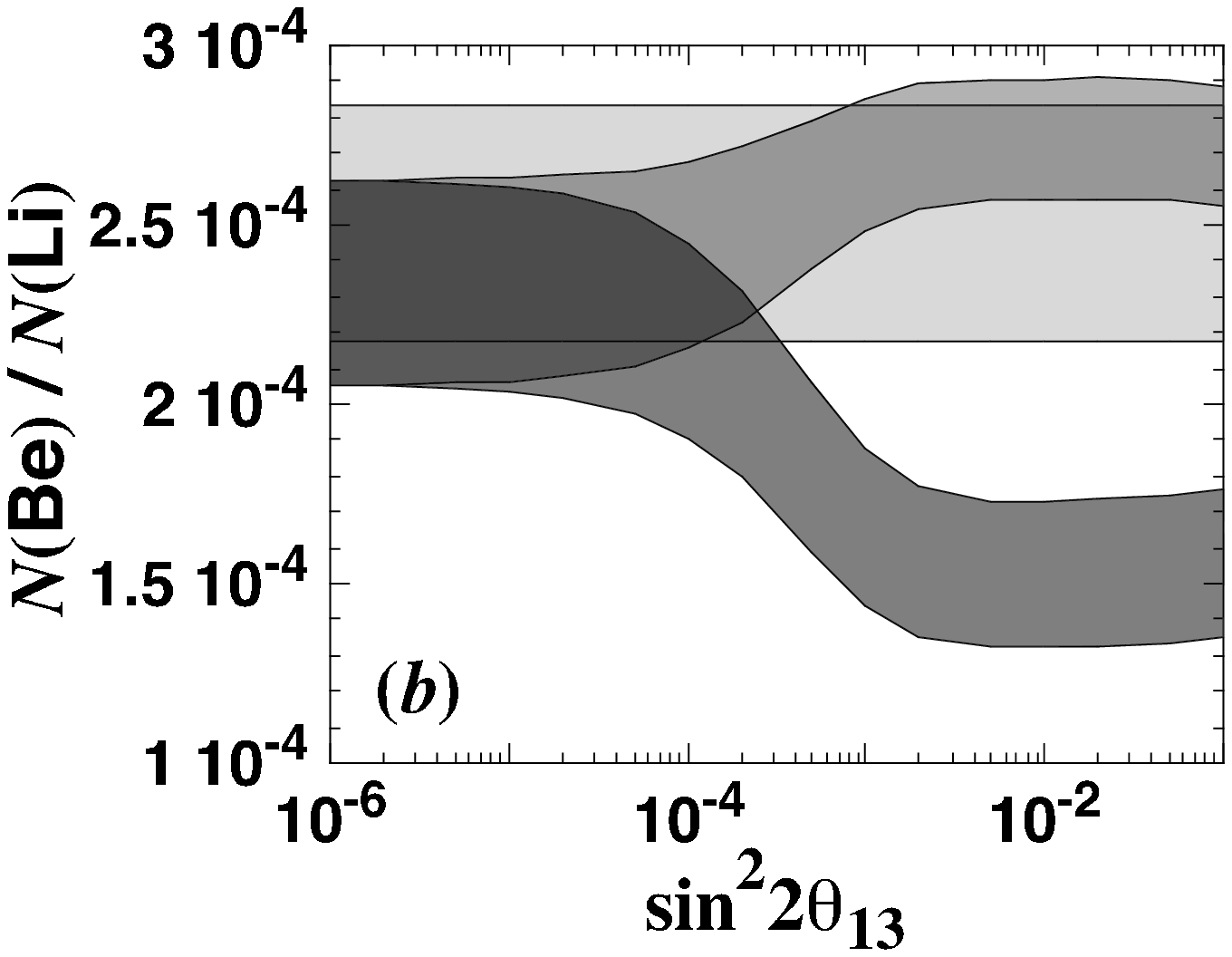}
\epsscale{.50}
\plotone{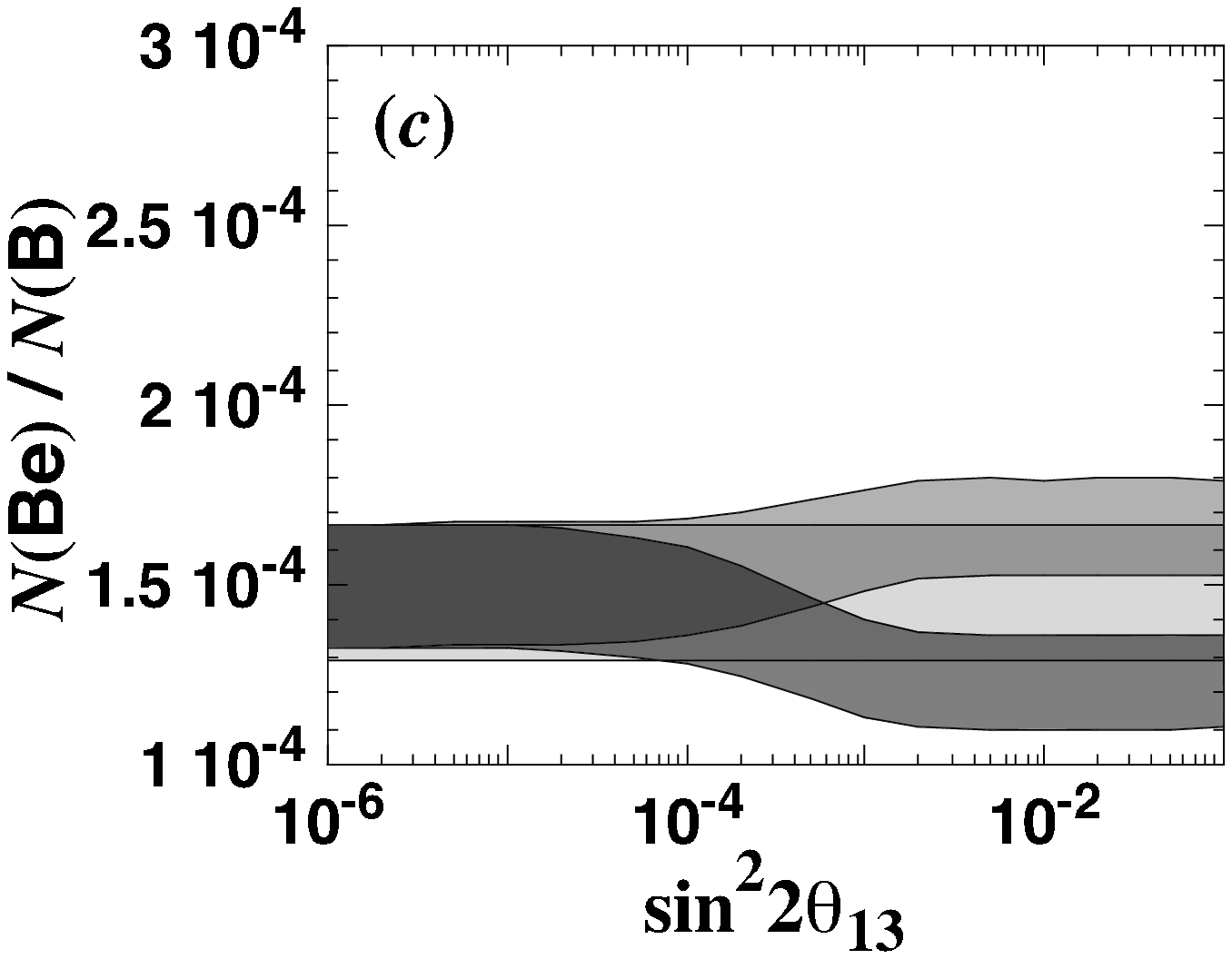}
\caption{
Elemental abundance ratios of ($a$) Li/B, ($b$) Be/Li, and ($c$) Be/B 
as a function of the mixing angle $\sin^22\theta_{13}$.
Dark and medium shaded regions correspond to normal and inverted mass
hierarchies, respectively. The lightly shaded region indicates the ratio 
obtained without neutrino oscillations.
Each range is drawn using the results of models 1, 2, LT, and ST.
}
\label{Elemratio}
\end{figure*}

In a normal mass hierarchy, the $^7$Li/$^{11}$B ratio depends on 
$\sin^22\theta_{13}$. For the non-adiabatic H resonance, the $^7$Li/$^{11}$B 
ratio lies between 0.64 and 0.66; the range is slightly larger than the case
without neutrino oscillations. This slight increase is due to the fact that 
the $^7$Li yield is larger even in non-adiabatic H resonance.
For an adiabatic H resonance, where $\sin^22\theta_{13} \ge 0.002$, 
the $^7$Li/$^{11}$B ratio is in the range of $0.78-0.83$.
The increase is slightly smaller than the one found in \citet{yk06a}.
The value of 0.78 corresponds to model 1.
As discussed in \citet{yk06b}, the variation of the $^7$Li/$^{11}$B ratio 
for a given value of $\sin^22\theta_{13}$ is mainly due to the uncertainties 
of $T_{\nu_e}$ and $T_{\bar{\nu}_e}$.
If the uncertainty of $T_{\nu_e}$ and $T_{\bar{\nu}_e}$ becomes small, 
the range of $^7$Li/$^{11}$B in adiabatic resonance becomes small.

In an inverted mass hierarchy, the $^7$Li/$^{11}$B ratio is not 
distinguishable from the one with normal mass hierarchy or without 
neutrino oscillations.
If the H resonance is non-adiabatic, the $^7$Li/$^{11}$B ratio is 
identical in normal and inverted mass hierarchies.
The $^7$Li/$^{11}$B ratio for an adiabatic H resonance is smaller 
than the one for a non-adiabatic resonance.

For completeness, we also show the elemental abundance ratios of light 
elements in Figure~\ref{Elemratio}.
The Li/B ratio is almost identical to the $^7$Li/$^{11}$B ratio.
This is because most of Li and B are produced as $^7$Li and $^{11}$B, 
respectively. The Be/Li and Be/B ratios are much smaller than the Li/B 
ratio because the Be yield is much smaller than those of Li and B.
The Be/Li ratio shows a dependence on mass hierarchies in the case of
$\sin^22\theta_{13} \ga 2 \times 10^{-3}$.
The Be/Li ratio is smaller than $1.8 \times 10^{-4}$ in a normal mass 
hierarchy. On the other hand, it is larger than $2.1 \times 10^{-4}$ in 
an inverted mass hierarchy. While the $^7$Li abundance increases, the 
$^9$Be abundance becomes slightly smaller due to the destructive reaction 
$^9$Be($p,\alpha)^6$Li. The contribution of $^{12}$C($\nu_e,e^-x)^9$Be does 
not affect the $^9$Be yield because the cross section is very small, even with
the enhancement of the average $\nu_e$ energy (see Figures~\ref{crosssc12sfo}
and~\ref{crosssc12mk}).

The Be/B ratio has a small dependence on mass hierarchies and mixing
angle $\theta_{13}$. The variation of the ratio due to these parameters 
is roughly equal to the uncertainty resulting from the neutrino temperatures.
From the viewpoint of elemental abundance ratios, the Li/B and Be/Li ratios 
depend on mass hierarchies and the mixing angle $\theta_{13}$.

We have shown above that the $^7$Li/$^{11}$B ratio in a normal
mass hierarchy and adiabatic H resonance 
($\sin^22\theta_{13} \ga 2 \times 10^{-3}$) is larger than the one obtained in
the other cases.
This increase is attributed to the effect of neutrino oscillations, and 
remains after taking into account uncertainties in neutrino energy spectra.
Therefore, we confirm with the new $\nu$-process cross sections that the
$^7$Li/$^{11}$B ratio is a promising probe of oscillation parameters.
If we find that the yields of $^7$Li and $^{11}$B produced in supernovae
require a $^7$Li/$^{11}$B ratio larger than 0.78, the mass hierarchy should
be normal, and $\sin^22\theta_{13}$ should be larger than $2 \times 10^{-3}$.
We expect that $^7$Li and $^{11}$B will eventually be detected in stellar 
material, indicating traces of SN material. Supernova remnants are also
promising candidates for this type of abundance constraint on neutrino physics.
There have been attempts to find stars with excesses in $^{11}$B, 
which would provide evidence for direct pollution with supernova ejecta 
\citep[e.g.,][]{rd98,pd98,pd99}.
The $^7$Li/$^{11}$B ratio of pre-solar grains from SNe might also provide
useful information \citep[for presolar grains; e.g.,][]{la05}. 
On the other hand, there are still theoretical uncertainties regarding 
neutrino energy spectra and stellar evolution.
The reduction of these uncertainties will bring about a stronger constraint
on neutrino oscillation parameters.

The observation of supernova neutrino signals just after a supernova 
explosion is one of the most promising methods to constrain unknown
neutrino oscillation parameters.
It is expected that SuperKamiokande will detect more than 1000 neutrinos
if a supernova explodes at the Galactic center \citep[e.g.][]{fl05}.
If detailed energy spectra of the neutrinos emitted from the supernova
are theoretically predicted, the analysis of the observed neutrino spectra
will constrain the oscillation parameters \citep[e.g.]{ds00,tw01}.
The time evolution of the neutrino signal may reveal the change of the
neutrino spectra due to the supernova shock propagation
\citep{ts03,tk04,fl05,km08}.
SuperKamiokande detects electron-type antineutrinos, so that the 
enhancement of the supernova $\bar{\nu}_e$ signal detected by SuperKamiokande
will be evidence for an inverted mass hierarchy and relatively large
value of $\theta_{13}$ ($\sin^22\theta_{13} \ga 10^{-3}$).
The enhancement of the $^7$Li/$^{11}$B ratio will be evidence for 
a normal mass hierarchy and relatively large value of $\theta_{13}$.
Therefore, these two constraints complement each other.

\section{Discussion}

\subsection{Temperatures of the $\nu_e$ and $\bar{\nu}_e$ neutrinos}

The temperatures of the $\nu_e$ and $\bar{\nu}_e$ neutrinos are less 
sensitive to the yield constraints than those of the
$\nu_{\mu,\tau}$ and $\bar{\nu}_{\mu,\tau}$, $T_{\nu_{\mu,\tau}}$ neutrinos.
The temperatures $T_{\nu_e}$ and $T_{\bar{\nu}_e}$ are smaller than
$T_{\nu_{\mu,\tau}}$ if neutrino oscillations are not taken into 
account.
On the other hand, the enhancement of the light element yields by 
neutrino oscillations do depend on $T_{\nu_e}$ and $T_{\bar{\nu}_e}$.
The production through charged-current reactions is more enhanced
when the temperature difference of $\nu_e$ and $\nu_{\mu,\tau}$ or
$\bar{\nu}_e$ and $\nu_{\mu,\tau}$ is larger.
\citet{yk06b} showed that the $^7$Li/$^{11}$B ratio exhibits significant
variation from different values of $T_{\nu_e}$ and $T_{\bar{\nu}_e}$ even 
for a fixed value of $T_\nu$.

Yields of species mainly produced through charged-current
$\nu$-process reactions can be used to constrain the temperatures 
$T_{\nu_e}$ and $T_{\bar{\nu}_e}$. Isotopes of special importance 
are $^{138}$La and $^{180}$Ta \citep{ga01,rh02,hk05}.
The main production process of $^{138}$La and $^{180}$Ta is
$^{138}$Ba($\nu_e,e^-)^{138}$La and $^{180}$Hf($\nu_e,e^-)^{180}$Ta,
respectively.
For $^{180}$Ta, about half of the yield is produced through
$^{181}$Ta($\gamma,n)^{180}$Ta and $^{181}$Ta($\nu,\nu'n)^{180}$Ta.
The GT strength distributions in $^{138}$La and
$^{180}$Ta were recently obtained experimentally \citep{ba07}.
If the temperature of $\nu_e$ could be constrained from the yields of
$^{138}$La and $^{180}$Ta and their observed abundances, the effect
of neutrino oscillations on $\nu$-process nucleosynthesis could be 
evaluated more precisely.

\subsection{Uncertainties in the Rates of $^7$Li and $^{11}$B Production
Reactions}

In the He/C layer, almost all $^7$Li are produced through
$^3$H($\alpha, \gamma)^7$Li and $^3$He($\alpha, \gamma)^7$Be
after the production of $^3$H and $^3$He through the $\nu$-process.
About 60\% of $^{11}$B is also produced through 
$^7$Li ($\alpha, \gamma$) $^{11}$B in the inner region of the He/C layer.
Therefore, the uncertainty of the rates of these reactions affects the yields
of $^7$Li and $^{11}$B.
We discuss this uncertainty briefly.

We adopted the rates of the three reactions from NACRE compilation 
\citep{aa98}.
The rates of $^3$H($\alpha, \gamma)^7$Li and 
$^3$He($\alpha, \gamma)^7$Be are in very good agreement with those
in \citet{cf88}.
For $^7$Li($\alpha, \gamma)^{11}$B, the NACRE compilation
showed slightly larger rate ($\sim 20\%$ in the temperature range between
$2 \times 10^8$ and $4 \times 10^9$ K) than in \citet{cf88}.
The NACRE compilation also showed uncertainties of their reaction rates;
these are 11\%, 17\%, and 16\% for $^3$H($\alpha, \gamma)^7$Li,
$^3$He($\alpha, \gamma)^7$Be, and $^7$Li($\alpha, \gamma)^{11}$B,
respectively.
Therefore, the $^7$Li and $^{11}$B yields would have the uncertainty of
about 17\% due to the uncertainty of the reaction rates.
The uncertainty of the $^7$Li/$^{11}$B ratio would be smaller, because more
than half of the $^{11}$B is produced through the common production sequence
of $^7$Li in the He/C layer.

\subsection{Neutrino-Neutrino Interactions}

It has been pointed out that neutrino-neutrino interactions in regions 
just above a proto-neutron star change neutrino flavors and thus affect 
the neutrino energy spectra.
These interactions contribute diagonal and off-diagonal potential to the
flavor-basis Hamiltonians.
This potential plays a complicated role in flavor exchange due to
the momentum transfer by the interactions.
Analytical evaluation have been carried out in some special cases
\citep[e.g.,][]{qf95,pr02,fq06}.
The change of the neutrino energy spectra for two neutrino flavors 
by neutrino-neutrino interactions has been investigated in neutrino driven 
winds \citep[e.g.,][]{df06a,df06b} and in the density profile of an exploding 
supernova \citep[e.g.,][]{fl07}.
The neutrino-neutrino interactions in three neutrino flavors have very
recently been investigated \citep{df08a,df08b,ep08}.
It has been proposed that these interactions affect the efficiency of 
r-process nucleosynthesis \citep{by05}.
Neutrino-neutrino interactions may also change the locations of resonance
to deeper regions, which could affect the $^7$Li and $^{11}$B yields
in supernovae.

\section{Conclusions}

We evaluated the neutrino-nucleus reaction cross sections for 
$^4$He and $^{12}$C using new shell-model Hamiltonians.
These cross sections are important for the yields of the light
elements Li, Be, and B produced through the $\nu$-process in SNe.
We investigated the nucleosynthesis of the light elements in a SN 
model corresponding to SN 1987A using these new cross sections.
We investigated the dependence of the light element yields on 
cross sections, neutrino energy spectra, and neutrino oscillations.
We obtained the following results.

1. The neutrino-nucleus reaction cross sections are evaluated using
WBP and SPSDMK Hamiltonians for $^4$He and SFO and PSDMK2 Hamiltonians
for $^{12}$C. Main production channels of the $^{12}$C reaction are 
the nuclei $n$, $p$, $^4$He, $^{11}$B, and $^{11}$C. Production of
$^6$Li, $^9$Be, $^{10}$Be, and $^{10}$B is also important.

2. For a given neutrino temperature set, the yields of $^7$Li and $^{11}$B
with the cross sections of WBP+SFO model are larger than those of the
cross sections in \citet{wh90}.
This is mainly due to larger cross sections of $^4$He($\nu,\nu'p)^3$H
and $^4$He($\nu,\nu'n)^3$He.

3. Larger yields of $^6$Li and $^9$Be result from the new cross sections, with
the exception of the yield of $^{10}$B, which is reduced. 
These changes reflect the difference of the cross sections of 
the $\nu$-process for $^{12}$C and deuteron production branches of $^4$He.
Radioactive $^{10}$Be is produced with abundance levels similar to $^6$Li.
The channel for producing $^6$Li and $^{10}$Be have been evaluated in the
$\nu$-process of $^{12}$C.

4. The larger cross sections slightly decrease the range of acceptable neutrino
temperature, as constrained by the  $^{11}$B abundance evolution during GCE.
The range of the neutrino temperature is
$4.3 {\rm MeV} \la T_{\nu_{\mu,\tau}} \la 6.5 {\rm MeV}$ for WBP+SFO model and
$4.0 {\rm MeV} \la T_{\nu_{\mu,\tau}} \la 6.0 {\rm MeV}$ 
for the SPSDMK+PSDMK2 model.

5. The dependence of the $^7$Li and $^{11}$B yields on neutrino oscillation
parameters, such as mass hierarchy and the mixing angle $\theta_{13}$,
is not changed by the new cross sections. Yield enhancements are smaller.

6. The $^7$Li/$^{11}$B abundance ratio depends on mass hierarchy and the
mixing angle $\theta_{13}$, even when considering uncertainties of neutrino
temperatures and the total neutrino energy.
For a normal mass hierarchy and $\sin^22\theta_{13} \ga 2 \times 10^{-3}$,
i.e., adiabatic H resonance, the $^7$Li/$^{11}$B ratio is larger than 0.78.
In the case of an inverted mass hierarchy or the case without neutrino 
oscillations, the $^7$Li/$^{11}$B ratio is smaller than 0.61.
Smaller uncertainty of neutrino temperatures extends the difference in
the $^7$Li/$^{11}$B ratio.

7. The Be/Li abundance ratio is very small. 
A decrease in the Be/Li ratio is seen for a normal mass hierarchy and 
$\sin^22\theta_{13} \ga 2 \times 10^{-3}$.

\acknowledgments

We would like to thank Koichi Iwamoto, Ken'ichi Nomoto, and Toshikazu 
Shigeyama for providing the data for the internal structure of progenitor
model 14E1 and for helpful discussions.
Numerical computations were in part carried out on general common use computer
system at Center for Computational Astrophysics, CfCA, of National
Astronomical Observatory of Japan. 
This work has been supported in part by the Ministry of Education, Culture, 
Sports, Science and Technology, Grants-in-Aid for Young Scientist (B) 
(17740130) and Scientific Research (C) (17540275, 18540290, 18560805), 
Mitsubishi Foundation, and the JSPS Core-to-Core Program, International 
Research Network for Exotic Femto Systems (EFES).

\clearpage

\end{document}